\documentclass[11pt,preprint]{aastex} 

\usepackage{graphicx}
\usepackage{dcolumn}
\usepackage{bm}
\usepackage{amsfonts}
\usepackage{color}
\usepackage{amsmath}
\usepackage{here}

\begin{document}

\shorttitle{Impact of Neutrino Opacities on Supernova Simulations}
\shortauthors{Kotake et al.}

\title{Impact of Neutrino Opacities on Core-Collapse 
Supernova Simulations}

\author{Kei Kotake\altaffilmark{1},
Tomoya Takiwaki\altaffilmark{2}, Tobias Fischer\altaffilmark{3}, 
 Ko Nakamura\altaffilmark{1}, \\
and Gabriel Mart\'inez-Pinedo\altaffilmark{4,5}
\affil{$^1$Department of Applied Physics, Fukuoka University,
 8-19-1, Jonan, Nanakuma, Fukuoka, 814-0180, Japan}
\affil{$^2$Division of Theoretical Astronomy/Center for Computational Astrophysics, National Astronomical Observatory of Japan, 2-21-1, Osawa, Mitaka, Tokyo, 181-8588, Japan}
\affil{$^3$Institute for Theoretical Physics, University of Wroc{\l}aw, plac Maksa Borna 9, 50-204 Wroc{\l}aw, Poland}
\affil{$^4$GSI Helmholtzzentrum f{\"u}r Schwerioneneforschung, Planckstra{\ss}e 1, 64291 Darmstadt, Germany}
\affil{$^5$
Institut f{\"u}r Kernphysik (Theoriezentrum), Technische Universit{\"a}t Darmstadt, Schlossgartenstra{\ss}e 2, 64289 Darmstadt, Germany}}

\begin{abstract}
 Accurate description of neutrino opacities is central both to the core-collapse 
 supernova (CCSN) phenomenon and to the validity of the explosion mechanism itself. 
In this work, we study in a systematic fashion the role of a variety of well-selected
 neutrino opacities in CCSN simulations where multi-energy, 
three-flavor neutrino transport is solved by the isotropic diffusion 
source approximation (IDSA) scheme. To verify our code,
 we first present results from one-dimensional (1D) simulations
 following core-collapse, bounce, and 
up to $\sim 250 $ ms postbounce of a $15 M_{\odot}$ star using a standard
set of neutrino opacities by Bruenn (1985). 
Detailed comparison with published results 
supports the reliability of our three-flavor 
IDSA scheme using the standard opacity set. We then investigate in 1D 
simulations how the individual opacity update 
 leads to the difference from the base-line run with the 
 standard opacity set. By making a detailed comparison with previous work, we check
 the validity of our implementation of each update in a step-by-step manner.
 Individual neutrino opacities with the largest impact on the overall evolution
 in 1D simulations are selected for a systematic comparison in our 
two-dimensional (2D) simulations. Special emphasis is devoted to the 
criterion of explodability in the 2D models. We discuss the implications of these results 
as well as the limitations and requirements for 
 future towards more elaborate CCSN modeling.
\end{abstract}

\maketitle

\section{Introduction}
\label{intro}
 A core-collapse supernova (CCSN) is triggered when the core of a massive star 
becomes gravitational unstable, 
predominantly due to electron captures on protons bound in heavy iron-group nuclei. The collapse 
proceeds supersonically until the central density exceeds nuclear saturation density, 
when the repulsive nuclear force balances gravity such that the core bounces back 
with the formation of a hydrodynamics shock wave. This bounce shock propagates quickly 
to radii on the order of 100~km. However, the shock eventually turns into an 
accretion front due to losses from neutrinos, when propagating across the 
neutrinosphere of last scattering, as well as the continuous dissociation 
of infalling heavy nuclei from the still gravitationally unstable layers 
above the stellar core. The revival of this standing accretion shock 
is subject to the so-called supernova problem, i.e. the liberation of 
energy available at the interior of the proto-neutron star (PNS), 
the central hot and compact object, into a thick layer of accumulated material 
behind the shock front (for details, see \citet{Janka07} for review).

 Neutrinos play crucial roles during all phases of CCSNe - 
 stellar core collapse, core bounce, postbounce  mass accretion, 
onset of the explosion and PNS deleptonization, until cooling of neutron
 stars (NSs) (e.g., \citet{woosley02,LM03,janka17,yakov04} for detailed reviews). 
 In particular, during the stellar core collapse nuclear electron-capture rates 
determine the deleptonization \citep[cf.][]{hix03,langanke03}, which in turn defines 
the location of the bounce shock.
 After bounce, the huge gravitational energy stored in the 
proto-neutron star (PNS) is almost completely carried away by neutrinos. 
A tiny fraction of the 
 streaming neutrinos deposits energy in the postshock material
 via weak interactions, leading to an explosion
 in the neutrino-driven mechanism of CCSNe
 \citep{Colgate66,Wilson85,Bethe85}. 

Multi-dimensional (multi-D) hydrodynamic instabilities play a crucial role in the 
neutrino mechanism. Non-linear turbulent 
flows associated with convective overturn and the standing-accretion-shock-instability 
(SASI, \citet{blondin03}) increase the neutrino heating efficiency 
in the gain region, essentially aiding the explosion (see \citet{janka17a,bernhard16,
hix16,thierry15,burrows13,Kotake12_ptep} for reviews). 
In fact, a growing number of neutrino-driven models
 have been reported so far in self-consistent two-dimensional (2D) 
simulations, which has supported the validity of the multi-D neutrino-driven mechanism
(e.g., \citet{Marek09,Suwa10,BMuller12b,Dolence14,Nakamura15,Pan16,oconnor15,
Burrows16,Summa16,nagakura17,bernhard17}).

This success is, however, highlighting new questions. 
The most challenging self-consistent three-dimensional (3D)
 simulations with spectral neutrino 
transport have failed to produce explosions for $11.2$, $20.0$, and 
$27.0 M_{\odot}$ progenitors (\citet{Hanke13,Tamborra14}, see, however,
 \citet{Roberts16} for an exploding 27 $M_{\odot}$ model). In a few
 successful cases, the explosions are more delayed
 in 3D than in 2D (e.g., \citet{Lentz15} and \citet{melson15b}),
 leading to smaller explosion energies in 3D 
 compared to 2D (\citet{Takiwaki14}). A few exceptions from this trend 
have been reported for 9.6 and 11.2 $M_{\odot}$ stars
 \citep{melson15a,bernhard15}. However, these two progenitors close to
 the low-mass end of the SN progenitors, may be rather peculiar in the sense that 
 the 9.6 $M_{\odot}$ star has a tenuous envelope and
 explodes even in the one-dimensional (1D) simulation, that the 11.2 $M_{\sun}$ star is seemingly 
very marginal to produce 3D explosion \citep{bernhard15,takiwaki12,Takiwaki14} or 
not \citep{Hanke13} with its progenitor's compactness parameter being smallest
 among 101 solar-metallicity progenitors in \citet{woosley02}.

One of the prime candidates to enhance the 
 "explodability" is to update neutrino physics in the multi-D models.
 \cite{horowitz02} pioneeringly pointed out that the contribution of 
 strange quarks to neutrino-nucleon scattering can affect neutrino opacity by
 $\sim 10 - 20 \%$ (see also \citet{kolbe98}). In fact, \citet{melson15b} obtained 3D explosions of a $20 M_{\odot}$ star only when the strageness effect was taken into account\footnote{
In order to clearly see the effect, \citet{melson15b} chose a relatively high 
strangeness contribution ($g^s_a = - 0.2$) to the axial vector coupling constant 
($g_a \approx 1.26$) compared to the constraint ($g^s_a \lesssim - 0.1$) proposed by 
\citet{hobbs16}.}. \citet{Burrows16} reported in their 
 2D self-consistent simulations of a 20 $M_{\odot}$ star that  
 many-body corrections to neutrino-nucleon scattering (e.g.,
 \citet{horowitz17,BS99,BS98} albeit in the different context) 
make explosions easier, increase explosion energy, and shorten the time to explosion.

Other intriguing possibilities to impact the CCSN explodability
 include general relativity
 (GR, e.g., \citet{BMuller12b,KurodaT12,KurodaT16,Roberts16}), 
stellar rotation (e.g., \citet{yamasaki08,Marek09,Suwa10,Nakamura14,takiwaki16,summa17,remi17}), 
magnetic fields (e.g., \citet{Kotake06,endeve12,jerome15,masada15,martin17}), 
and inhomogeneities in the progenitor's burning shells 
(e.g., \citet{couch_ott,couch15,bernhard15,abdikamalov16,bernhard17}). 

Joining in the effort to update neutrino physics in 
CCSN codes, we investigate in this study impact of neutrino opacities in 1D and 2D
 core-collapse simulations where three-species neutrino transport is solved by the 
isotropic diffusion source approximation scheme (IDSA, \citet{idsa}). We 
 first start with 1D simulations where we use the same input physics and the 
 same equation of state 
 (EOS) as those in \citet{Liebendorfer05}. 
In the seminal work,
  detailed comparison between the two reference codes between 
{\tt Agile-BOLTZTRAN} \citep{Liebendorfer04,tony93c} and {\tt VERTEX(-PROMETHEUS)} \citep{Rampp02} 
was made. Their results are available online\footnote{We use data downloadable 
 from \hbox{http://iopscience.iop.org/0004-637X/620/2/840/fulltext/datafiles.tar.gz}},
 so that we are able to make a comparison with their data set. In the original IDSA
 scheme \citep{idsa}, a base-line set of neutrino opacities \citep{Bruenn85} 
(often referred to as the Bruenn rate) is used.
 Following the implementation schemes of microphysical update 
 in the literature (e.g., \citet{Buras06a,tobias09,fischer16}), we study
 how individual update in the neutrino opacity leads to differences from
 the base-line run with the Bruenn rate. 
From these systematic 1D runs, we could have a guess which 
update could potentially help (or harm) the 
 explodability in multi-D models. Then we move on to perform 
2D simulations where several sets of
 neutrino opacities are included. This is because a full investigatation of 
the individual rates is currently too computationally expensive to do
 in multi-D simulations.

This paper is organized as follows. In Section \ref{sec2}, we summarize
 numerical methods including model setup, neutrino opacities, and the
 three-flavor IDSA scheme. In Section \ref{sec3},
 we first compare our results with \citet{Liebendorfer05} (Section \ref{sec3-1}).
 Then we proceed to study impact of the individual (updated) rates in 1D runs
 one by one from Section \ref{sec3-2} to \ref{set6}.
Then it may be interesting to see which expectations regarding the explodability of the 
1D models survive when we perform 2D simulations (Section \ref{sec4}).
We summarize our results and discuss its implications in Section \ref{sec6}.

\section{Numerical methods}\label{sec2}

\subsection{Model setup}\label{sec2-1}
In our 1D simulations, we employ a standard 15$M_{\odot}$ progenitor
 \citep[``s15s7b2'' in][]{WW95}, following the work by \citet{Liebendorfer05}. 
To see effects of individual neutrino rates, we follow the dynamics starting 
 from core collapse, through bounce, up to $\sim 500$ ms postbounce (pb) in each 1D 
run. 
In our 2D runs, we choose a 20$M_{\odot}$ 
progenitor model of \citet{Woosley07} that has been widely used 
 in recent multi-D simulations (e.g., \citet{melson15b,Burrows16,Bollig17}). 
This progenitor is characterized by high explodability where the neutrino-driven 
shock revival was obtained around $\sim 250$ ms (pb) at the earlist 
in the literature
(e.g., \citet{melson15b,Summa16,oconnor15}).

Our non-relativistic hydrodynamics code employs a high-resolution 
shock capturing scheme with an approximate Riemann solver of \citet{hlle} 
(see \citet{Nakamura15} for more details).
Self-gravity is computed by a monopole 
approximation with an approximate treatment of GR gravity by the effective
 potential of Case A of \cite{Marek06}. Our 1D and 2D runs are 
computed on a spherical polar grid with a resolution of 
 $n_r = 512$, and $n_r \times n_{\theta} $ = $512 \times 128 $, respectively.
 Non-equally spacial radial zones covers from the center to an outer boundary 
of $5\times 10^8$ cm. The radial grid is chosen such that the resolution $\Delta
 r$ is better than 250m in the PNS interior and
 typically better than 1km outside the PNS. 
Seed perturbations for aspherical 
instabilities are imposed by hand at 10 ms after bounce by introducing
 random perturbations of $1\%$ in velocity behind the stalled shock.
For the spectral transport, we use 20 logarithmically spaced energy bins 
ranging from 3 to 300 MeV. In our multi-D runs, we take into account the 
 energy feedback from nuclear burning processes into hydrodynamic evolution
 by solving a 13-species $\alpha$-nuclei network (see \citet{Nakamura14a} for details).

Throughout the paper, we use the EOS of \citet{LSEOS} (LS).
Only in Section \ref{sec3-1}, we set the incompressibility 
parameter of $K=180$ MeV (LS180) for the sake of comparison 
with \citet{Liebendorfer05}, whereas in all the other models, 
we set $K=220$ MeV (LS220) that can 
account for the 2 $M_{\odot}$ NS mass measurements \citep{Demorest10,antoniadis}\footnote{
A detailed comparison of the role of the nuclear EOS, including LS, in CCSN 
simulations was reported in \cite{hempel12} and \citet{fischer14}.}.

\subsection{Neutrino opacities} \label{2-2}
 Regarding neutrino opacities, our base-line model (set1, see Table \ref{table1}) 
employs the standard weak interaction set given 
in \citet{Bruenn85} plus nucleon-nucleon bremsstrahlung 
\citep{Hannestad98} (see also \citet{RamppPhd} for detailed implementation schemes).
 Note in set1, ion-ion correlations for neutrino scattering on heavy nuclei \citep{Horowitz97} 
and the correction form factor \citep{tony93c,RamppPhd} are also included. 
In 1D runs, all the following update is basically added individually to the set1.
 
 In set2 (see Table \ref{table1}), electron capture (EC) rate on nuclei in set1 
 \citep{FFN} is replaced with the currently most elaborate one by \citet{juoda} which is a
 significant extension of the EC rate by \citet{langanke03} (see Section \ref{sec2}). 
In set3, electron neutrino pair annhilation into $\mu/\tau$ neutrinos 
(set3a in Table \ref{table1}) and $\mu/\tau$-neutrino 
scattering on electron (anti)neutrinos (set3b) \citep{Buras03} 
is added to set1, respectively (see Appendices
 \ref{appA} and \ref{appB} for details). In set4a, medium modification to 
electron(/positron)
 capture reactions on proton(/neutron) are taken into account 
\citep{GMP12,roberts12,hempel15,roberts17} at the mean-field level \citep{reddy98} 
(see Section \ref{set4}). Set4b includes medium dependent suppresion 
of Bremsstrahlung (\citet{fischer16}, their Equation (11)).

In set5a, inelastic contributions and weak magnetism corrections are included following
 \citet{horowitz02} for the charged current absorption and neutral current scattering 
 processes. Set5b includes correction to the effective nucleon mass \citep{reddy99}.
  Following \citet{Buras06a} (their Equation (A.1)), we replace
 the nucleon mass ($m_{N}$) with the density-dependent 
 nucleon mass ($m^*_{N}(\rho)$), which 
 accordingly changes the neutrino opacities.

In set6a, quenching of the axial-vector coupling constant at high densities 
\citep{carter02} (e.g., Equation (A.9) in \citet{Buras06a}) is included but 
using more recent fitting formula
 (e.g., Equation (8) in \citet{fischer16}). In set6b, we employ the formulas 
suggested by \citet{horowitz17} for the neutral-current axial 
response that accounts for virial effects at low density and many-body correlations 
at high densities (their Equations (36-39)). Finally in set6c, 
a strangeness-dependent contribution to the axial-vector coupling constant 
\citep{horowitz02} with $g^s_A = - 0.1$ \citep{hobbs16}
 is considered for neutrino-nucleon scattering.
  
Note that even the full set in Table \ref{table1} is no way complete. Inclusion of 
 muons significantly effects explodability \citep{Bollig17} and a proper treatment of 
nucleon kinematics \citep{reddy98} is not taken in account in our full set 
(roles of nuclear de-excitation \citep{fischer13} and light nuclear clusters \citep{sumi08} 
are also not included yet). 
These updates are another major undertaking, which we leave 
for future work.

\begin{table}[htpb]
\begin{center}
\begin{tabular}{cl l}
\hline\hline
Model & Weak Process or Modification & References\\
\hline
set1 &$\nu_e\, n \rightleftharpoons e^- \,p$ &  \cite{Bruenn85} \\
     &$\bar{\nu}_e\,p \rightleftharpoons e^+\,n$ &   \cite{Bruenn85}\\
     &$\nu_e\,A' \rightleftharpoons e^-\,A$ &  \cite{Bruenn85} \\
     &$\nu\,N \rightleftharpoons \nu\, N$ &  \cite{Bruenn85} \\
     &$\nu\,A \rightleftharpoons  \nu\,A$   &  \cite{Bruenn85},\cite{Horowitz97} \\
     & $\nu\,e^{\pm} \rightleftharpoons \nu\, e^{\pm}$   &  \cite{Bruenn85} \\
     & $e^-\,e^+ \rightleftharpoons \nu\, \bar\nu$ &  \cite{Bruenn85}  \\
     & $NN \rightleftharpoons \nu \bar\nu NN$  & \cite{Hannestad98} \\
\hline
set2 &$\nu_e\,A \rightleftharpoons e^-\,A'$ &  \cite{juoda} \\
\hline
set3a & $\nu_e + \bar\nu_e \rightleftharpoons \nu_x +\bar\nu_x$ & \cite{Buras03,tobias09} \\
set3b & $\nu_x + \nu_e (\bar{\nu_e}) \rightleftharpoons \nu'_x + \nu'_e (\bar{\nu}'_e)$ &\cite{Buras03,tobias09} \\
\hline
set4a & $\nu_e\, n \rightleftharpoons e^- \,p$,\,\, $\bar{\nu}_e\,p \rightleftharpoons e^+\,n$& 
\cite{GMP12} \\
set4b & $NN \rightleftharpoons \nu \bar\nu NN^*$  & \cite{fischer16} \\
\hline
set5a &$\nu_e\, n \rightleftharpoons e^- \,p$,\,\, $\bar{\nu}_e\,p \rightleftharpoons e^+\,n$,
$\nu\,N \rightleftharpoons \nu\, N$ & \cite{horowitz02} \\
set5b & $m_{N}  \rightarrow m_{N}^{*}$ & \cite{reddy99} \\
\hline
set6a & $g_A \rightarrow g^*_A$ & \cite{fischer16} \\
set6b &  $\nu\,N \rightleftharpoons \nu\, N$ (Many-body and Virial corrections) & 
 \cite{horowitz17} \\
set6c & $\nu\,N \rightleftharpoons \nu\, N$ (Strangeness contribution) & \cite{horowitz02} \\
\hline
\end{tabular}
\caption{Summary of neutrino opacity input in our 1D runs with the references.
 The symbols $e^-$, $e^+$, $n$, $p$, and $A$ denote electrons, positirons, free neutrons
 and protons, and heavy nuclei respectively; the symbol $N$ means $n$ or $p$.
 $m_N$ denotes nucleon mass and the quantity with $*$ indicates the one with in-medium correction. $\nu$ in the neutral current 
reactions represents all species of neutrinos ($\nu_e,\bar\nu_e,\nu_{x}$) with $\nu_x$ representing heavy-lepton neutrinos ($\nu_{\mu},\nu_{\tau}$) and their antiparticles. 
}
\label{table1}
\end{center}
\end{table}

\subsection{Three-flavor IDSA scheme} \label{sec2-3}

The IDSA scheme splits the neutrino distribution fuction ($f$) 
into two components ($f = f^{\rm t} + f^{\rm s}$) with $f^{\rm t}$ and $f^{\rm s}$
representing streaming and trapped neutrinos, both of which 
are solved using separate numerical techniques (see \citet{idsa} for detail).
In the original (two-neutrino-flavor) IDSA scheme, a steady-state approximation 
($\partial {f^{\rm s}(\epsilon)}/(\partial t) = 0$) is assumed for the streaming
 neutrinos where $\epsilon$ represents
 the neutrino energy in the comoving frame. 
Then one should deal with
a Poisson-type equation to find the solution of $f^s$ 
(e.g., Equation (10) in \citet{idsa}). This is relatively 
computationally expensive especially 
 in multi-D simulations.

To get around the problem,
 we directly solve the evolution of the streaming neutrino (e.g., 
Equation (1) of \citet{Takiwaki14}). In
 this work, we further incorporate GR effects approximately following \citet{Rampp02,oconnor15}) as,
\begin{eqnarray}
\frac{\partial \mathcal{E}^{\rm s}}{c\partial {t}} &+&
\frac{1}{r^2}\frac{\partial}{\partial r}\,\alpha r^2 \mathcal{F}^{\rm s}
= \mathcal{S}[{\hat{j}}, {\hat{\chi}}, {\Sigma}] - \alpha \mathcal{F}^{\rm s} 
\frac{\partial \phi/c^2}{\partial r},\label{eq:fs-evol}\\
\mathcal{E}^{\rm s} &\equiv& \frac{\epsilon^3}{(2 \pi \hbar c)^3}\, \frac{1}{2}\int d{\mu}\,{f}^{{\rm s}},\\
\mathcal{F}^{\rm s} &\equiv& \frac{\epsilon^3}{(2 \pi \hbar c)^3}\, \frac{1}{2}\int \mu d{\mu}\,{f}^{{\rm s}},\\
\mathcal{S} &\equiv& - \alpha \left( \hat{j}+\hat{\chi}\right)\mathcal{E}^{\rm s}+\Sigma,
\end{eqnarray}
where $\mathcal{E}^{\rm s}$ and $\mathcal{F}^{\rm s}$ corresponds to the 
 radiation energy and flux of the streaming particle, and $\mathcal
S$ represents the source term that is a functional 
of the effective neutrino emissivity ($\hat{j}$), absorptivity ($\hat{\chi}$), and the 
isotropic diffusion term ($\Sigma$) all defined in the 
laboratory (lab) frame, respectively. Note $\phi$ is the gravitational potential and 
 $\alpha = \exp(\phi/c^2)$ is the GR correction where $c$ is the speed of light. 
For closure, we use a prescribed relation between the radiation energy and
 flux as ($\mathcal{F}^{\rm s}/\mathcal{E}^{\rm s}= \frac{1}{2}(1 +\sqrt{1-
[{R_{\nu}}/\max\left(r,R_\nu\right)]^2})$ with $R_\nu$ being
 the radius of an energy-dependent scattering sphere 
(see Equation (11) in \citet{idsa}). 
Since the cell-centered value of the flux, $\mathcal{F}^{\rm s}$, is
obtained by the prescribed relation,  the cell-interface 
value is estimated by the first-order upwind scheme assuming that 
the flux is out-going along the radial direction. 
With the numerical flux, the transport equation 
 of $\mathcal{E}^{\rm s}$ (Equation (1)) is now expressed in a hyperbolic form.
 The velocity dependent terms ($O(v/c)$) are only
 included (up to the leading order) in the trapped part of the
 distribution function (Eq.
  (15) in \citet{idsa}).

 For the three-species neutrinos considered in this work ($\nu \in \nu_e, \bar\nu_e, \nu_{x}$), we take into account the collisional kernels up to 
the zeroth-order expansion with respect to the scattering angle 
(for example, $\Phi^{p/a}_{0,{\rm TP}}$ in the case of neutrino pair production from pair annihilation (TP), see Equation (C62) \citep{Bruenn85})\footnote{Consideration up to
 the first-order angular expansion is technically possible, which however makes 
long-term IDSA simulations unstable at this stage.}. To be more specific, 
neutrino-electron scattering (NES) and TP of the Bruenn 
rate both of which were neglected in the 
 original IDSA scheme are now added to the effective emissivity and absorptivity in the 
 source terms ($\mathcal{S}$ and $\Sigma$) as,
\begin{eqnarray}
 \hat{j} + \hat{\chi} &=& j(\epsilon) + \frac{1}{\lambda(\epsilon)} - A^0_{\rm NES}(\epsilon) - A^0_{\rm TP}(\epsilon),
\label{eq1}\\
\Sigma &=& \min\{\max[\alpha_{\rm diff} + \alpha(\hat{j} + \hat{\chi})\mathcal{E}^s),0], \,\alpha
\,\frac{\epsilon^3}{(2 \pi \hbar c)^3} (j(\epsilon) + C^0_{\rm NES}(\epsilon) + C^0_{\rm TP}(\epsilon))\},
\label{eq2}
\end{eqnarray}
where  $j(\epsilon)$ and $\frac{1}{\lambda(\epsilon)}$ represents the emissivity and 
 absorptivity of charged current interactions (e.g., for the first three-line reactions in 
Table \ref{table1}, see also Equations (A12) of \citet{Bruenn85}), the exact
 expression of $A^{0}_{\rm NES}$, $A^{0}_{\rm TP}$, $C^{0}_{\rm NES}, C^{0}_{\rm TP}$ is given in Equation (A34), (A43), (A36), and (A45) in \citet{Bruenn85} and 
of $\alpha_{\rm diff}$ is in Equation (7) in \citet{idsa}, respectively. In this work, 
heavy-lepton neutrino emission from TP, Bremsstrahlung, electron pair neutrino 
annihilation (10th line reaction in Table \ref{table1}),
 and by neutrino-neutrino scattering (11th in Table \ref{table1}) are all treated 
 as the effective emissivity and absorptivity up to the zeroth moment of the
 neutrino production kernels (like by adding terms of $A^0$ and $C^0$ to
 Equations (\ref{eq1}) and (\ref{eq2})). As we will show later, this approximation 
works well at least in the postbounce accretion phase.
However, consideration of higher moments \citep{Pons98}, full set of 
 velocity dependent terms (equivalently treatment of full energy-group couplings in the 
 transport equations), and neutrino-flavor 
coupling is surely needed for more sophisticated simulations (e.g., 
\citet{Rampp02,Thompson03,Liebendorfer04,Sumiyoshi05,hubney07,Buras06a,lentz12a,BMuller12a,KurodaT16,nagakura17}). 

In our multi-D simulations, 
we apply a ray-by-ray approach where the neutrino transport is solved along a 
given radial direction assuming that the hydrodynamic medium for the direction is 
spherically symmetric. This also remains to be updated with more advanced schemes
 (e.g., \citet{skinner,nagakura17}).

\section{1D Results}\label{sec3}

Following the seminal code comparison work by \citet{Liebendorfer05}, we first make a
 quick comparison between our results from the three-flavor IDSA scheme and the results 
 from the two reference codes, {\tt Agile-BOLTZTRAN} \citep{Liebendorfer04} and 
{\tt VERTEX} \citep{Rampp02}. 
{\tt Agile-BOLTZTRAN} solves the full GR neutrino Boltzmann equation 
 with the $S_n$ method in spherically symmetric Lagrangian mesh, 
whereas VERTEX is an Eulerian code that solves the moment equations of a model 
Boltzmann equation by the VEF method in the Newtonian hydrodynamics plus a modified GR
 potential with Case R of \citet{Marek06}. 

\subsection{Comparison with 1D base-line simulations}\label{sec3-1}

\begin{figure}[H]
\begin{center}
\includegraphics[width=160mm]{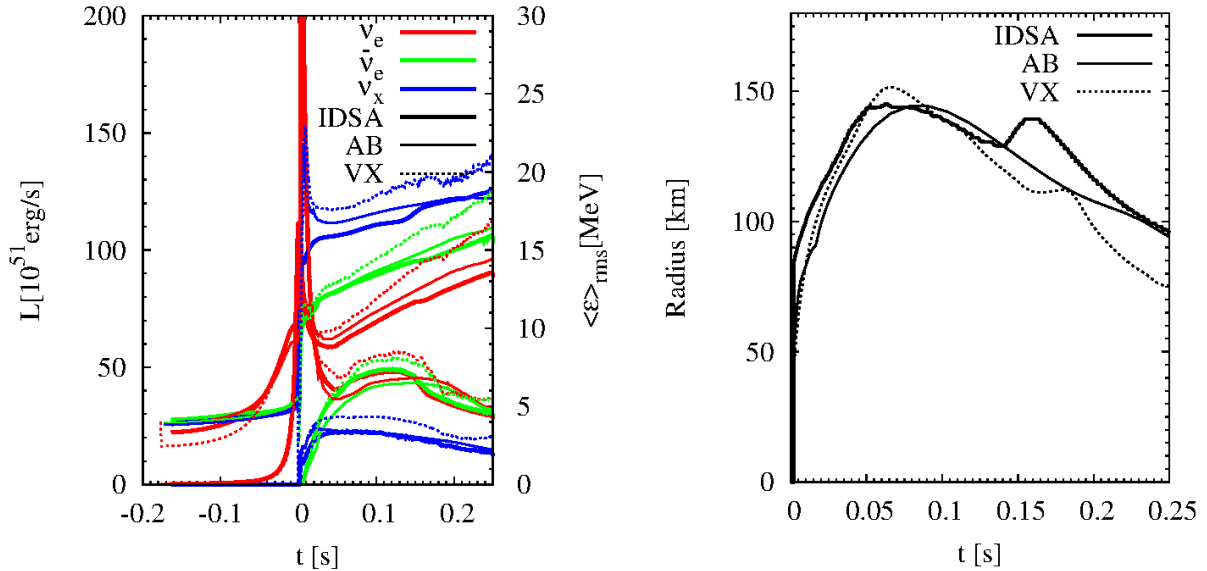}
\end{center}
  \caption{Left panel shows comparison of neutrino luminosities and rms energies as
a function of postbounce time (similar to panel (b) in \citet{Liebendorfer04}). The values are sampled at a radius of 500 km
 in the lab frame. The line width distinguishes between 
 the results IDSA (thick lines), {\tt Agile-BOLTZTRAN} (labeled as "AB", thin lines),
 and {\tt VERTEX} (labeled as "VX", dotted lines). Thick red, green, and blue line
 denotes $\nu_e$, $\bar{\nu}_e$, and $\nu_x$, respectively. Same as the left panel, but 
the right panel shows comparison of the accretion shock front between IDSA (thick line), AB (thin line), and VX (dotted line).}
\label{f1}
\end{figure}

We first present Figure \ref{f1} that is plotted in a similar way as
 Figure 10 in \citet{Liebendorfer05}, showing
 the comparison of key neutrino quantities 
(left panel) and the shock radius (right panel) between IDSA (thick lines), 
{\tt Agile-BOLTZTRAN} (labeled as "AB" in short, thin lines), and {\tt VERTEX} (labeled as "VX", dotted lines), respectively. Note in the left panel that all luminosities ($L_{\nu}$) 
and root-mean-squared (rms) energies ($\langle \epsilon^2 \rangle^{1/2}$) 
are sampled at a radius 500 km in the lab frame. The data from the two reference
 codes are originally given 
in the fluid frame, which is converted by the 
following relations $L_{\nu} =  L^{\rm fluid}_{\nu} (1 + v_r/c)/(1 -v_r/c)$ with $v_r$ 
the radial velocity, $\langle \epsilon^2 \rangle^{1/2} = (\langle \epsilon^2 
\rangle^{\rm fluid})^{1/2} W (1 + v_r/c)$ with $W = 1/\sqrt{1 - (v_r/c)^2}$ 
(e.g., Equations (56)-(58) in \citet{BMuller10}). Here we take $v_r = -0.06 c$ that is 
the (average) infall velocity at 500 km over the entire 250 ms postbounce 
 in \citet{Liebendorfer05}.
Improvements to {\tt VERTEX} 
after \citet{Liebendorfer05} and the follow-up detailed comparison work \citep{BMuller10} 
 suggests that the data from {\tt Agile-BOLTZTRAN} are currently one of 
 the best reliable ones.

From the left panel of Figure \ref{f1}, one can see that the neutrino properties from
IDSA are much closer to those in AB ({\tt Agile-BOLTZTRAN}) than in VX 
({\tt VERTEX}). Here it should be emphasized that the higher neutrino luminosities and rms energies of VX are now lowered to meet closely with those of AB by changing the approximate GR 
treatment from Case R to Case A \citep{Marek06}, the latter of which is employed in this work. In the sense, the qualitative agreement of the three codes is convincing, 
which we will 
explain more in detail below. 

More quantitatively, the peak luminosity during the electron neutrino burst (red lines,
 out of the scale of the left plot)
 is higher by ($\sim 8 \%$) for IDSA compared to that of 
AB ($L_{\nu_e}^{\rm peak} = 3.3 \times 10^{52}$ erg/s), whereas the half-width of the peak
($\sim 6$ ms) agrees well with each other. After the deleptonization burst, the 
luminosity of $\nu_e$ (red think line) and $\bar\nu_e$ (green thick line) of IDSA
 are higher (maximally by $10 \%$) than those of AB till the first $\sim$ 160 ms 
after bounce. As already pointed out by \citet{BMuller10} and \citet{GR1D}, this
 is most likely to come from the higher resolution at the shock front 
in the Eulerian codes compared to the Lagrangian code of AB. 
After the neutronization burst, the maximum luminosity of $\nu_e$ and $\bar \nu_e$ 
deviates maximally by $\sim 8 \%$ between IDSA and AB. And the luminosity of each 
 neutrino species between IDSA and AB points to a converged value towards
the final simulation time (250 ms after bounce in this comparison).

The $\nu_x$ luminosity agrees quite well between IDSA (blue thick line) and AB (blue 
thin line) over the entire 250 ms postbounce. On the other hand, IDSA fails to reproduce 
the spike in the $\nu_x$ rms energy near bounce ($t \sim 0$) peaking at
$\sim 24$ MeV, which is present in both AB and VX (blue thin and blue dotted line). 
This is one of the limitation of the IDSA scheme 
(M. Liebend\"orfer in private communication), which attempts to bridge the streaming 
 and trapped neutrinos by the prescribed isotropic diffusion source term. 
For $\nu_e$ and $\bar\nu_e$, the
 neutrino sphere(s) is well defined as the decoupling surface from 
 thermal equilibrium mediated by charged current interactions. The IDSA works
 well to capture this energy sphere. For $\nu_x$, on the other hand, not the 
energy sphere(s) but the scattering sphere(s) is the decoupling surface
 \citep{raffelt12}. 
In the transition region from the energy to scattering spheres, Doppler-shift 
 terms (as well as the gravitational redshift) play an essential role 
 in accurately determining the neutrino spectrum. By design, the IDSA (in the current form)
 cannot treat the highly complex transport phenomena appropriately. Moreover, 
the energy-redistribution in the neutrino phase space such as due to 
neutrino-electron scattering cannot be treated accurately in the current 
effective emissivity/absorptivity approaches (see Section \ref{sec2-3}).
 All of these simplifications should potentially lead to the missing of the spike 
in the $\nu_x$ energy near bounce. 
  
The $\bar\nu_e$ (rms) energy (green thick line) is in good agreement with AB over the 
 entire 250 ms, although IDSA underpredicts the $\nu_e$ energy (red thick line)
 by $\sim 6 \%$ compared to AB.
 After bounce, IDSA underestimates the $\nu_x$ energy (blue thick line) by
 $\lesssim 10 \%$ compared to AB till $\sim$ 160 ms after bounce, then matches 
closely to AB during the simulation time. The transition timescale ($\sim$ 160 ms)
 corresponds to the time
 when the silicon (Si)-rich shell
 accretes through the shock (e.g., middle left panel of Figure \ref{f5}).
 This can be seen as a hump (solid black line) 
in the right panel of Figure \ref{f1}, which leads to a drop in the $\nu_e$ and 
 $\bar\nu_e$ luminosities (see red and green solid lines in the left panel).
Note that the hump is missing in AB (right panel), which was
 due to the artificial diffusion introduced in the 
adaptive gridding of AB as already discussed 
 in \citet{Liebendorfer05}.

More detailed code comparison not only between AB, VX, and IDSA, but also 
 including Fornax \citep{skinner}, FLASH \citep{oconnor15}, GR1D \citep{GR1D} codes 
 is currently in progress (O'Connor et al. in preparation). 
We leave more detailed comparison to the forthcoming work.
Given our approximate treatment of GR, neglect of energy-bin couplings, 
and the partial implementation of the Doppler-shift terms,
 it may not be surprising that IDSA has $10 \%$ 
levels of mismatch with the full-GR and full-Boltzmann result of {\tt Agile-BOLTZTRAN}.
   In fact, such
 discrepancies (given the use of similar level of approximation) has been also 
observed in the literature (e.g., \citet{GR1D,oconnor15}). 
Having overviewed the validity and limitation of the 
 current IDSA scheme, we shall move on to focus on the microphysical update 
in the following sections.

\begin{figure}[H]
\begin{center}
\includegraphics[width=70mm]{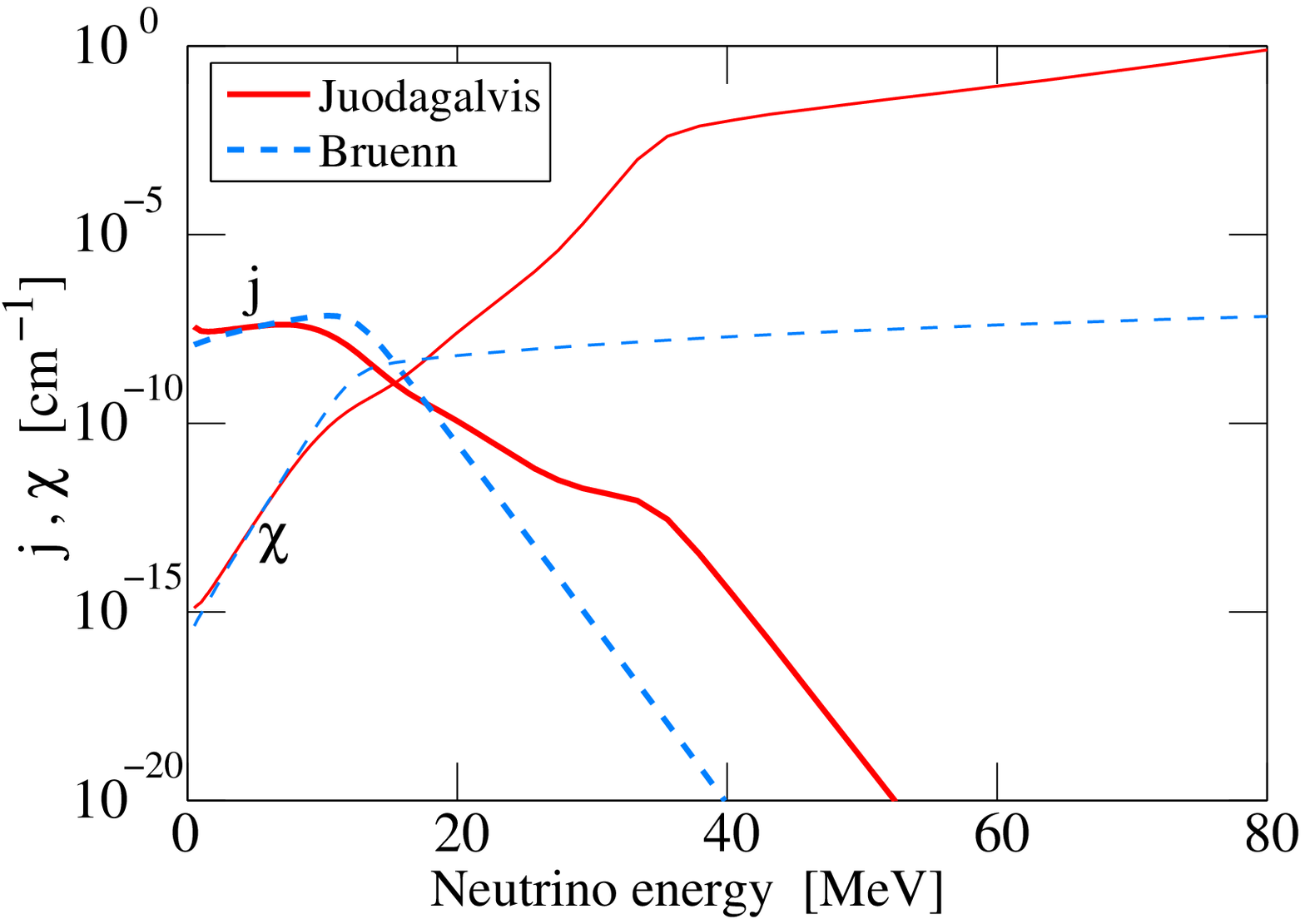}
\mbox{\raisebox{-7mm}{\includegraphics[width=85mm]{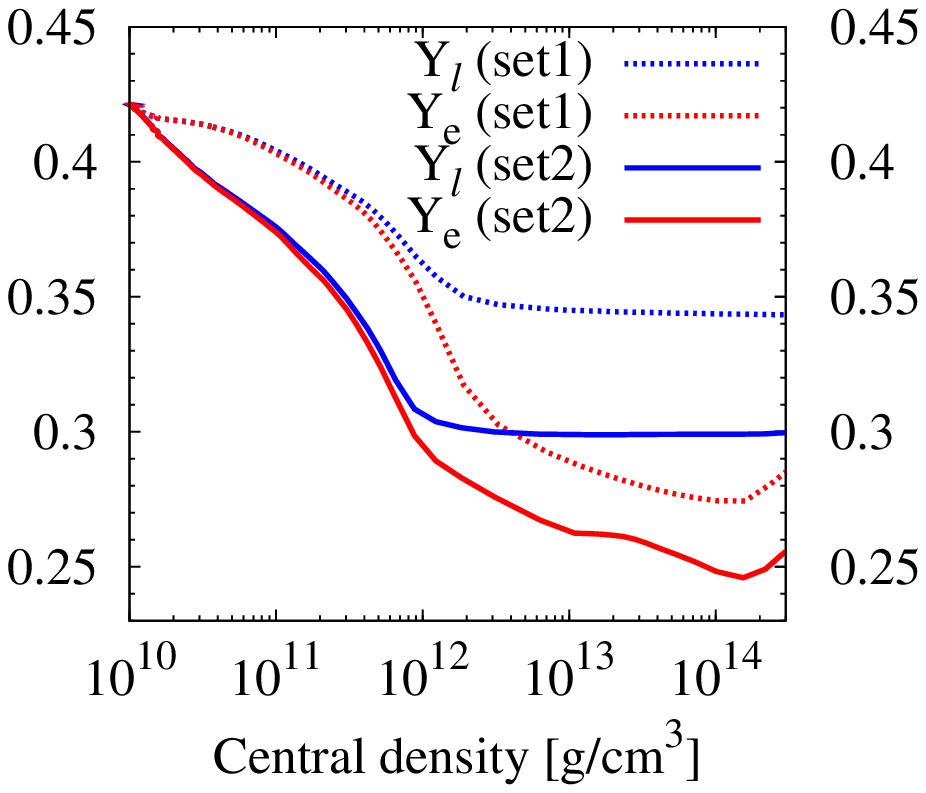}}}
\end{center}
  \caption{Left panel show comparison of emissivity ($j$) and absorptivity 
($\chi$) of electron capture on heavy nuclei at a thermodynamic condition of 
$\rho = 10^{11}\,{\rm g}~{\rm cm}^{-3},\,Y_e = 0.45, T = 10^{10} \,{\rm K}$ (corresponding
 to $\mu_e$ = 18.2 MeV) between the Bruenn rate (blue dashed line) and the rate 
 by \cite{juoda} (red solid line), respectively. Here $\rho$, $T$, $Y_e$, and $\mu_e$ denotes density, temperature, electron fraction, and electron chemical potential, respectively. The absorptivity (equivalently $1/\lambda^{(a)}$ in \cite{Bruenn85}, e.g., the Equation (C29))
 is calculated by a detailed balance. The right panel shows comparison of $Y_e$ 
and lepton fraction ($Y_l$) as a function of the central density 
between set1 (dashed lines) and set2 (solid lines) using
 either the Bruenn or Juodagalvis rate, respectively. }
\label{f2}
\end{figure}

\subsection{Improved electron capture rate on heavy nuclei (set2)}\label{sec3-2}

We start to describe our first update in neutrino opacity, which is electron 
capture on heavy nuclei. Based on detailed shell-model calculations
 by \citet{Langanke00}, \citet{langanke03} showed that not only a generic 
$0f_{7/2} \rightarrow 0f_{5/2}$ Gamow-Teller (GT) transition, but also
additional GT transitions, forbidden transitions, and thermal unblocking 
 play a crucial role, which makes electron capture on heavy nuclei
 dominant over that on protons in the later phases in the core-collapse phase.
 
We employ a tabulated electron capture rate on heavy nuclei by \citet{juoda} who 
has extended significantly the covered mass range of nuclides ($\sim 2700$) 
compared to that of \citet{Langanke00} ($\sim 100$). To calculate the needed 
abundances of the heavy nuclei, 
a Saha-like nuclear-statistical-equilibrium (NSE) is assumed. The table 
by \citet{juoda} then provides an average neutrino emissivity per heavy
  nucleus. Following \citet{hix03}, one can calculate the 
 full neutrino emissivity by the product of this average neutrino emissivity 
and the number density of heavy nuclei calculated by the employed EOS (here for 
 LS EOS)\footnote{See \citet{sullivan} for a more detailed, nucleus-by-nucleus 
investigation of electron capture on heavy nuclei in the CCSN context.}.

The left panel of Figure \ref{f2} compares
 the full neutrino emissivity/absorptivity ($j, \chi$, red solid line)) 
with that of 
  the Bruenn prescription (blue 
dashed line, e.g., Equation (C27) of \citet{Bruenn85}) for a given thermodynamic 
condition in the core-collapse phase. From the panel, one can see 
a significant enhancement of the neutrino emissivity in the Juodagalvis
 rate (red solid line) compared to the Bruenn rate (blue dashed line) 
for neutrino energy above $\sim 18$ MeV. This high electron capture rate 
 is also seen as a steeper slope in $\chi$ than that of the Bruenn rate 
($\chi \propto \epsilon^2$ with $\epsilon$ representing the neutrino energy).

 The right panel of Figure \ref{f2} compares $Y_e$ and lepton fraction ($Y_l$) 
as a function of the central density ($\rho_c$) between set1 (dashed line) 
and set2 (solid line) with the Bruenn or Juodagalvis rate, respectively.
Deleptonization ends approximately at $\rho_c = 2 \times 10^{12}\,{\rm g}\,{\rm cm}^{-3}$, which marks onset of the neutrino trapping \citep{Sato75}. The central
 $Y_l \sim 0.34$ at bounce (see at the right edge of blue dashed curve) 
using the Bruenn rate is lowered about $\sim 10 \%$ ($Y_l \sim 0.3$, 
blue solid curve) with the Juodagalvis rate. For set1,
 the evolution of $Y_e$ and $Y_l$ matches quite nicely with AB \citet{Liebendorfer05} (see their Figure 7(b)). For set2, our results are quantitatively very 
 close to those of \citet{hix03} where the LMP rate (\citet{langanke03})
 was implemented in the AB run using the same $15 M_{\odot}$ progenitor model \citep{WW95}.
 
\begin{figure}[htbp]
\begin{center}
\includegraphics[width=100mm]{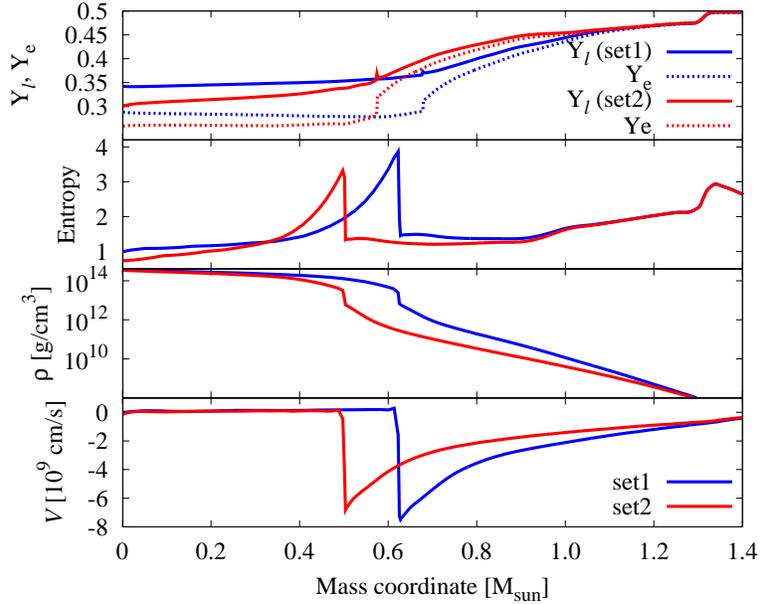}
\end{center}
  \caption{$Y_e$, entropy, density, and velocity profiles as functions
 of the enclosed mass at core bounce. The blue and red solid curve denotes
 the 1D run with the Bruenn rate (set1) or Juodagalvis rate (set2), respectively. }
\label{f3}
\end{figure}

\begin{figure}[H]
\begin{center}
\includegraphics[width=79mm]{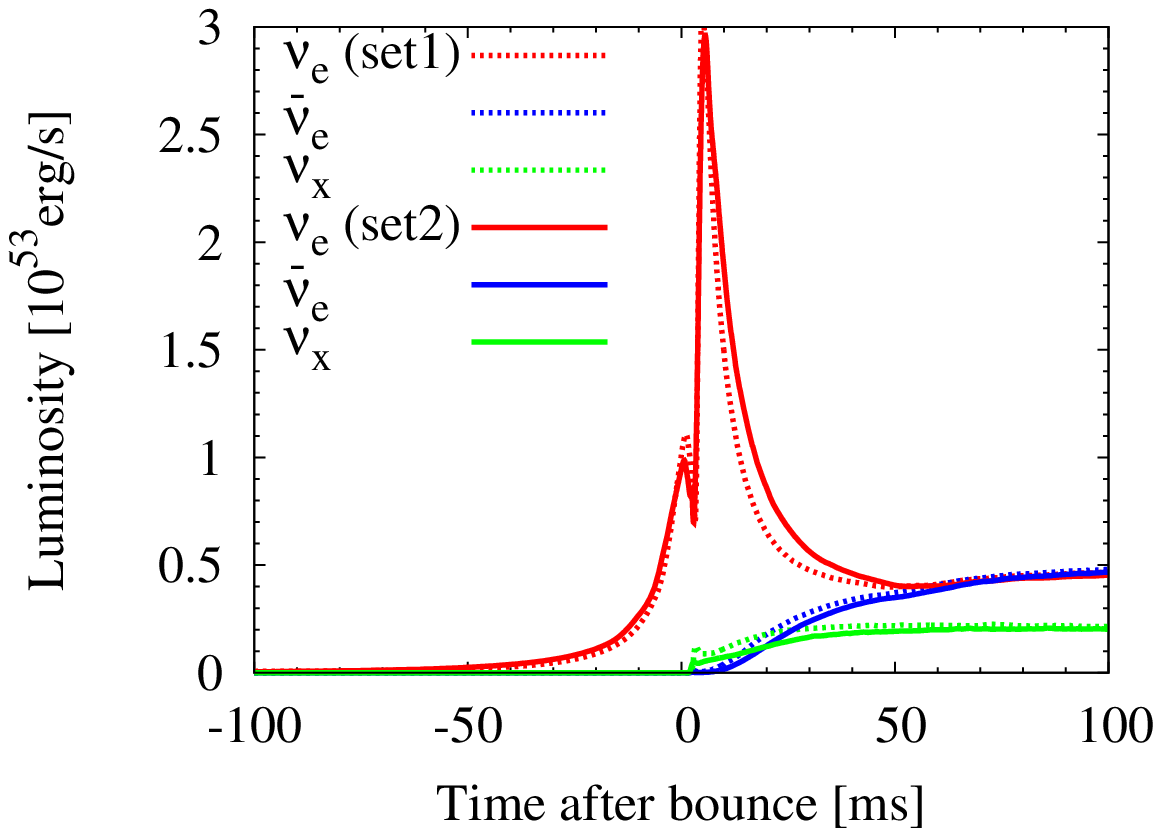}
\includegraphics[width=79mm]{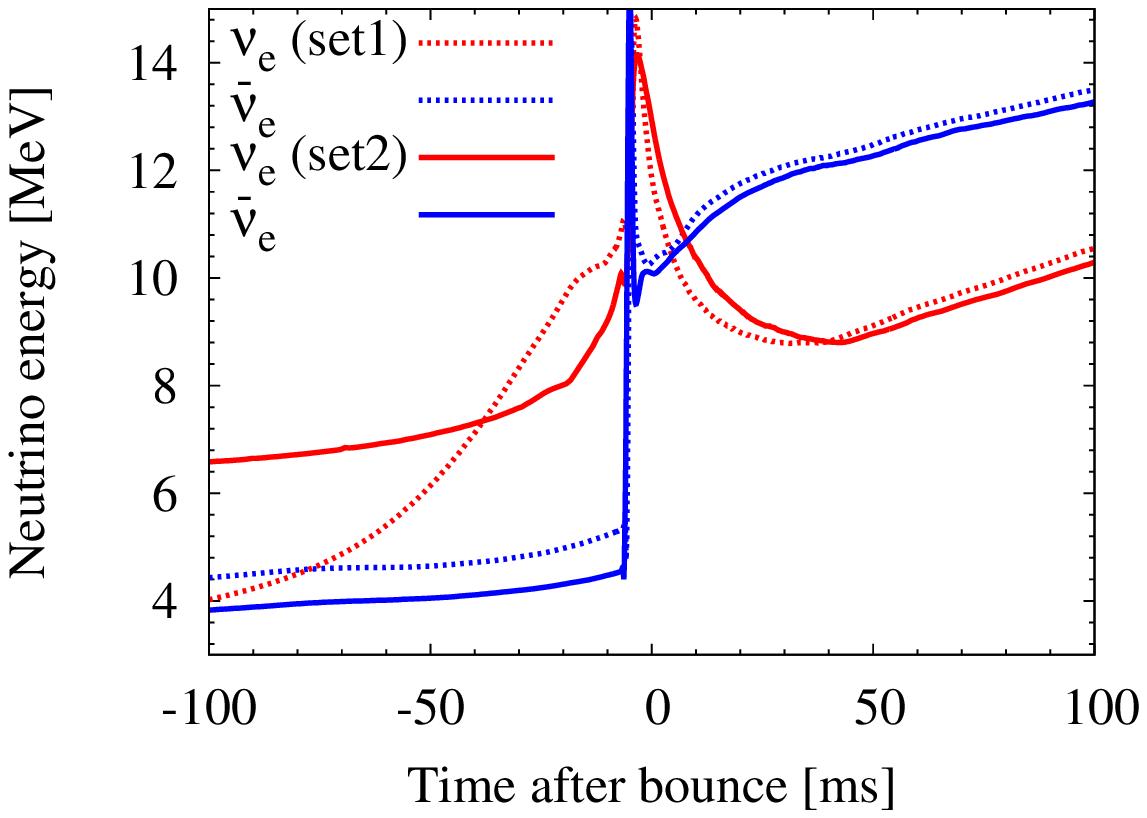}
\end{center}
  \caption{Comparison of the 
neutrino luminosity (left panel) and the rms energy (right panel) measured 
at a 500 km in the lab frame.}
\label{f4}
\end{figure}

For making the comparison easy, Figure \ref{f3} and \ref{f4} are plotted 
in a similar way to Figure 1 and 2 in \citet{hix03} (see also \citet{GMP06}). 
Note that the LS180 was used in \citet{hix03}, however, 
the differences with the difference $K$ of LS EOS are only a few percent
 around core bounce and less than $\sim 10 \%$ in the first 200 ms after bounce 
\citep{Thompson03,BMuller10}. So we consider that the 
different choice of $K$ does not significantly affect the aim of comparison here.

In fact, Figure \ref{f3} shows nice agreement with Figure 1 of \citet{hix03}.
From the top panel, the use of the improved electron capture rate 
 leads to $\sim 10 \%$ reduction in the central $Y_l$ and $Y_e$ compared 
 to those with the Bruenn rate. The above match suggests that hydrodynamics 
impacts between the LMP and Juodagalvis rate should be fairly small. 
From the velocity profile (bottom panel), one can see that 
the mass of the unshocked, homologous core at bounce is reduced about 
$\sim 20 \%$ from $\sim 0.62 M_{\odot}$ for set1 (see discontinuety in the
 velocity plot) to $\sim 0.5 M_{\odot}$ for set2.
 As already pointed out by 
 \citet{hix03}, the $10 \%$ reduction of $Y_l$ leads to $\sim 20 \%$ reduction 
 in the homologous core because the (Chandrasekhar) mass of the unshocked core 
scales as $\langle Y_l\rangle^2$. The smaller entropy and central density 
(second and third panel in Figure \ref{f2}) for set2 compared to set1 is 
 also quantitatively consistent with \citet{hix03}.

Figure \ref{f4} also supports correct implementation of the Juodagalvis
 rate in our code. The left panel shows that because of 
the enhanced electron capture rate 
and the resulting smaller radius at the shock formation (the bottom 
 panel of Figure \ref{f3}), the shock breakout is slightly delayed 
 and the duration becomes slightly longer
(a few ms) for set2 (red solid line) compared to set1 (dashed red line). 
In accordance with \citet{hix03}, this feature is also seen 
 in other neutrino flavors near core bounce ($\lesssim 50$ ms postbounce,
 see green and blue curves in the left panel). An overall trend in the rms neutrino 
energy (right panel of Figure \ref{f4}) both in pre- and post-bounce phase 
is also in line with \citet{hix03}; the $\nu_e$ rms energy is as much 
as $\lesssim$ 1 MeV smaller for set2 than set1 
over the first 50 ms after bounce (compare
 red solid with red dashed line), but lower thereafter. The difference of $\bar\nu_e$ 
 is minute compared to that of $\nu_e$.

\begin{figure}[H]
\begin{center}
\includegraphics[width=79mm]{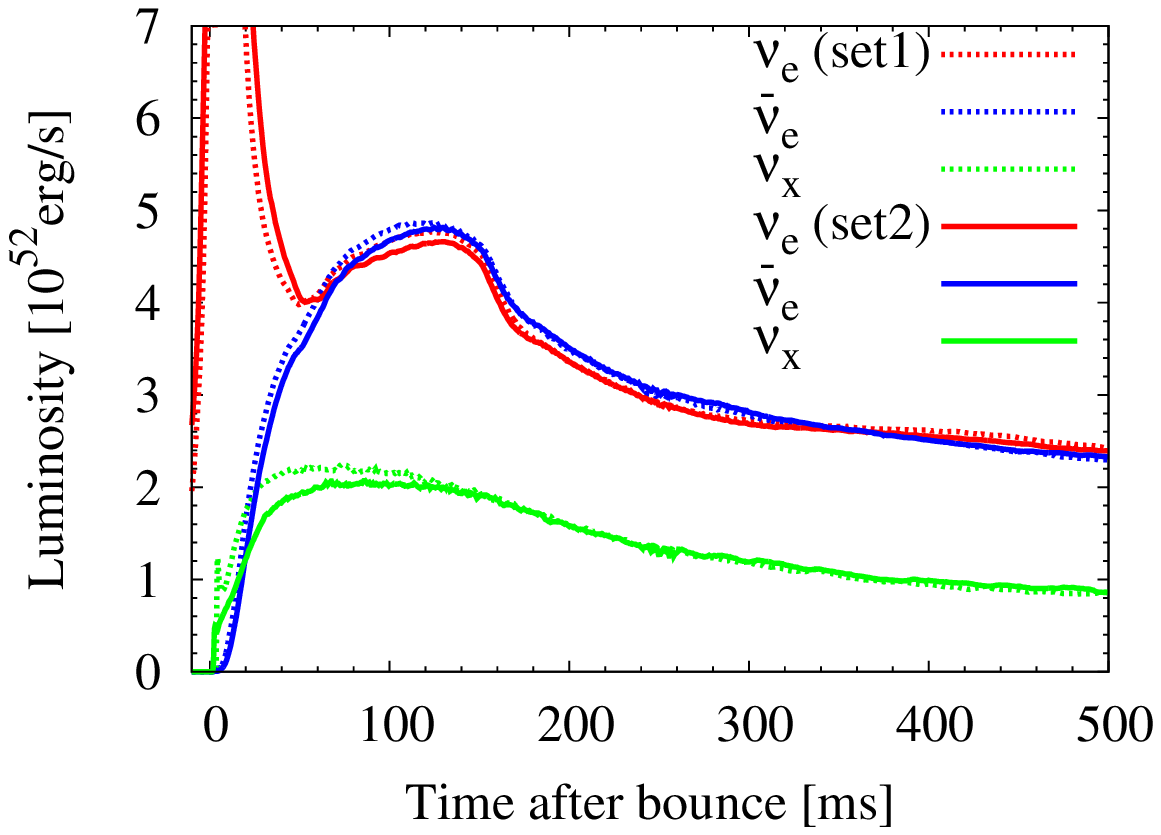}
\includegraphics[width=79mm]{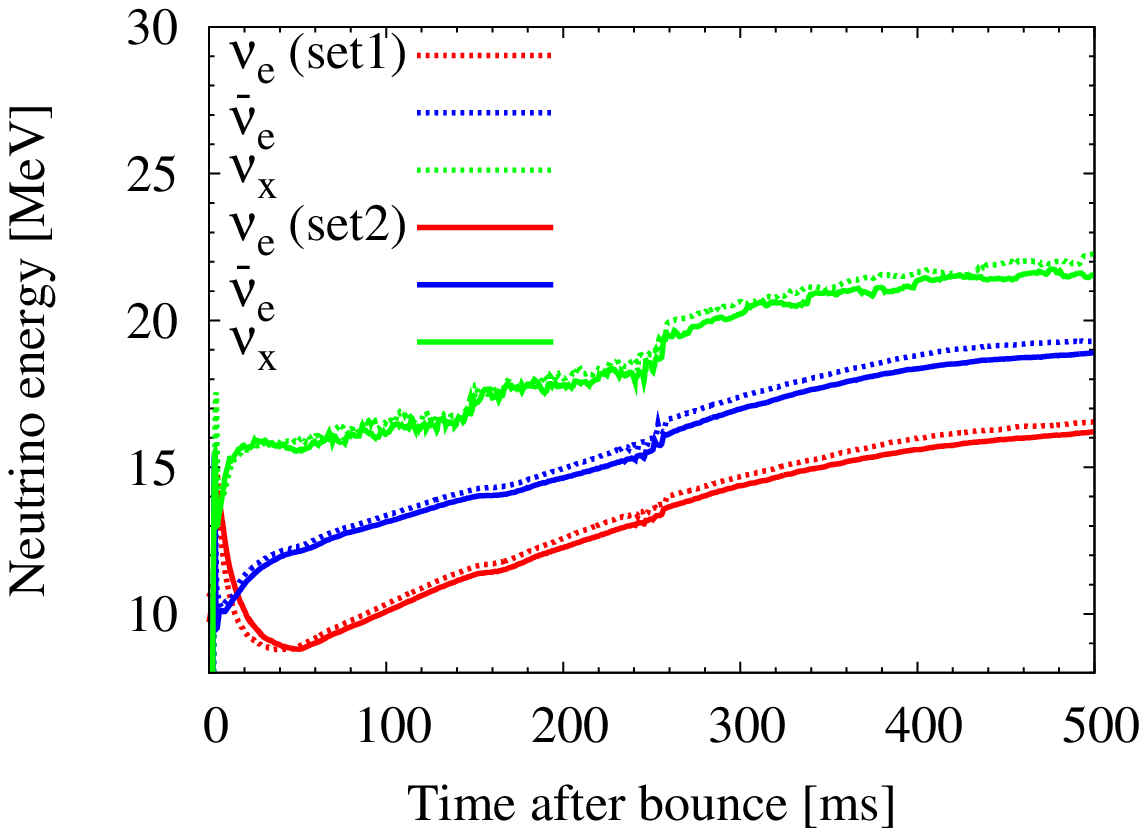}
\includegraphics[width=79mm]{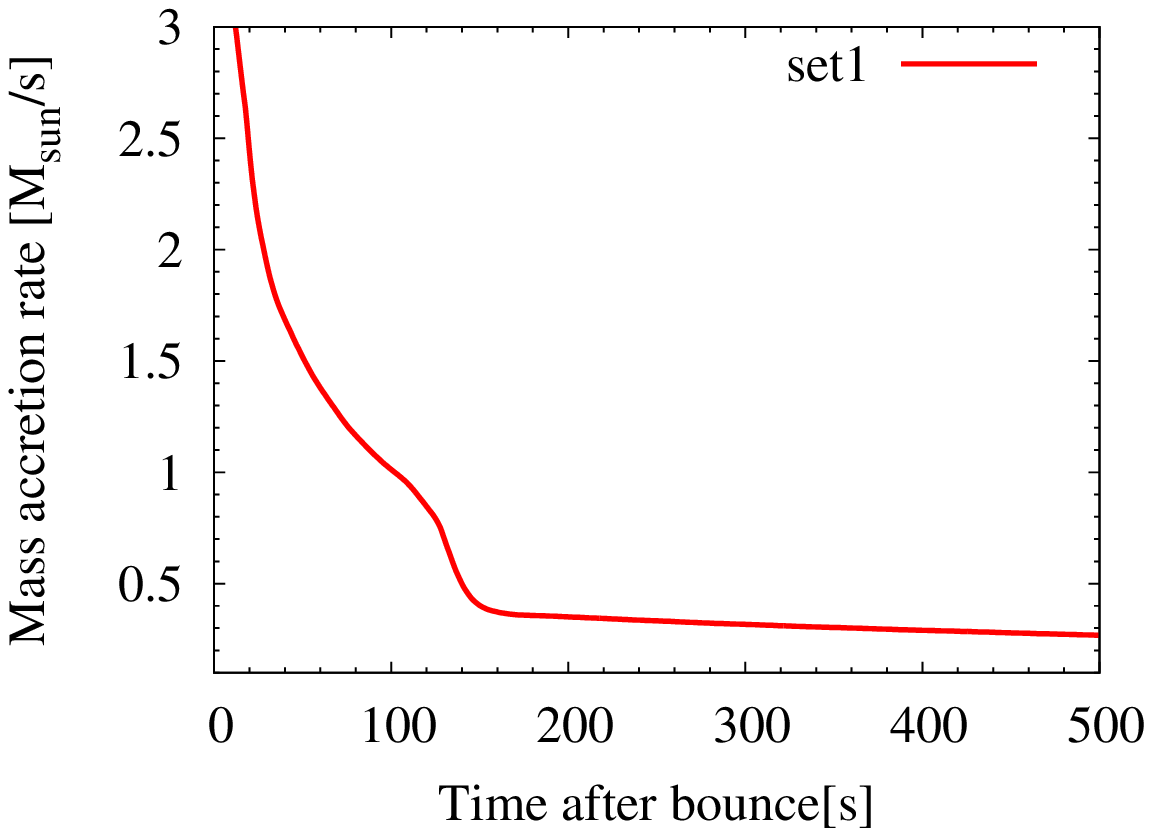}
\includegraphics[width=79mm]{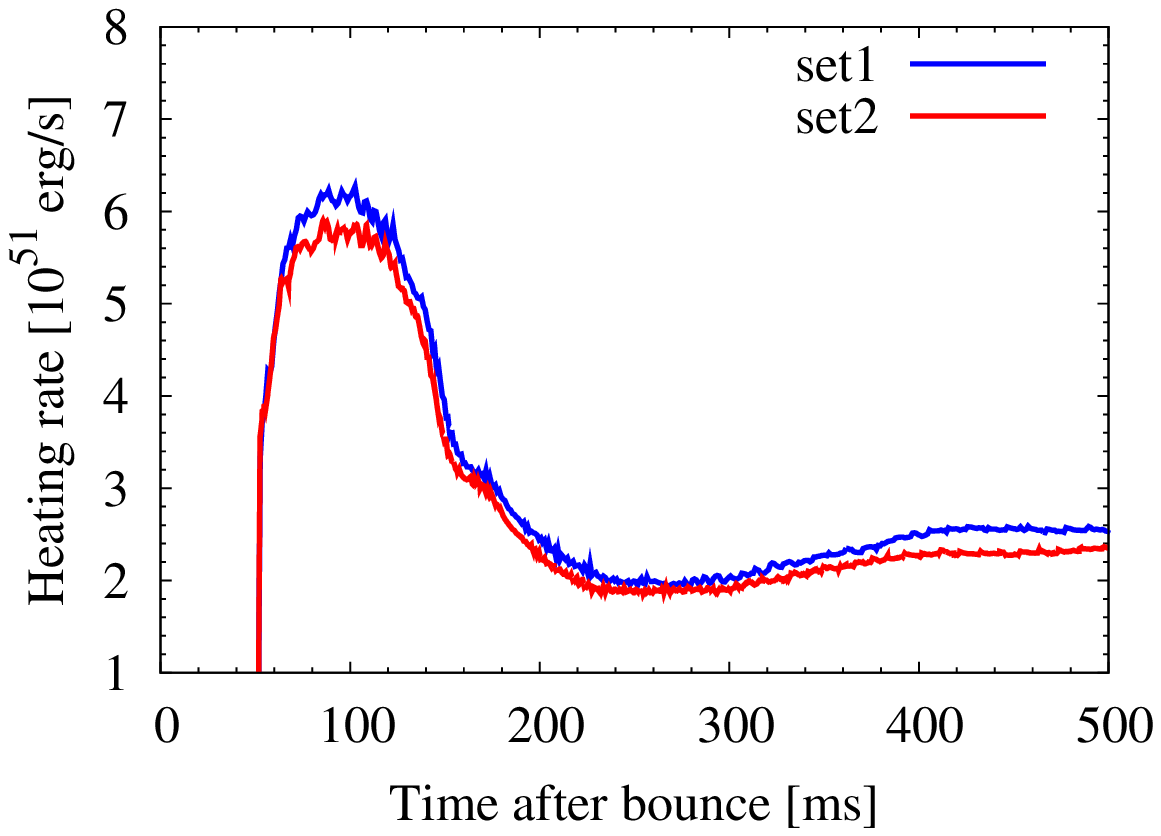}
\end{center}
  \caption{Comparison of several key quantities in 1D simulations 
with neutrino physics set1 or set2. In the top panels, we show the neutrino luminosities ({\it left}) and 
 the neutrino rms energy ({\it right}) measured at a radius of 500 km in the lab frame.
The bottom left panel shows the mass-accretion rate evaluated at 500 km.
The bottom right panel compares the net heating rate integrated over the gain region.}
\label{f5}
\end{figure}

Figure \ref{f5} summarizes several key quantities for comparison between set1 and 
 set2 over the first 500 ms after bounce\footnote{We choose the (fiducial) final
 computational time as 500 ms because the shock revival mostly occurs within
 this timescale in multi-D models that are trending towards explosion 
(e.g., \citet{Bruenn13,Lentz15,Bollig17}).}. The top left panel shows 
that the largest difference is a $\lesssim 15 \%$ reduction of 
$\nu_x$ luminosity of set2 (solid green line) compared to set1 (dashed green line) 
in the first $\sim$ 140 ms after bounce, but vanishes thereafter. 
The bottom left panel shows that the drop in the mass accretion rate 
($\sim$ 140 ms) corresponds to the epoch when the Si-rich layer is passing 
through the shock. This leads to a dorm-like shape in 
the $\nu_e$ and $\bar \nu_e$ luminosity (see red and blue curves in the top left panel), 
the peak of which is at around 
 the $\sim$ 140 ms after bounce. In the first 160 ms after bounce,
  the Si-rich layer has entirely passed though the shock as indicated 
 by a low mass accretion in the bottom left panel. 
 At this phase, one can 
see the $\sim 5 \%$ reduction of the $\nu_e$ and $\bar \nu_e$ luminosity in set2 
compared to set1. The rms energy (top 
right panel) for all the neutrino flavors becomes 
 smaller (maximally by $\sim$ 0.4 MeV) in set2 (solid lines) compared to set1 
(dashed line). 
The smaller homologous mass at bounce for set2 could potentially 
lead to smaller gravitational energy release in the postbounce phase compared to set1,
 which is reconciled with the above trends. The bottom right panel shows that
 the net heating rate for set2 (red line) is smaller 
by $\lesssim 7 \%$ compared to set1 (blue line). 

 Regarding the update in this section, one would
 imagine from the 1D comparison that the improved electron capture rate on nuclei
 might weaken "explodability" in multi-D simulations. The readers would see 
whether this expectation is correct or not
 in Section \ref{sec4} (2D results).



\subsection{Electron Neutrino Pair Annhilation (set3a) and 
$\nu_x + \nu_e (\bar\nu_e)$ Scattering (set3b) }\label{nupair}

In this section, we first focus on set3a (see Table \ref{table1}) where 
electron neutrino pair annhilation
 process ($\nu_e + \bar\nu_e \rightleftharpoons \nu_x +\bar\nu_x$, for short,
 we call this as the "nupair" process in this section) is added to set1.
\citet{Buras03} were the first to point out that 
as a source for $\nu_x$ the nupair reaction 
is always more important than the traditional electron-positron pair annihilation process
 ($e^-\,e^+ \rightleftharpoons \nu_x\, \bar\nu_x$, for short, we call this as 
the "eepair" process in the following).
The implementation scheme of the nupair process in IDSA
is given in Appendix \ref{appA}.

\begin{figure}[H]
\begin{center}
\includegraphics[width=81mm]{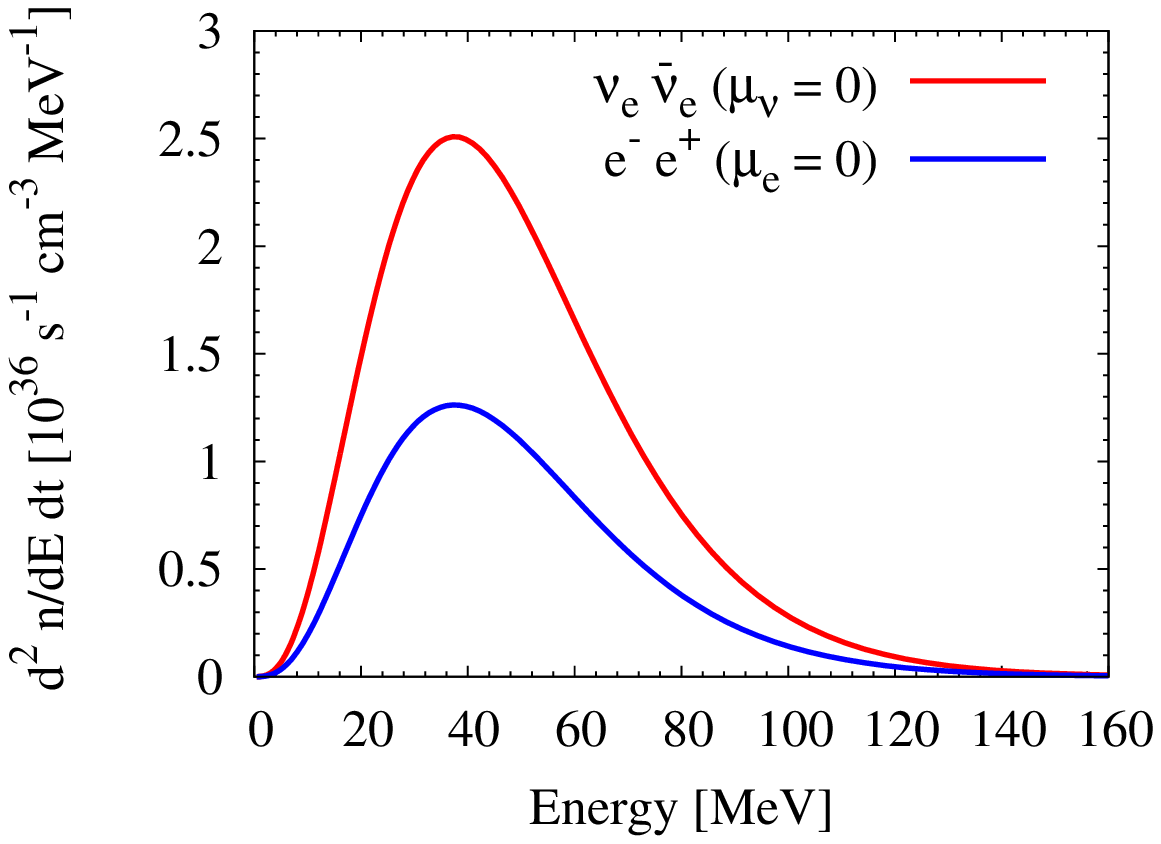}
\includegraphics[width=81mm]{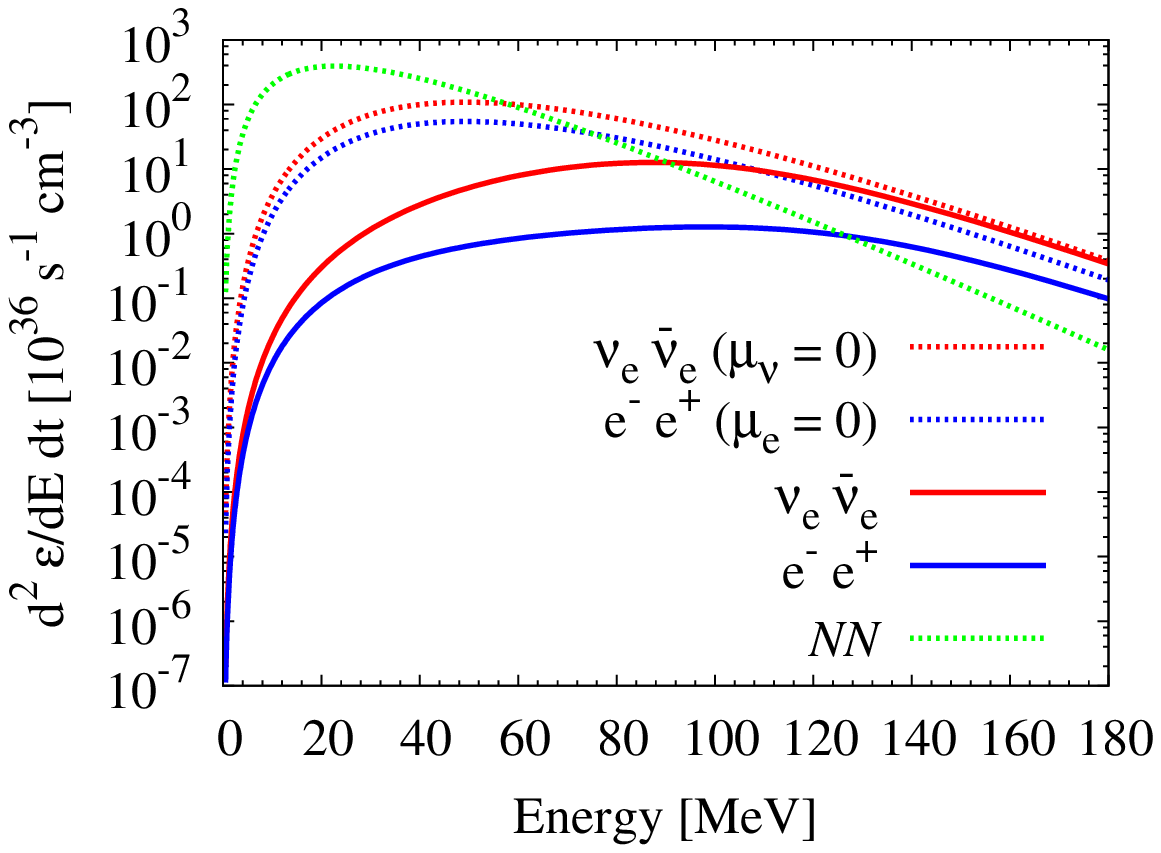}
\end{center}
  \caption{The left panel compares neutrino-pair production number spectra (Equation 
(\ref{number_nupair})) as a function of neutrino energy for $T = 12$ MeV between
 the nupair and eepair process. Note that the chemical potential for electron and neutrinos is set to zero ($\mu_e = 0$ and $\mu_{\nu} =0$), corresponding to Figure of \citet{Buras03} (but with a different scale in the $y$ axis).
The right panel shows energy production rates (Equation (\ref{energy_nupair}))
 for the nupair and eepair processes
 including nucleon-nucleon Bremsstrahlung (labeled as ${\it "NN"}$ in the panel).
For the right panel, the following thermodynamic condition is chosen as
$\rho = 5\times 10^{13}\,\rm{g}/\rm{cm}^{-3}$, $Y_e = 0.3$, $T =12$ MeV, 
(corresponding to $\mu_e = 96$ MeV and $\mu_{\nu_e} = 123$ MeV), 
respectively. The right panel 
corresponds to Figure 16a of \citet{tobias09}, where some typos 
in the label of $y-$axis is now corrected.
Note in both of the panels that the final state Pauli blocking is neglected.}
\label{f6}
\end{figure}

In fact, the left panel of Figure \ref{f6} clearly shows the dominance of the nupair 
process (red solid line, labeled as "$\nu_e \bar\nu_e$") over the eepair process
(blue solid line, labeled as "$e^- e^+$"). In this panel, the chemical potential for 
electron and neutrinos is set to zero ($\mu_e = 0$ and $\mu_{\nu} =0$), 
following Figure 3 of \citet{Buras03}. 
The peak in the spectra of the nupair process is about two times higher than 
 that of the eepair process at the neutrino energy of $\sim 40$ MeV. 
As explained in \citet{Buras03}, this simply comes from the different weak 
coupling constants between the two processes. The production kernels have a 
similar form as, 
$\Phi^{p}= (C_{\rm V} + C_{\rm A})^2 J^I + (C_{\rm V} - C_{\rm A})^2 J^{II} $ (e.g.,
 Equation (C63) in \citet{Bruenn85}). For the eepair process, 
$C_{\rm V} = -1/2 + 2 \sin^2 \theta_{w}$ and $C_{\rm A} = -1/2$
(with the weak mixing angle; $\sin^2 \theta_w = 0.23$), whereas $C_{\rm V} = 1/2 
$ and $C_{\rm A} = 1/2$ for the nupair process. Given that $J^I \sim J^{II}$, the two-times difference can be 
readily seen by putting these numbers in the above equation of $\Phi^p$.

The right panel of Figure \ref{f6} shows comparison between the energy 
production rates with (solid lines) and without chemical potentials
(dashed lines, labeled with $\mu_{\nu}$ = 0, $\mu_{e}$ =0). 
As a reference, the production rate 
of nucleon-nucleon Bremsstrahlung is also shown 
(labeled as ${\it "NN"}$).
In accordance with \citet{tobias09}, the chemical potentials make the spectra harder and the rate smaller.
 Quantitatively,
the $\eta$-paramter of $\sim 10$ in the right panel (e.g., for electron, $\eta_e = \mu_e/T = \sim 96{\rm MeV}/12{\rm MeV} = 8$, and electron-neutrino, $\eta_{\nu_e} = \mu_{\nu_e}/T = \sim 123{\rm MeV}/12{\rm MeV} \sim 10$). This leads to $\gtrsim 1/10$ reduction
 in the rates with the chemical potential than those without. This 
 is consistent with \citet{Buras03} (their Figure 2).
 Comparing to Figure 16a of \citet{tobias09}, 
the peak energy of Bremsstrahlung is around $\sim 30$ MeV, and the specral shape matches 
 well together.\footnote{All the related opacities in Table \ref{table1} are 
 compared with those in \citet{tobias09,tobias12,fischer16}.
 For the Bruenn rate, the opacity plots are already shown in \citet{KurodaT16}.}

\begin{figure}[H]
\begin{center}
\includegraphics[width=81mm]{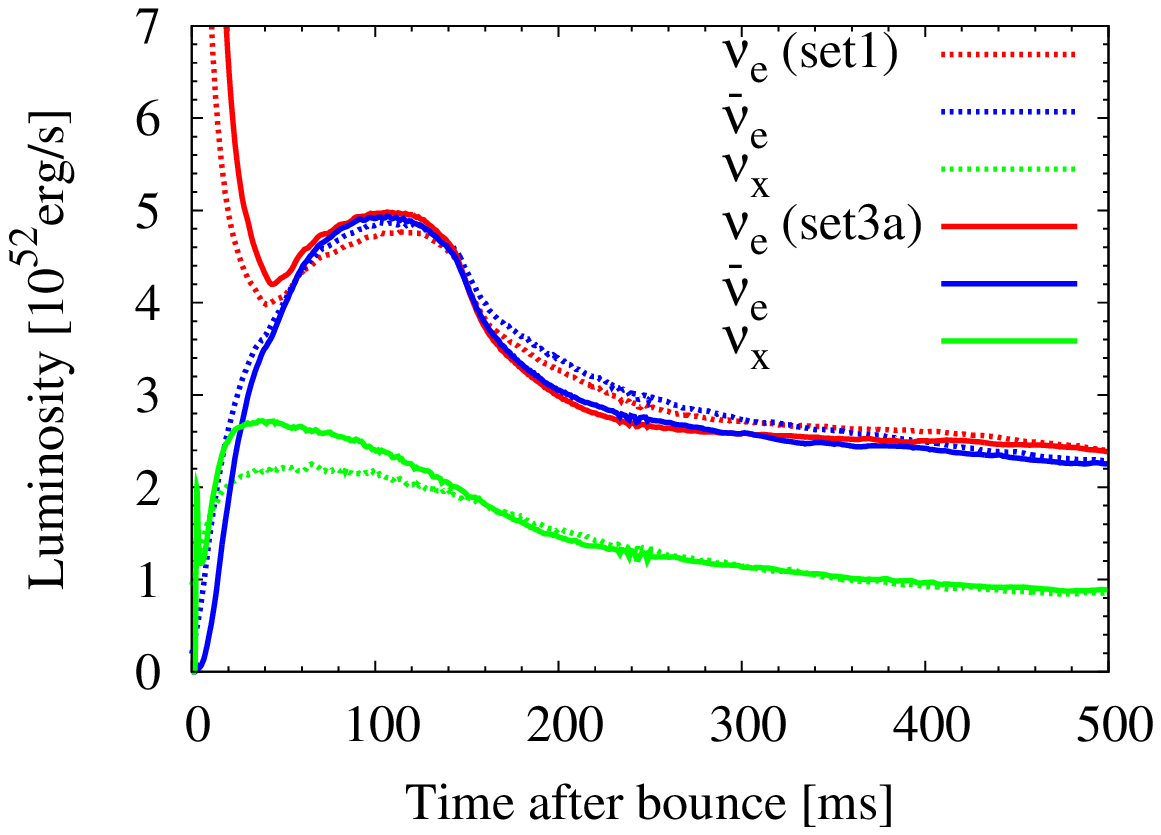}
\includegraphics[width=81mm]{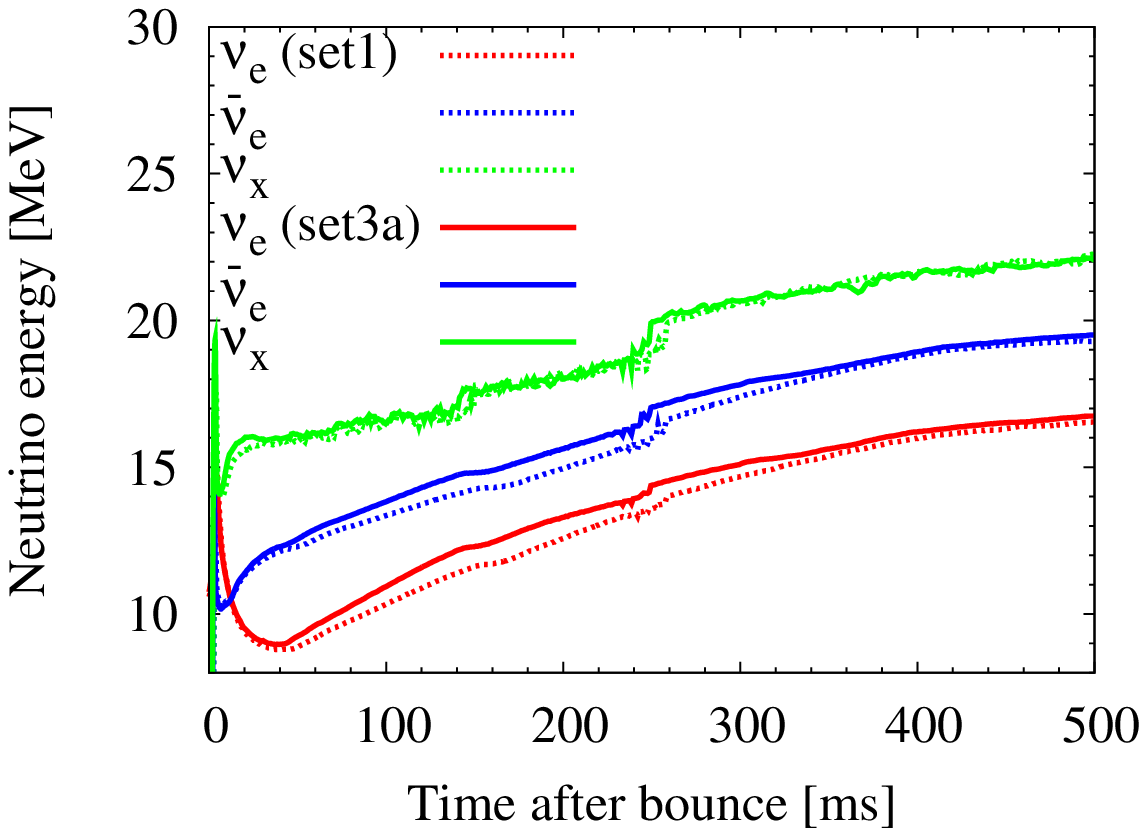}
\includegraphics[width=80mm]{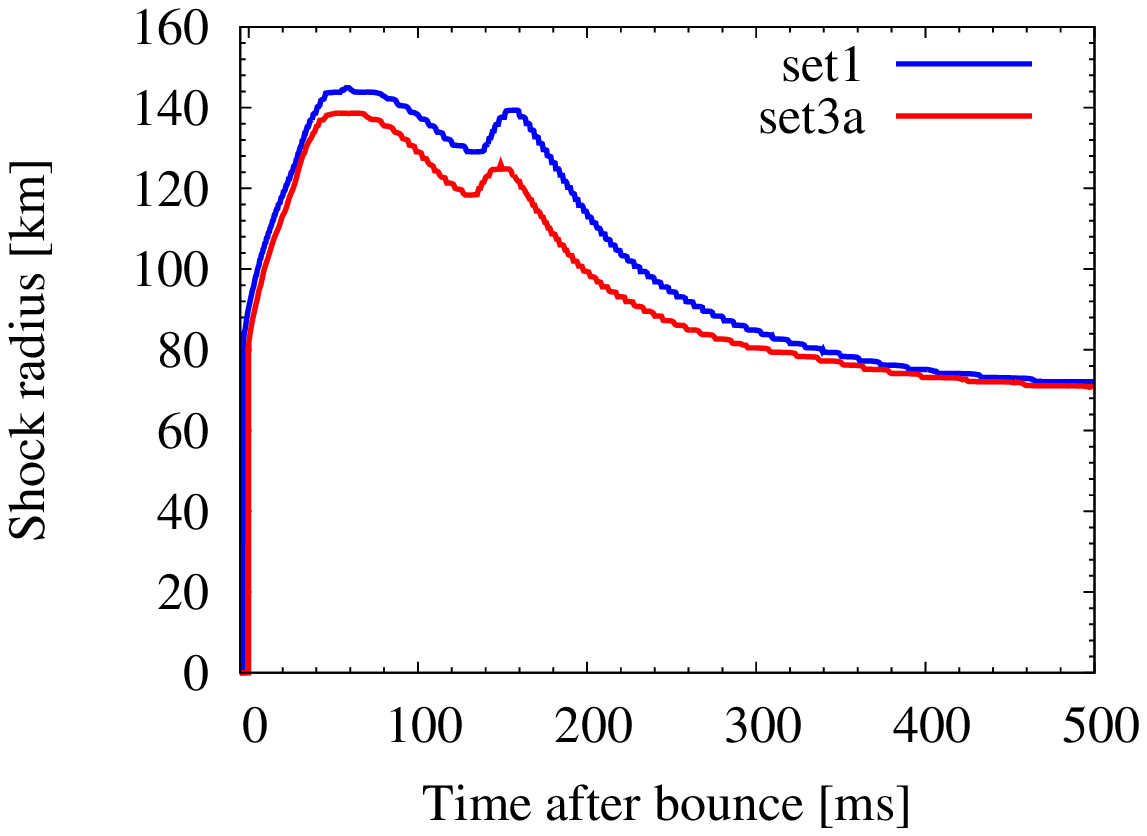}
\includegraphics[width=80mm]{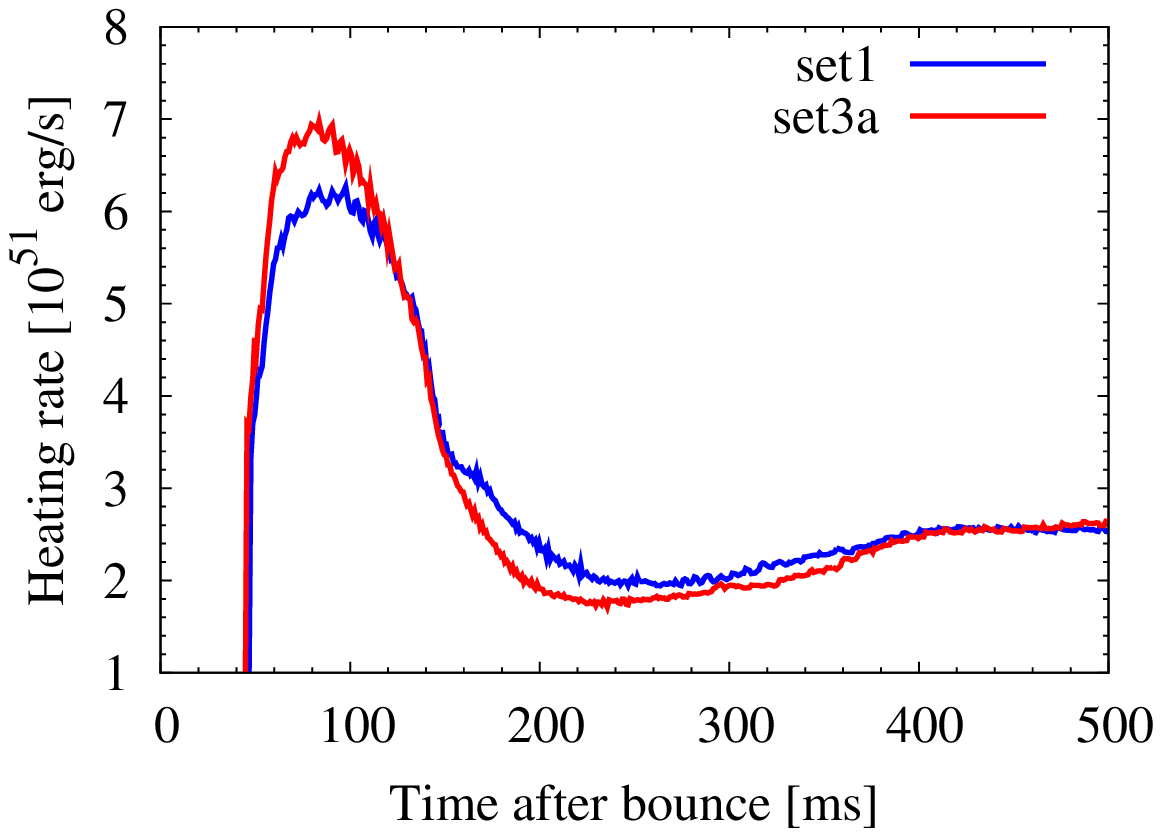}
\end{center}
  \caption{Same as Figure \ref{f5} but for the comparison between set3a and set1. The 
 bottom left panel compares the shock evolution.}
\label{f7}
\end{figure}

The top left panel of Figure \ref{f7} compares the neutrino luminosities 
 between set3a (solid lines) and set1 (dashed lines). Before the first
 160 ms after bounce when the mass accretion rate is still high (e.g., the bottom
 left panel of Figure \ref{f5}), the influence of the nupair process is 
 biggest for the $\nu_x$ luminosity ($\lesssim 20 \%$ increase from set1, compare
 green solid with green dashed line), 
followed in order by the $\nu_e$ luminosity ($\sim 4 \%$ increase, compare 
red solid line with red dashed line)
 and by the $\bar\nu_e$ luminosity ($\lesssim 2 \%$ increase, blue solid line with
 blue dashed line). Using the same $15 M_{\odot}$ progenitor, this is quantitatively
 very close to the results in \cite{Buras03} (see their Figure 7). 
 As is consistent with \citet{Buras03} and \citet{tobias09}, 
the additional source of $\nu_x$ increases the $\nu_x$ luminosity most significantly.
 In the accretion phase ($\lesssim 160$ ms postbounce), the nupair process 
also leads to higher rms neutrino energies (top right panel of Figure \ref{f7}).
 This is due to the enhanced cooling which makes the positions of the neutrino 
spheres and the PNS formed deeper inside (e.g., Figure 17 of \citet{tobias09} of the corresponding 
{\tt Agile-BOLTZTRAN} run but for a $40 M_{\odot}$ star).
After the accretion phase ($\gtrsim 160$ ms postbounce), the 
luminosities of $\nu_e$ and $\bar\nu_e$ become
 smaller than those in set1 (top left panel of Figure \ref{f7}), 
As already pointed out by \citet{Buras03}, this is most likely because of
 the more compact neutrino spheres (e.g., smaller emission region) 
in response to the more accelerated PNS contraction. The (maximum) shock position of 
 set3a is more compact compared to set1 by $5 \sim 10 \%$, which is within the 
change seen in \citet{Buras03} and \citet{tobias09}. From the bottom right panel 
 of Figure \ref{f7}, it is interesting to note that the net heating rate of 
 set3a (red line) dominates over that of set1 (blue line) in the accretion 
 phase ($\lesssim 160$ ms postbounce), which reverses thereafter 
(until $\sim 400$ ms postbounce). This is in line with 
 the higher (and lower) luminosities of $\nu_e$ and $\bar\nu_e$ of set3a 
compared to set1 in the pre- and (post-) accretion phase, respectively 
as already mentioned above.   

\begin{figure}[H]
\begin{center}
\includegraphics[width=80mm]{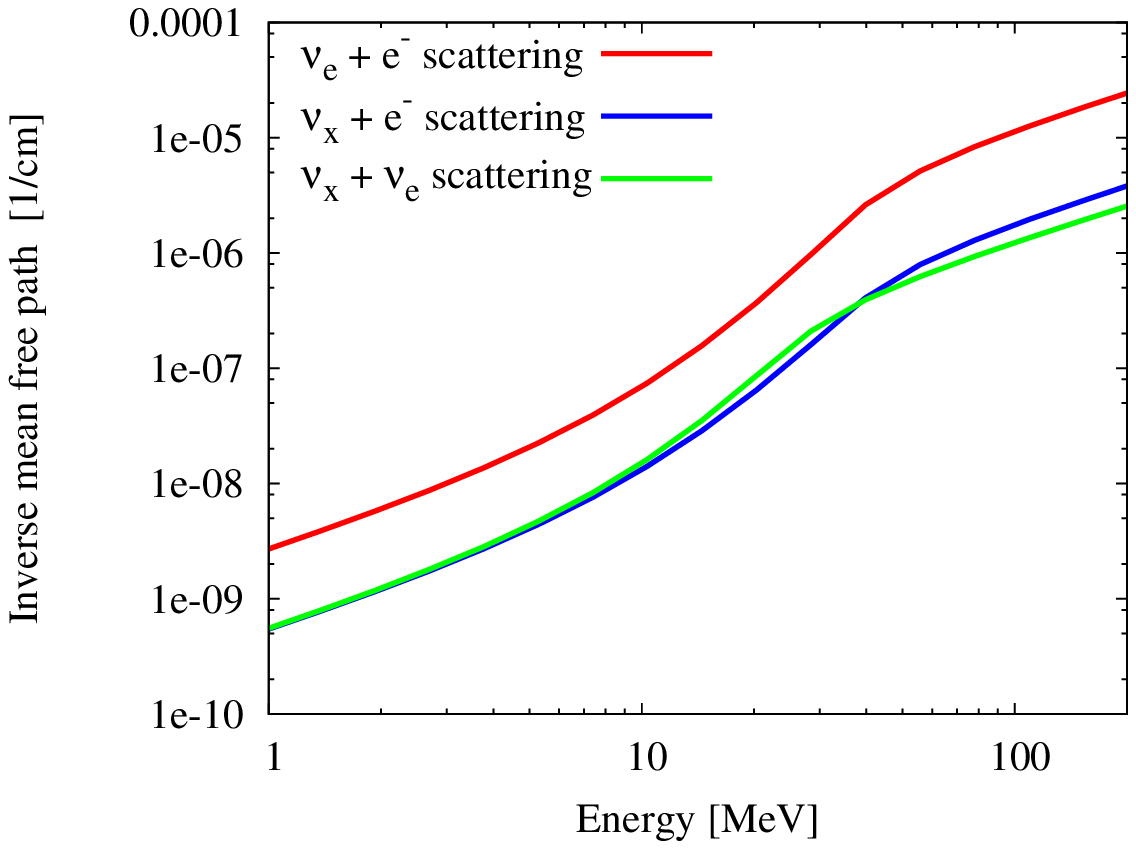}
\includegraphics[width=80mm]{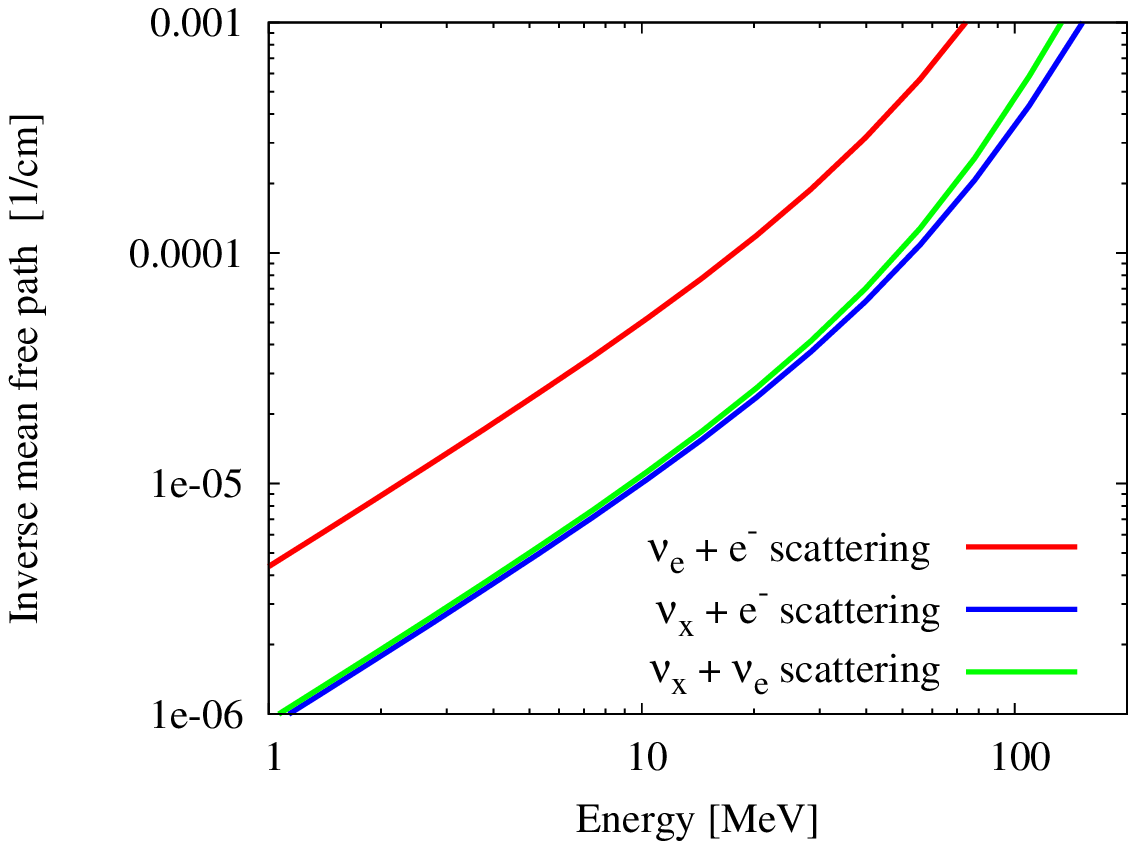}
\includegraphics[width=80mm]{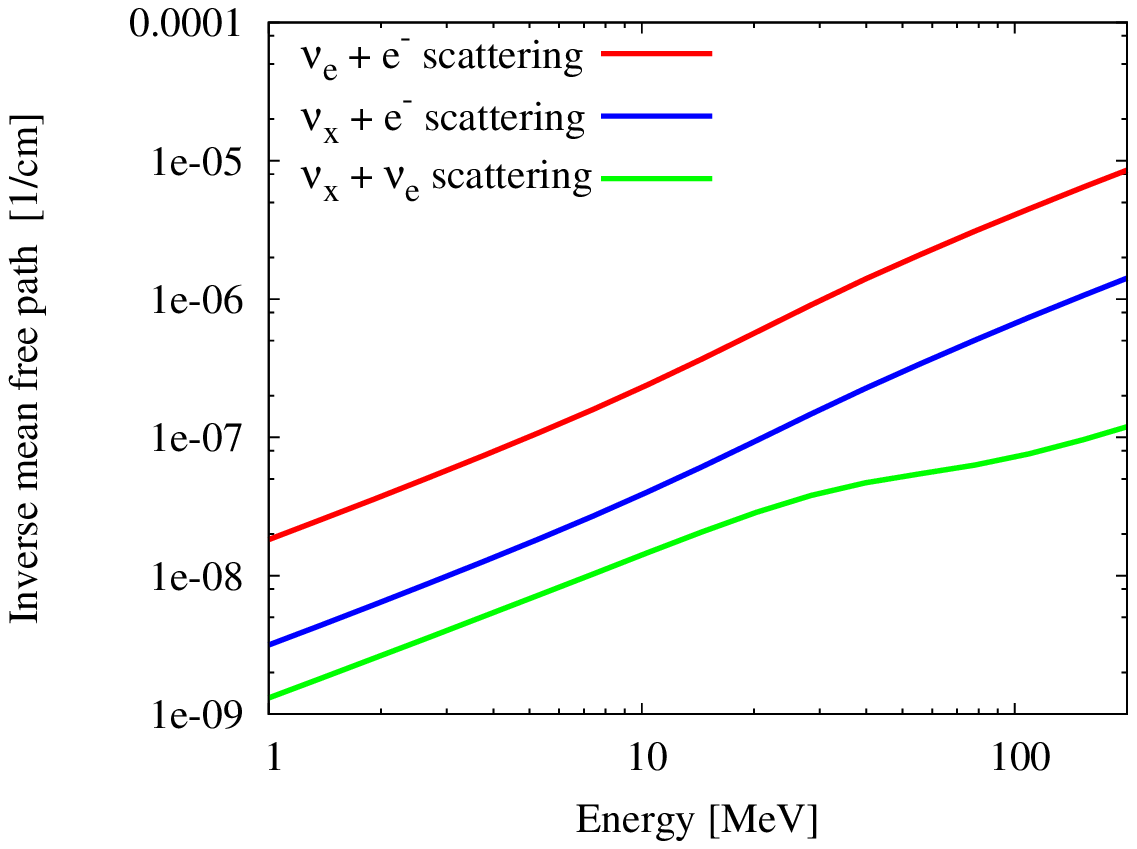}
\includegraphics[width=80mm]{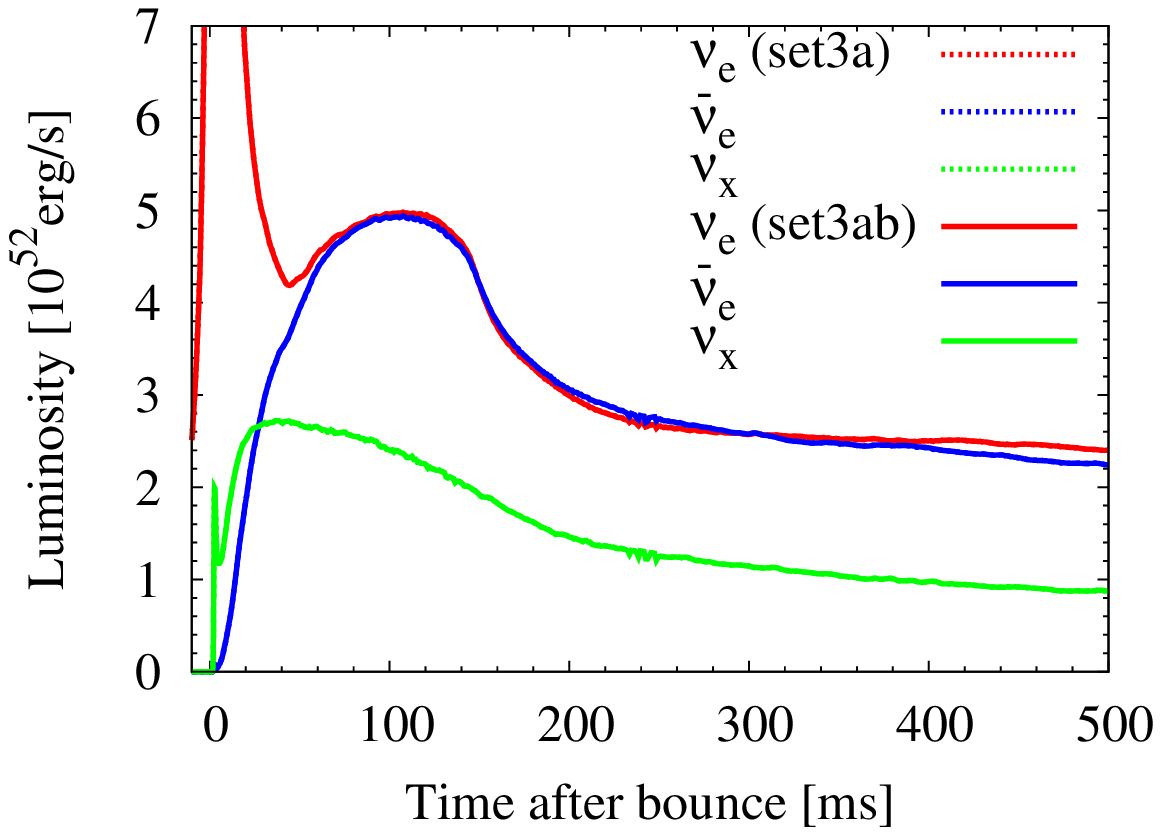}
\end{center}
  \caption{Inverse mean free path (e.g., Equation (\ref{IMFP})) 
as a function of neutrino energy 
for three typical conditions; near neutrino trapping (top left,
$\rho = 10^{12}\,{\rm g}\,\,{\rm cm}^{-3}$, $T = 1.76$ MeV, 
 and $Y_e = 0.35$), near core bounce (top right panel, $\rho = 3 \times 10^{14}\,{\rm g}\,\,{\rm cm}^{-3}$, $T = 12$ MeV, and $Y_e = 0.27$), and in the postshock region 
behind the shock (bottom left panel, $\rho = 1\times 10^{12}\,{\rm g}\,\,{\rm cm}^{-3}$, $T = 7$ MeV, and $Y_e = 0.10$), respectively. 
 Here Fermi-Dirac final state neutrino distributions are assumed.
The bottom right panel compares the neutrino luminosities between set3a and 
 set3ab. Note that set3ab is the model where $\nu_x + \nu_e(\bar\nu_e)$ scattering
 is added to set3a.}
\label{f8}
\end{figure}

As originally pointed out by \citet{Buras03}, 
the cross channel of 
 the nupair process, that is $ \nu_x + \nu_e/\bar\nu_e  \rightleftharpoons 
\nu_x + \nu_e/\bar\nu_e$, could be of comparable importance to 
 $\nu_x + e^{\pm}$ scattering. The top and bottom left panel 
of Figure \ref{f8} compares the (inverse) mean free path of
$\nu_e + e^-$ scattering (red line), 
$\nu_x + e^{-}$ scattering (blue line), and $\nu_x + \nu_e$ 
scattering (green line) for typical thermodynamics conditions 
in the supernova core, respectively. Note
 that the corresponding reactions with $\bar\nu_e$ and $e^{+}$ are not shown 
in the panels because they are much smaller compared to those with 
$\nu_e$ and $e^-$. In the prebounce phase, the top panels of Figure \ref{f8}
 show that $\nu_x \nu_e$ scattering (green line) 
is almost comparable to $\nu_x e^{-}$ scattering (blue line). But they
 play a minor role as a opacity (in the leptonic channels)
 because of the dominant contribution from $\nu_e\,e^{-}$ scattering (red line).  In the postbounce phase (bottom left panel of Figure \ref{f8}), 
 the dominance of $\nu_e\,e^{-}$ scattering is also unchanged, but
 the opacity of $\nu_x e^{-}$ scattering becomes higher than that of 
 the $\nu_x \nu_e$ scattering, as previouly shown in \citet{Buras03}.

The bottom right panel of Figure \ref{f8} 
compares the neutrino luminosities between set3a 
(dashed line) and set3ab (solid line). Note that set3ab is the run where 
neutrino-scattering scattering is added to set3a. The solid and dashed 
 lines are completely overlapped, which confirms the expectation
 that $\nu_x \nu_e$ scattering plays a very minor role at least over 
  the first 500 ms postbounce.



\subsection{Mean-field modifications (set4a and set4b)} \label{set4}

\citet{GMP12} and \citet{roberts12} clearly pointed out
 that medium effects \citep{reddy98} affect differently protons and neutrons 
(e.g., the reactions of set4a in Table \ref{table1}),
 leading to a significant impact on the neutrino 
luminosities and spectra especially in the PNS cooling phase
 (after the onset of an explosion). By definition,
  our (non-exploding) 1D simulation can cover only a pre-explosion phase.
 Having in mind future applications for a long-term 
evolution in multi-D (exploding) models, we explore in this section 
 the impact of the mean-field corrections on the charged-current opacities 
treated at the elastic level (\citet{GMP12,roberts12}).

\begin{figure}[H]
\begin{center}
\includegraphics[width=75mm]{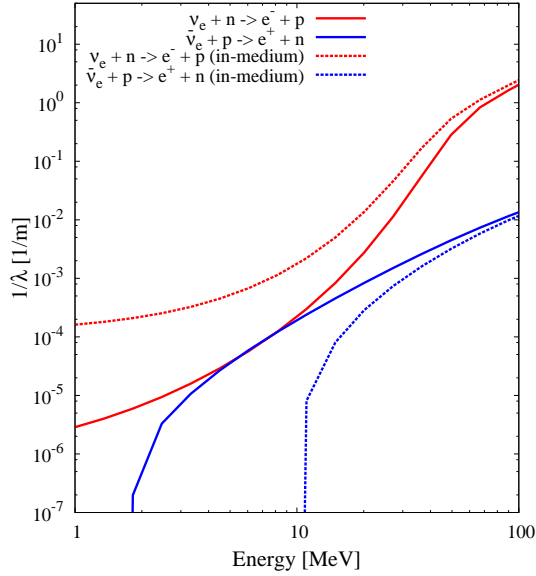}
\end{center}
  \caption{Inverse mean free path as a function of neutrino energy for $\nu_e$
 (red line) and $\bar\nu_e$ (blue line) with (dashed lines, labeled with "in-medium")
 and without (solid lines) mean-field corrections. A thermodynamics condition of
$T = 8$ MeV, $n_B = 0.02\,{\rm fm}^{-3}$, $Y_e = 0.027$ is chosen with $n_B$ the
 baryon number density, which corresponds to Figure 3 of \citet{roberts12}.
 Note that the nucleon potential difference is 
$\Delta U = 9$ MeV for the EOS used 
in \citet{roberts12}, whereas $\Delta U = 7.67$ MeV for LS220 in this work,
 leading to a slight difference quantitatively. }
\label{f9}
\end{figure}

  From Equation (3) of \citet{GMP12}, 
the opacity of $\nu_e$ absorption on neutron ($\nu_e\, n \rightarrow e^- \,p$)
 is expressed as,
\begin{equation}
\frac{1}{\lambda_{\nu_e}} \propto E_e^2
[1 - f_e(E_e)]\frac{n_n - n_p}{1 - \exp{\beta(\mu^0_p - \mu^0_n + \Delta U)}},
\label{in-medium3}
\end{equation}
where $E_e$ is the electron energy, $f_e$ is the electron distribution function, 
$n_{i}$ and $\mu^0_{i}$ is the number density and chemical potential (without rest mass) 
for $i = n,p$ (neutrons and protons) and $\beta$ is the inverse temperature, 
respectively. At the level of an
elastic approximation
\citep{reddy98,GMP12}, the following relation holds
\begin{equation}
E_{e} = E_{\nu_e} + Q + \Delta U,
\label{in-medium1}
\end{equation}
 where $E_{\nu_e}$ is the $\nu_e$ energy, $Q = m_n - m_p$ is the so-called $Q$ value with $m_i$ the rest mass for
 $i =n, p$, and $\Delta U = U_n - U_p$ is the difference of the mean-field potentials of neutrons and protons\footnote{Note for a neutron rich environment (like in the pre-explosion phase),
 $\Delta U > 0$ (e.g., \citet{roberts12}).}. 
 
From Equation (\ref{in-medium1}), $E_e^2$ in Equation (\ref{in-medium3})
 becomes larger due to $\Delta U$, which leads to increase in the $\nu_e$ opacity
 comparing to the free gas case ($\Delta U = 0$) at lower neutrino 
energies. At larger neutrino energies, the Pauli blocking disappears, which makes the opacity with and without the 
 mean-field effects approach each other closely (e.g., \citet{GMP14} for more detail).
 For $\bar \nu_e$, the 
positron energy becomes $E_{e^+} = E_{\bar \nu_e} - Q - \Delta U$. This leads 
 to the reduction of the opacity at lower neutrino energies. Note also 
 that the $Q$ value 
 of this reaction increases from $E_{\bar \nu_e} > Q$ to $E_{\bar \nu_e} > Q +
 \Delta U$.

Figure \ref{f9} is consistent with the above explanations, 
 which compares the inverse mean free path for 
$\nu_e$ (red lines) and $\bar\nu_e$ (blue lines) 
with (dashed lines) and without the mean-field corrections (solid lines).
 These features are also in good agreement with previous work 
(e.g., \citet{roberts12} and \citet{GMP12}).
 
\begin{figure}[H]
\begin{center}
\mbox{\raisebox{+1mm}{\includegraphics[width=81mm]{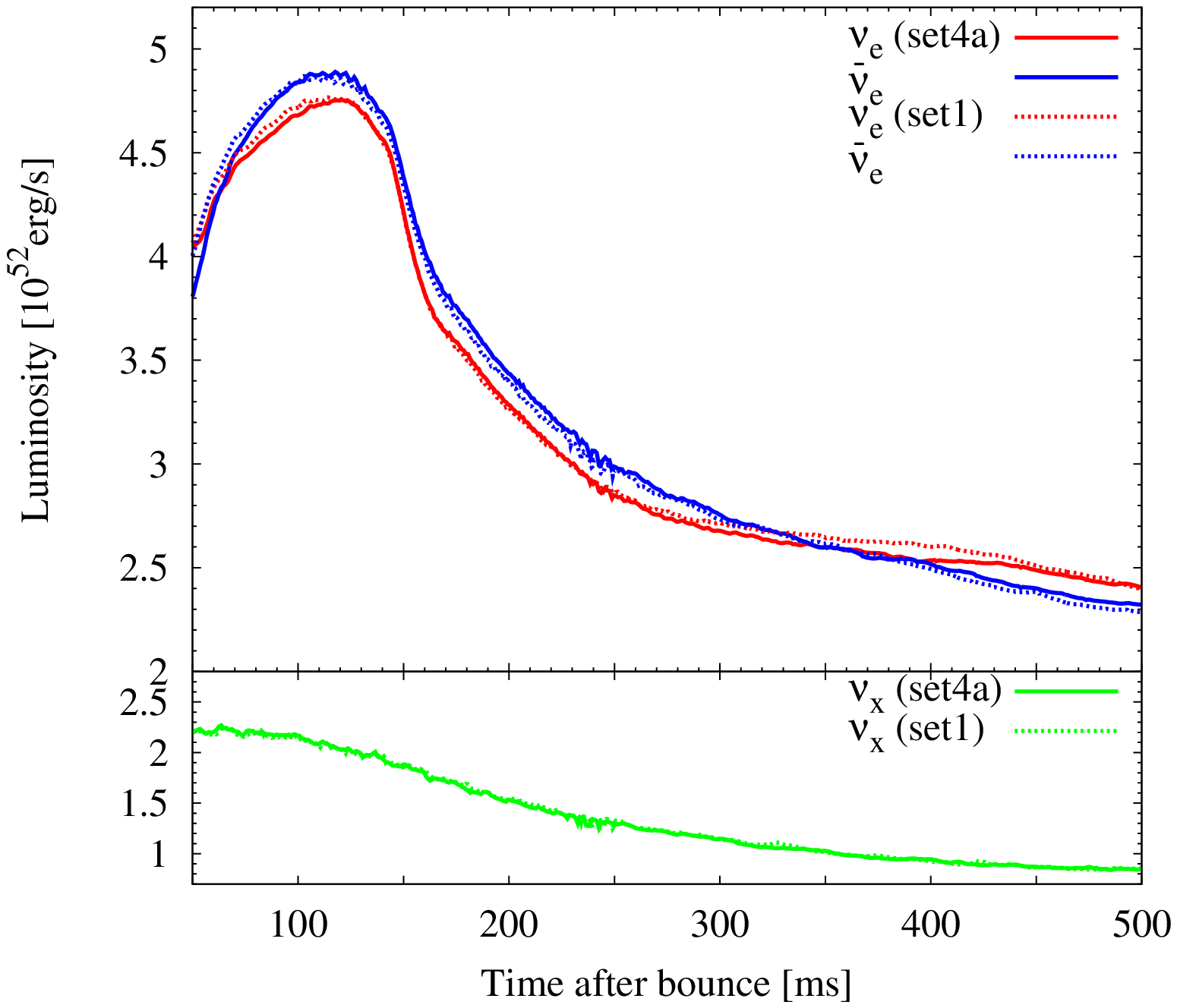}}}
\includegraphics[width=81mm]{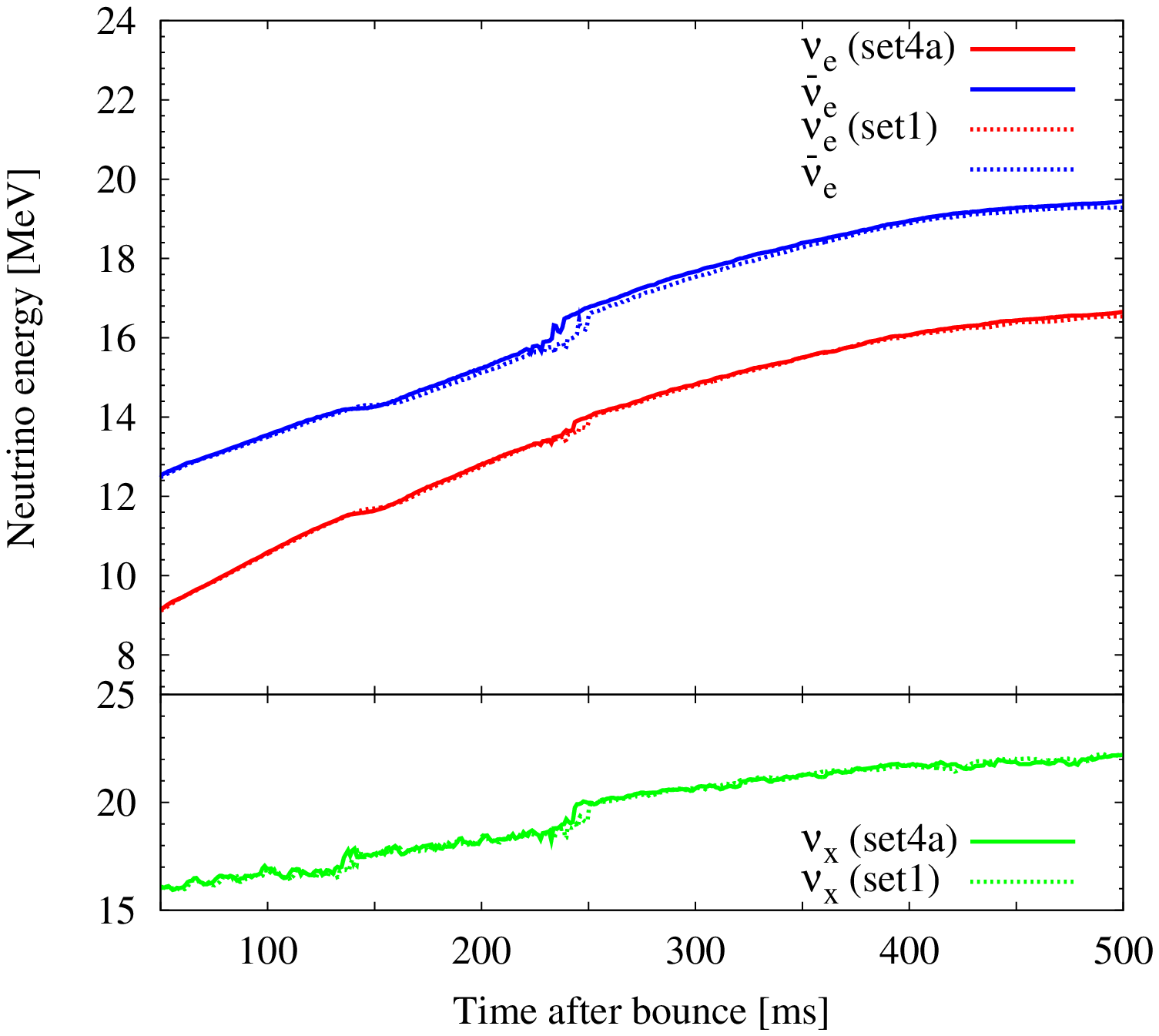}
\end{center}
  \caption{Comparison of $\nu_e$ and $\bar\nu_e$ luminosities and the rms 
 energies (upper part of the panels) and those of $\nu_x$ (lower part of the panels)
 between set4a (solid lines) and set1 (dashed lines), respectively.}
\label{f10}
\end{figure}

The upper panels of Figure \ref{f10} compare the $\nu_e$ and $\bar\nu_e$ 
luminosities (left panel) and the rms energies (right panel) between
 set4a (solid lines) and set1 (dashed lines), respectively. After the accretion 
phase ($\sim 160$ ms after bounce), the $\bar\nu_e$ luminosity for set4a 
(blue solid line) becomes slightly larger (by $\lesssim 1 \%$) compared to 
that of set1 (blue dashed line). More apparent difference can be seen by comparing
 the $\nu_e$ luminosity of set4a (red solid line) that of set1 (red dashed line) 
  (approximately $3 \sim 4 \%$ lower for set4a). First of all, 
these features are consistent with 
  the reduction of the $\bar\nu_e$ opacity (leading to higher $\bar\nu_e$ luminosity) 
and the increase of the $\nu_e$ opacity (lower $\nu_e$ luminosity) 
due to the mean-fields effects, as we mentioned above.
Regarding the rms neutrino energies (right panel), the mean-field effects increase
 the $\bar \nu_e$ energy by $\sim 20 \%$, but barely affect the $\nu_e$ energy, also the $\nu_x$ luminosities and the rms energy (compare green solid 
 lines with green dashed lines). 

The bigger mean-field effects observed in this study, such as 
on the $\nu_e$ luminosity compared to the $\bar\nu_e$
 luminosity, the same as for the $\bar \nu_e$ rms energy compared to the $\nu_e$ rms, 
 are consistent with \citet{horowitz12}. Note that \citet{horowitz12}
 observed more stronger impact of the mean-field effects, especially on the 
increase of the $\bar \nu_e$ rms energy and the reduction of the $\nu_e$ luminosity
 (see their Figure 4). The employed progenitor (15 $M_{\odot}$) and the EOS (LS220) are 
the same as those in this work. However, the quantitative differences from 
\citet{horowitz12} could originate from their use of $\Delta U$ obtained 
by a virial expansion calculation
  (not from the LS220 EOS as in this work), the inclusion of the weak magnetism
 correction (not included in our set4a and set1), and the GR hydrodynamics 
(essentially Newtonian hydrodynamics in this work).\footnote{Note also that 
comparison with \citet{GMP12} is more difficult because they focused on the 
 later postbounce evolution (after $\sim $500 ms) 
in the {\tt Agile-BOLTZTRAN} run of a different progenitor ($18 M_{\odot}$ star) using
 the different EOS \citep{Shen98}.}

\begin{figure}[H]
\begin{center}
\includegraphics[width=81mm]{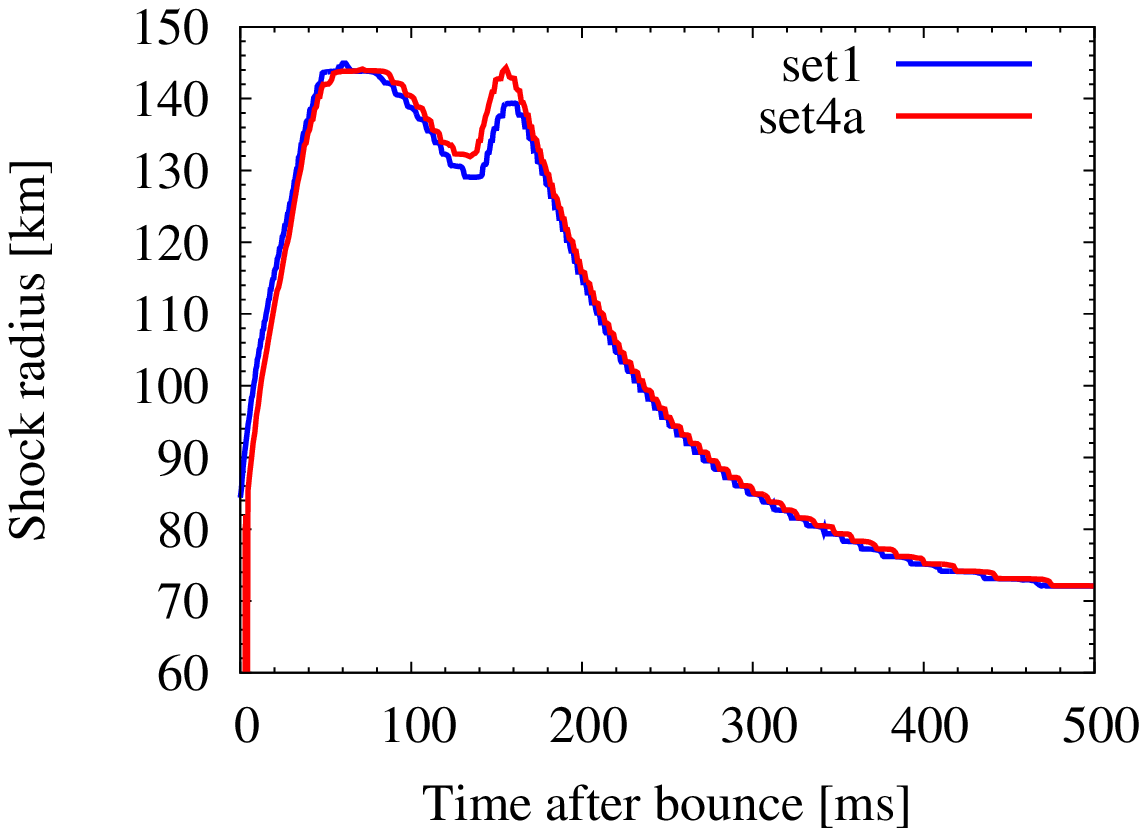}
\includegraphics[width=81mm]{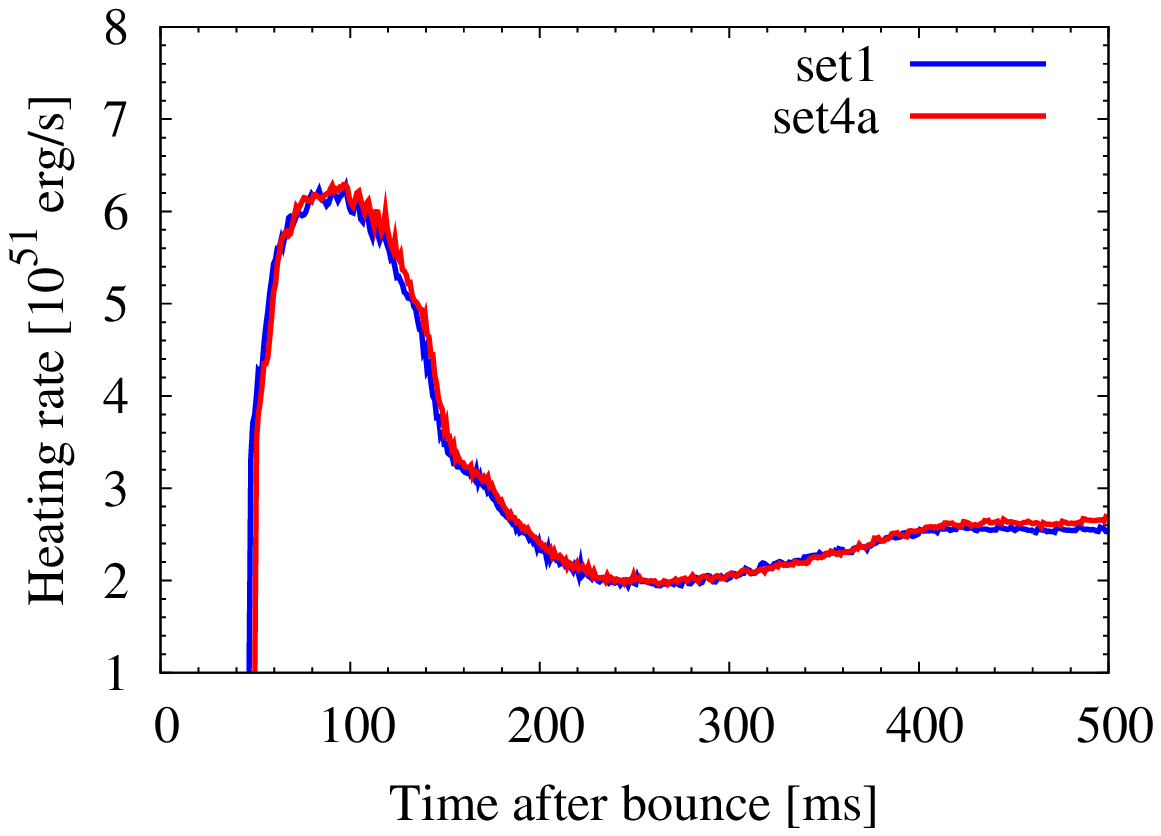}
\end{center}
  \caption{Comparison of the shock radius 
(left panel) and
 the net heating rate in the gain region (right panel) for set1 and set4a, respectively.}
\label{f11}
\end{figure}

The left panel of Figure \ref{f11} compares the shock radius between 
set4a and set1. The shock radius becomes 
 slightly bigger for set4a (red line) compared to set1 (blue line) 
 for a short period ($\sim 120 - 160$ ms after bounce) when the Si-rich layer is 
advecting through the shock, but the difference disappears thereafter. Note in the 
period that the $\bar\nu_e$ luminosity is bigger than the $\nu_e$ luminosity
 (left panel of Figure \ref{f10}). As mentioned above, the mean-field effects
 of set4a lead to higher $\bar\nu_e$ luminosity than set1, which is consistent
 with the bigger shock radius, albeit transietnly. 

The right 
 panel of Figure \ref{f11} compares the net heating rate in the gain region, suggesting
  that the mean-field effects, as previously reported 
(e.g., \citet{GMP12,horowitz12}), would not have a significant impact on the onset of 
an explosion. 
Similar comparison between set4b and set1 shows that
 the difference from set1 due to the in-medium suppression of Bremsstrahlung 
 (see Table \ref{table1}) is much smaller comparing to the mean-field effects mentioned above.
 The comparison plots between set4b (not shown) and set1 are almost completely overlayed 
(like in the middle left panel of Figure \ref{f12} or 
in the bottom right panel of Figure \ref{f8}). 
 It was shown (e.g., \citet{fischer16} and \citet{bartl16}) that 
 the medium suppression of Bremsstrahlung affects the neutrino properties 
 only clearly after the onset of an explosion and the following
 PNS cooling phase. In this respect, our results
 showing a negligle impact in the pre-explosion phase are in line with the literature.

\subsection{Weak Magnetism and Recoil (set5a), Nucleon Effective Mass (set5b)}
\label{set5}

\begin{figure}[H]
\begin{center}
\includegraphics[width=81mm]{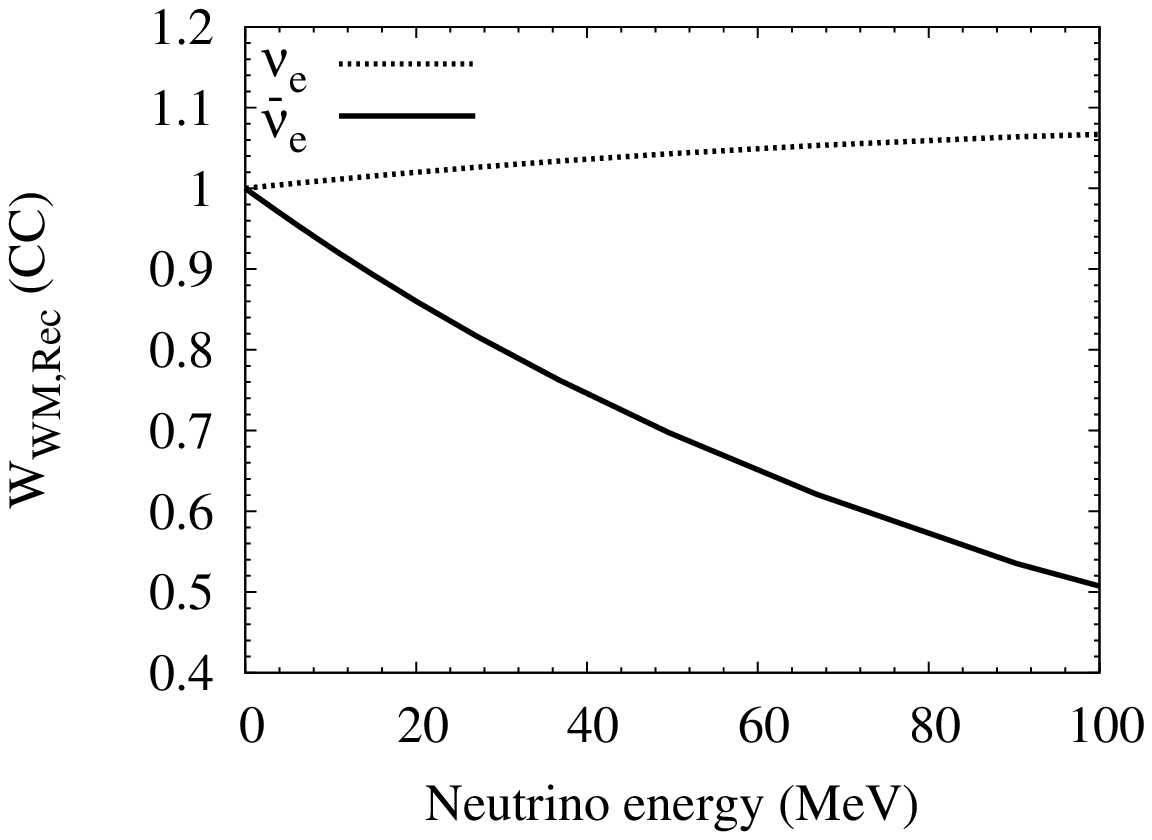}
\includegraphics[width=81mm]{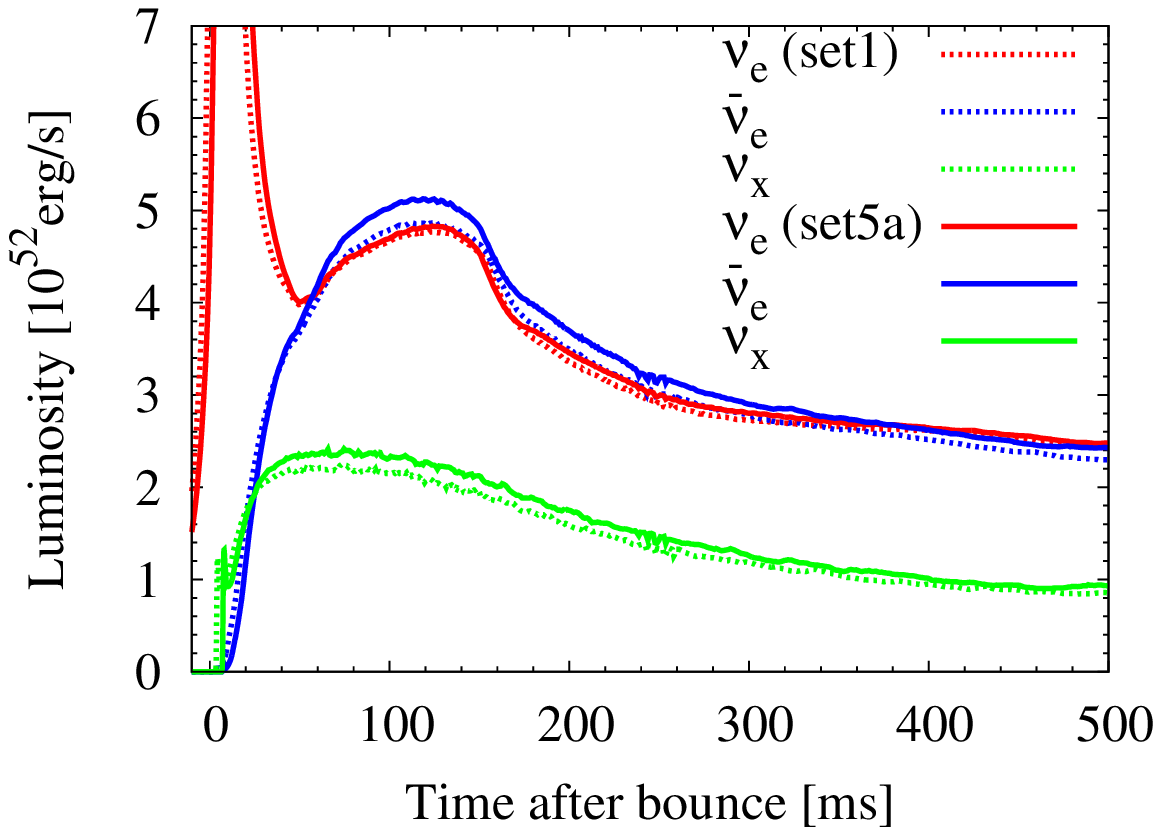}
\includegraphics[width=81mm]{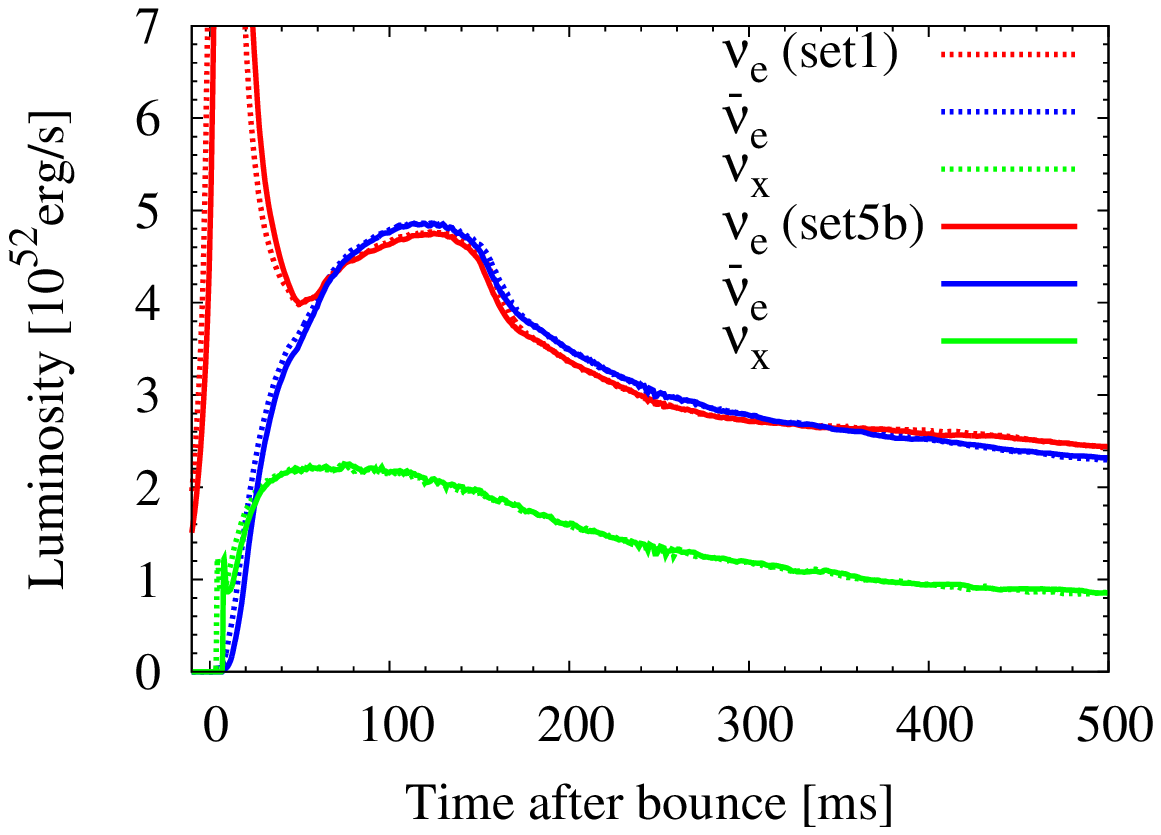}
\includegraphics[width=81mm]{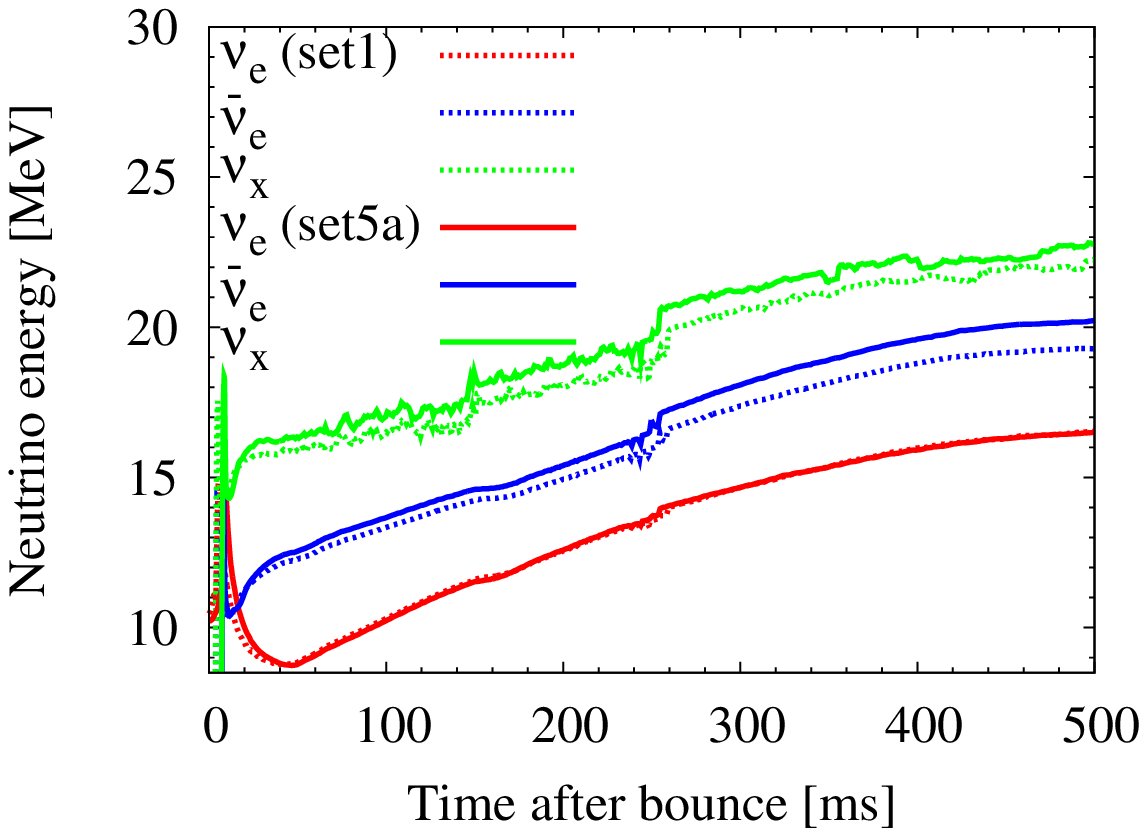}
\includegraphics[width=81mm]{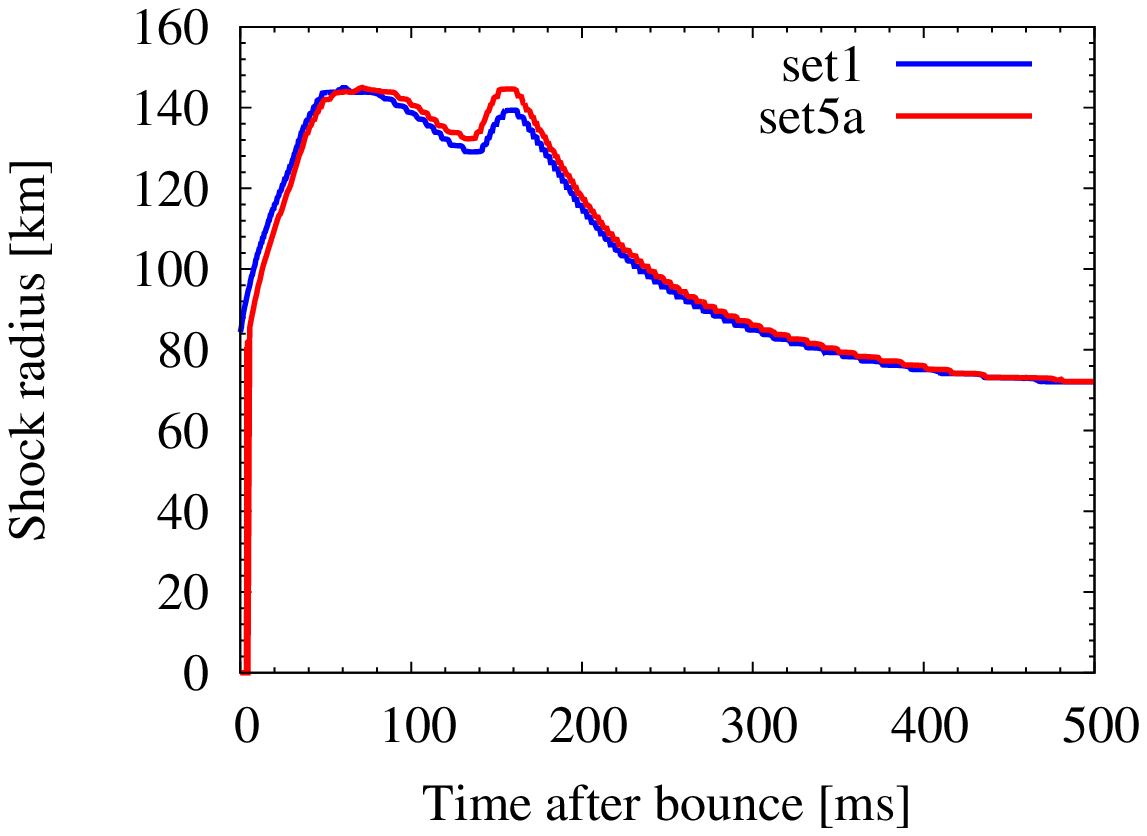}
\includegraphics[width=81mm]{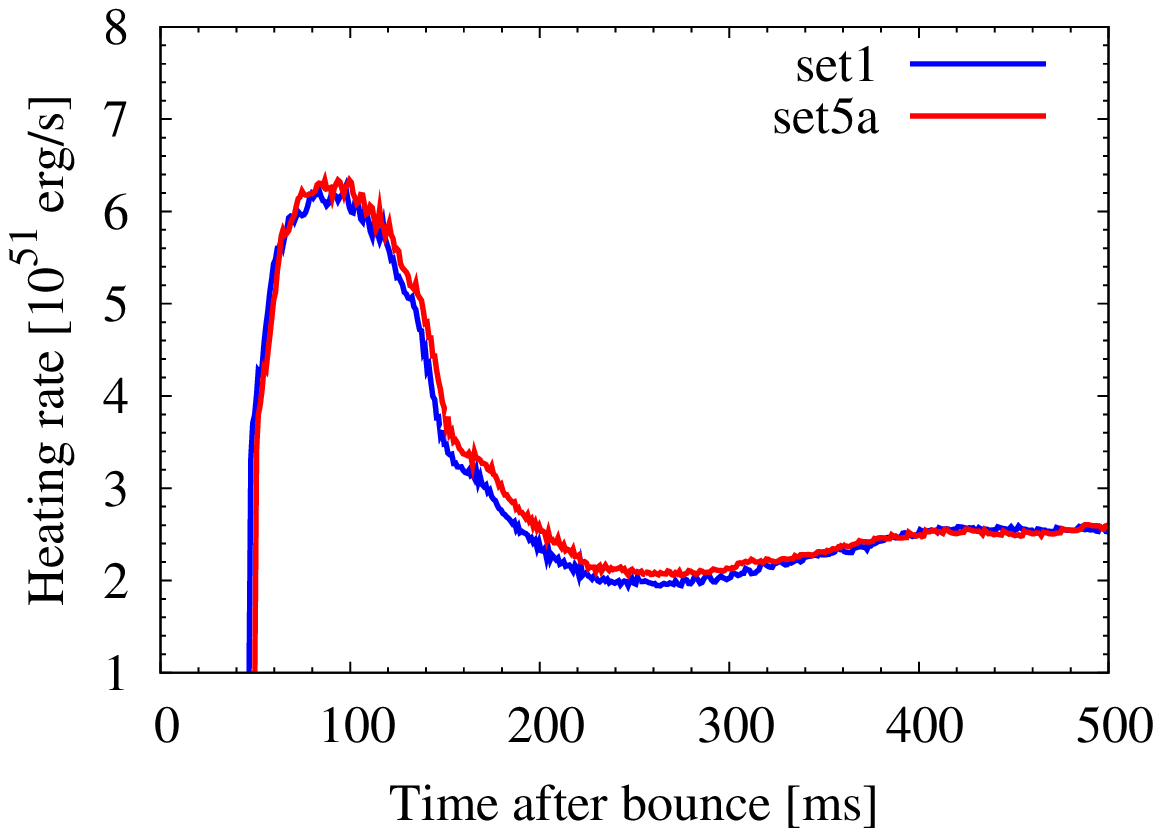}
\caption{The top left panel shows the correction factor of the charged 
 current cross section for $\nu_e$ (dashed line) and $\bar\nu_e$ (solid line)
 including weak magnetism and recoil as a function of neutrino energy.
Note that except for middle left panel (showing a negligible difference 
from set5b and set1), the other panels are comparison between set5a and set1.}
\label{f12}
\end{center}
\end{figure}

In order to take into account the effects of weak magnetism and recoil both on the 
 charged current (CC) and neutral current (NC) reactions, we follow
 \citet{horowitz02} (their Equations (22) and (32)). The top left panel 
 of Figure \ref{f12} correpsonds to Figure 1 \citep{horowitz02}, showing that the 
main effect from weak magnetism is to reduce $\bar\nu_e$ opacity (solid 
line) by a large amount ($\sim 15 \%$ reduction at a neutrino energy of 20 MeV).
In comparison, the $\nu_e$ opacity is enhanced only by a small amount (dashed
 line). Regarding NC reactions, Figure 2 of \citet{horowitz02} shows that 
the reduction of the opacity is slightly higher for $\nu p$ scattering than
 $\nu n$ scattering, and that the reduction of the $\bar\nu$ 
reactions is higher than the corresponding $\nu$ reactions 
($\sim 10 \%$ reduction for $\bar\nu$ at a neutrino energy of 20 MeV).
 
The top right panel of Figure \ref{f12} compares the neutrino luminosities 
between set5a (solid lines) and set1 (dashes line) 
with and without weak magnetism and recoil, respectively. One can clearly 
see the enhancement of the $\bar\nu_e$ 
 luminosity (blue solid line) for set5a, which is by $\sim 8 \%$ bigger than 
set1 (blue dashed line). This comes from the reduction of the $\bar \nu_e$ opacity 
as mentioned above. The difference, however, becomes very small after $\sim 340$ ms postbounce.  The $\nu_e$ luminosities (red solid line
 and red dashed line) are hardly affected, 
which is in line with the very small change in the $\nu_e$ opacity. The $\nu_x$ 
luminosity (green solid line and green dashed line) is enhaced up to about 
$\sim 10 \%$ for set5a compared to set1. Regarding the rms neutrino energies (middle
 right panel), the reduced opacities of $\bar\nu_e$ and $\nu_x$ result
 in the higher $\bar \nu_e$ (blue solid line) and $\nu_x$ energies (green solid lines) 
up to $\sim 1$ MeV, comparing to those 
of set1 (blue dashed line and green dashed line).

From the bottom left panel of Figure \ref{f12}, one can see that the shock radius of set5a is 
 transiently bigger than set1 for the epoch when the Si-rich shell is passing.
 This is most likely to come from the higher $\bar\nu_e$
 luminosity and energy (top right and middle right panel). In fact, the 
 net heating rate is slightly higher for set5a (red line) compared to set1 (blue line)
 up to the first $\sim 340$ ms after bounce. Thereafter, the $\bar \nu_e$
 luminosity with and without weak magnetism correction approach each other together 
 (top right panel). Set5b (effective mass correction, e.g., Table \ref{table1}) does not exhibit visible changes from set1 (as was 
 the case for set3b and set4b), only the comparison plot of the neutrino luminosities
 (middle left panel) is shown as a reference.
 


\subsection{Quenching of $g_a$ (set6a), Many-body Effect (set6b), Strangeness Contribution (set6c), and the Whole Set}
\label{set6}

Finally, the model series with "set6" include modifications to the 
 axial-vector currents in the weak interactions either from the in-medium effects 
(set6a), many-body effects (set6b), or strangeness-dependent contributions
 (set6c), respectively (see Section \ref{2-2} and Table \ref{table1} for details).

\begin{figure}[H]
\begin{center}
\includegraphics[width=81mm]{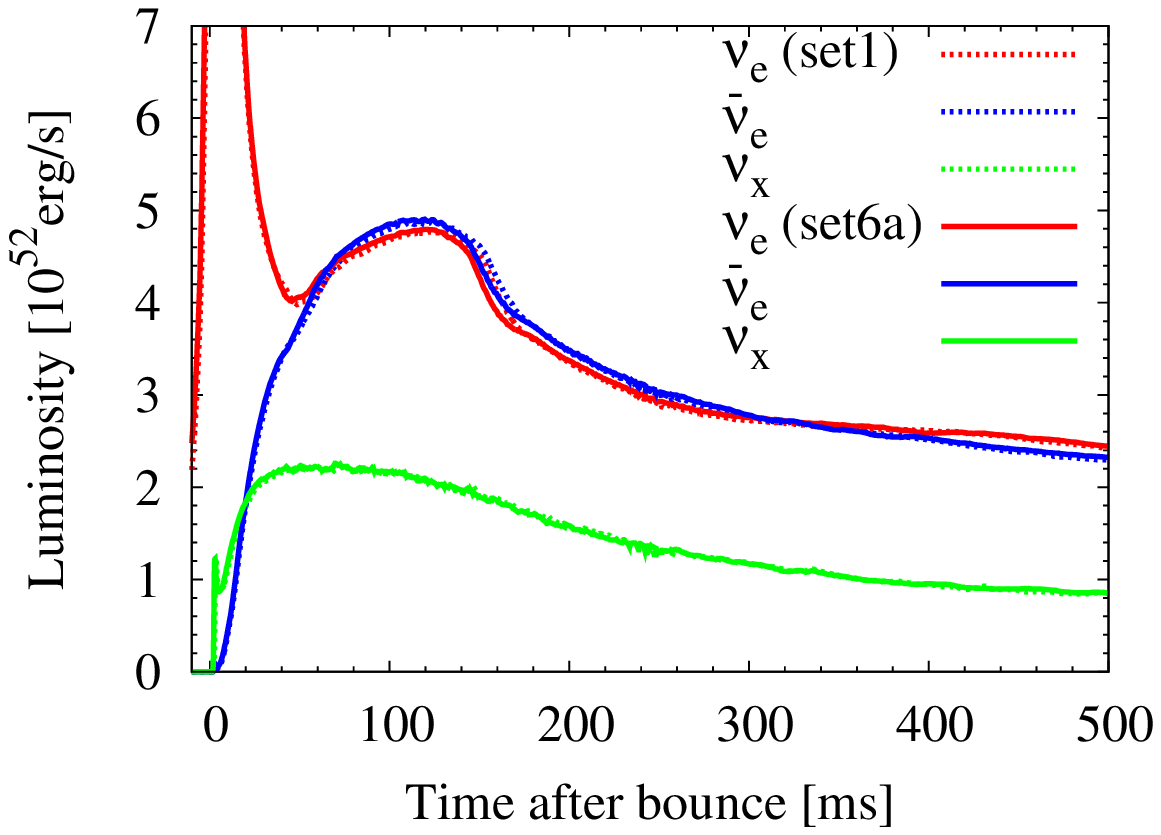}
\includegraphics[width=81mm]{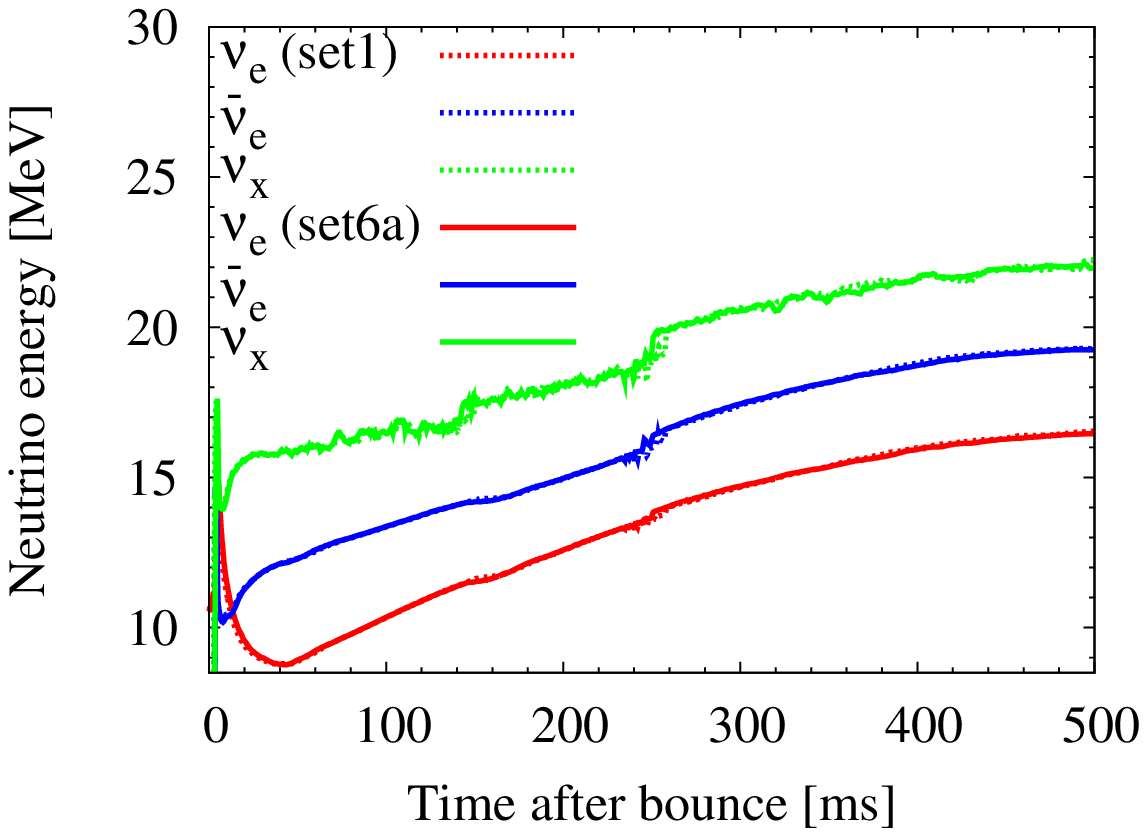}
\includegraphics[width=81mm]{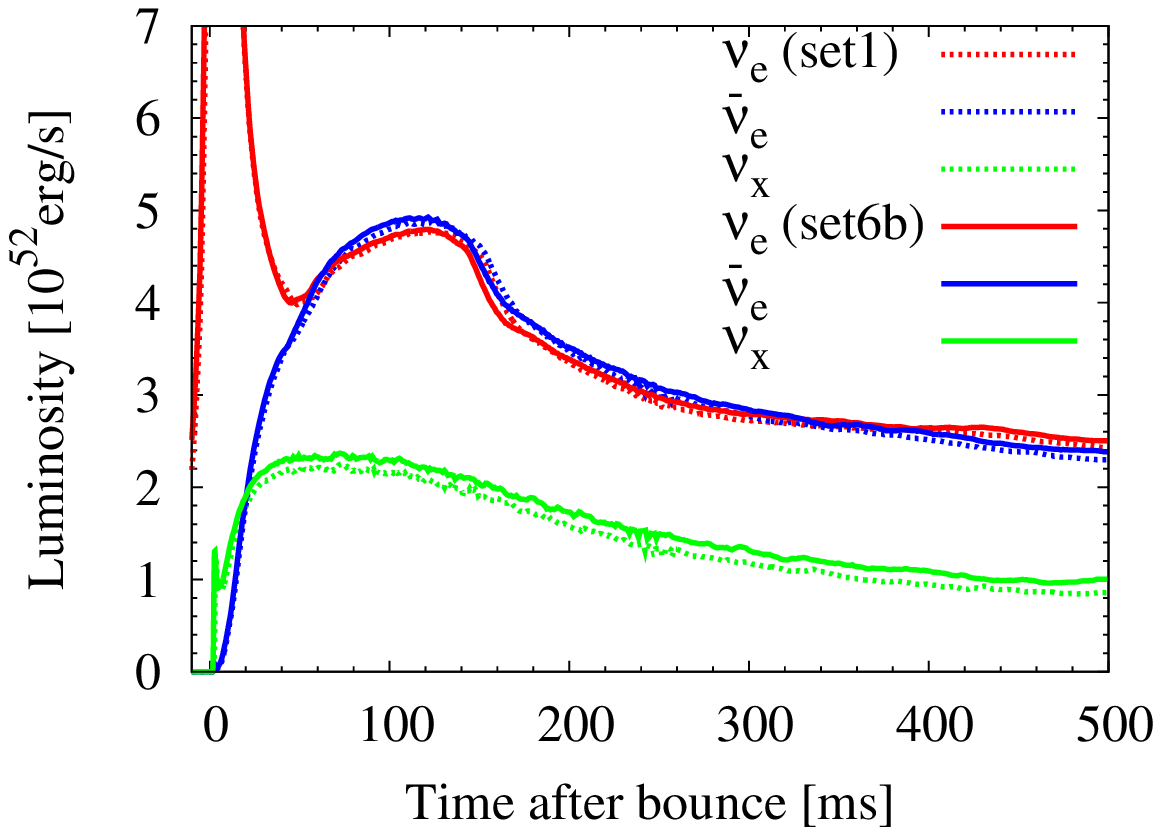}
\includegraphics[width=81mm]{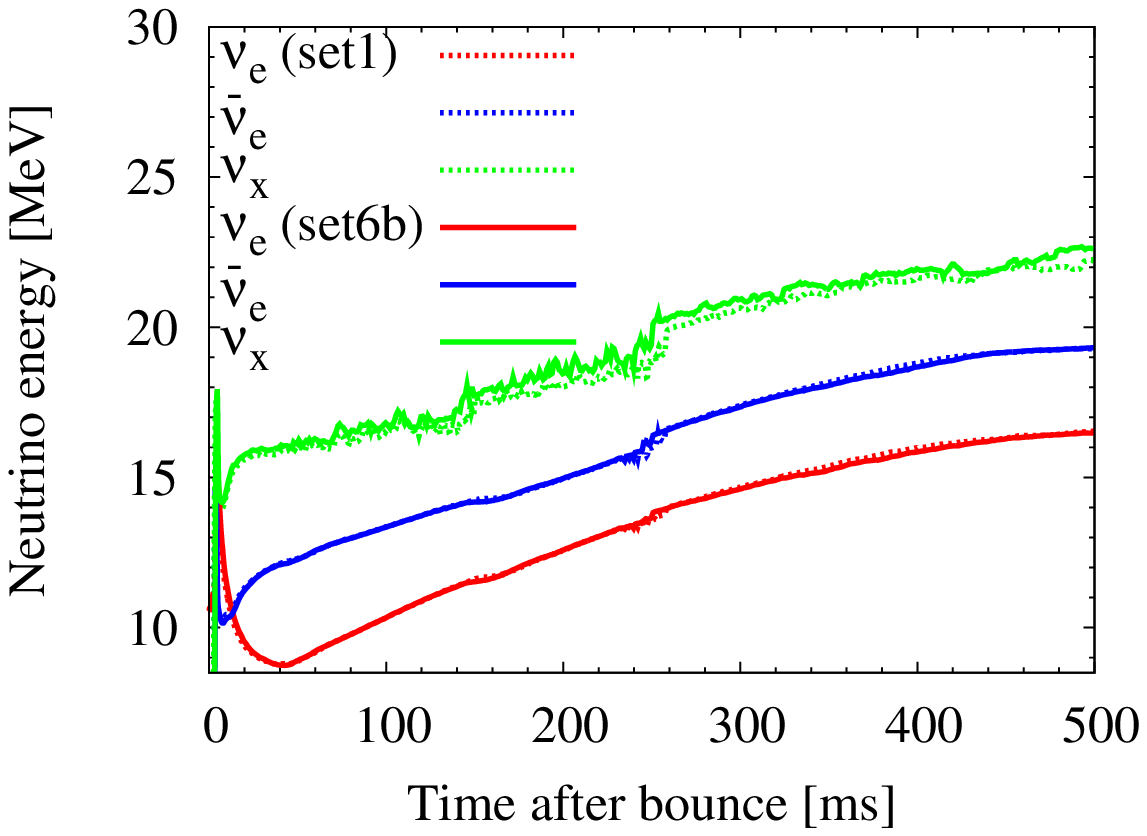}
\includegraphics[width=81mm]{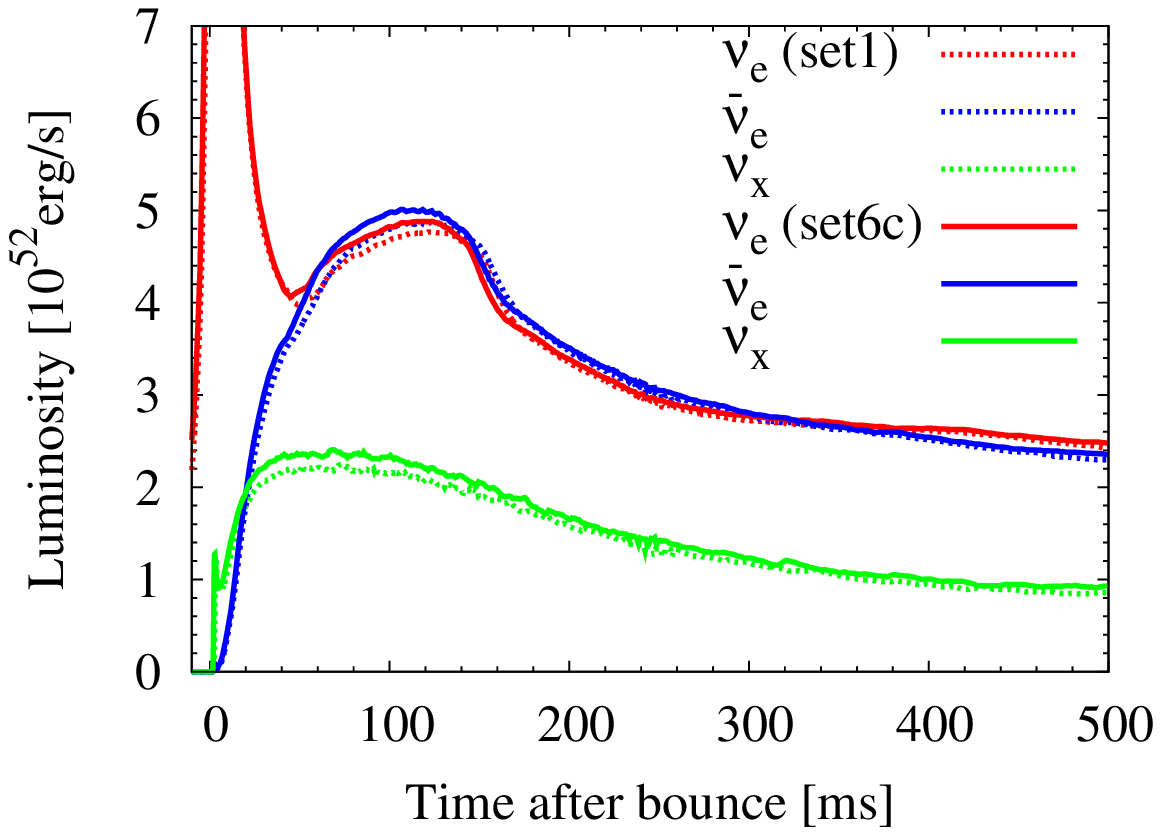}
\includegraphics[width=81mm]{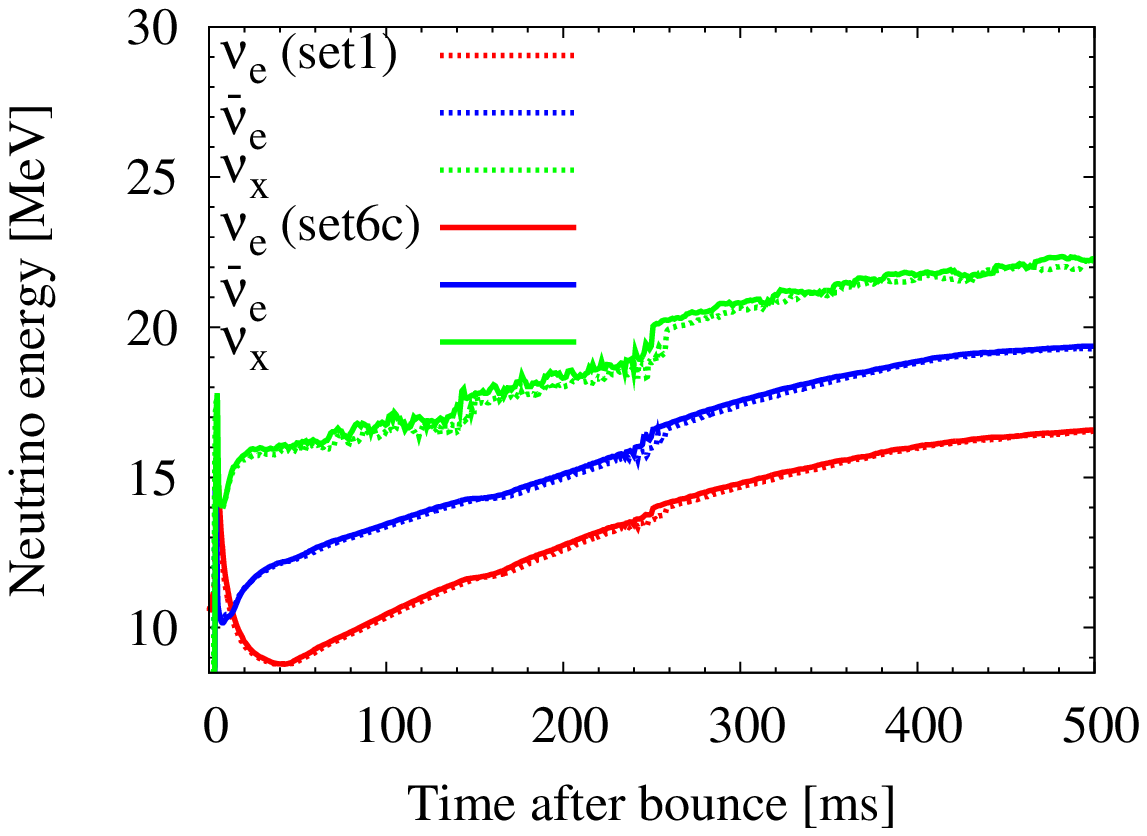}
\caption{Comparison of the neutrino luminosities (left panels) and rms neutrino 
 energies (right panels) between the set6 series (solid lines) with set1 (dashed lines),
 respectively.
}
\label{f13}
\end{center}
\end{figure}

From the top panels of Figure \ref{f13}, it is very hard to see significant
 differences between set6a and set1. This suggests
 that the quenching of $g_a$ plays a negligible role in the first 500 ms after bounce 
covered in our 1D run. For set6b, the left middle panel shows that the $\nu_x$ luminosity is 
higher by $\sim 10\%$ (green solid line) compared to set1 (green dashed line). 
The relative difference becomes larger in the later postbounce phase
predominantly because the many-body effects reduce the opacity of
 the $\nu N$ scattering at high densities \citep{horowitz17}. This is 
 also the case for the $\nu_e$ and $\bar \nu_e$ luminosities, where the
 luminosities become higher by $\sim 3 - 4 \%$ for set6b toward the final simulation time. The clearer
 impact of the many-body effects on $\nu_x$ compared to $\nu_e$ and $\bar\nu_e$ is 
 also seen in the middle right panel, showing an increase of $\sim 1$ MeV 
in the $\nu_x$ energy for set6b (green solid line) comparing to set1 (green dashed line).

The bottom left panel of Figure \ref{f13} shows a clear increase
 of the $\nu_e$ and $\bar{\nu}_e$ luminosities (by $\sim 4 \%$) 
for set6c compared to set1 in the first $\sim$ 160 ms after bounce.
At this epoch, the increase in the $\nu_x$ luminosity is more bigger ($\sim 9 \%$).
The bottom right panel shows that the strangeness effects lead to a 
 slight increase in the rms neutrino energies where the maxium upshift is 
$\sim 0.2$ MeV in the $\nu_x$ energy (green solid line and green dashed line).
 These trends with the strangess contribution are qualitatively 
consistent with \citet{melson15b}.
In the 3D full-scale simulations by \citet{melson15b}, they observed 
 much bigger effects from the strangeness effects, such as 
$\sim 30 \%$ and $10 -15 \%$ increase in the $\nu_x$ and $\nu_e$/$\nu_e$ luminosity,
 respectively, and $\sim 1$ MeV increase in the mean neutrino energies. Note
 in \citet{melson15b} that the use of the larger value of $g^s_a = - 0.2$
 and the choice of the more massive progenitor with the higher mass 
accretion rate (a $20 M_{\odot}$ star) could potentially lead to the more clearer 
impact of the strangeness effect comparing to those in this work (see also 
 \citet{Bollig17} for 2D results using $g^s_a = - 0.1$).

\begin{figure}[H]
\begin{center}
\includegraphics[width=81mm]{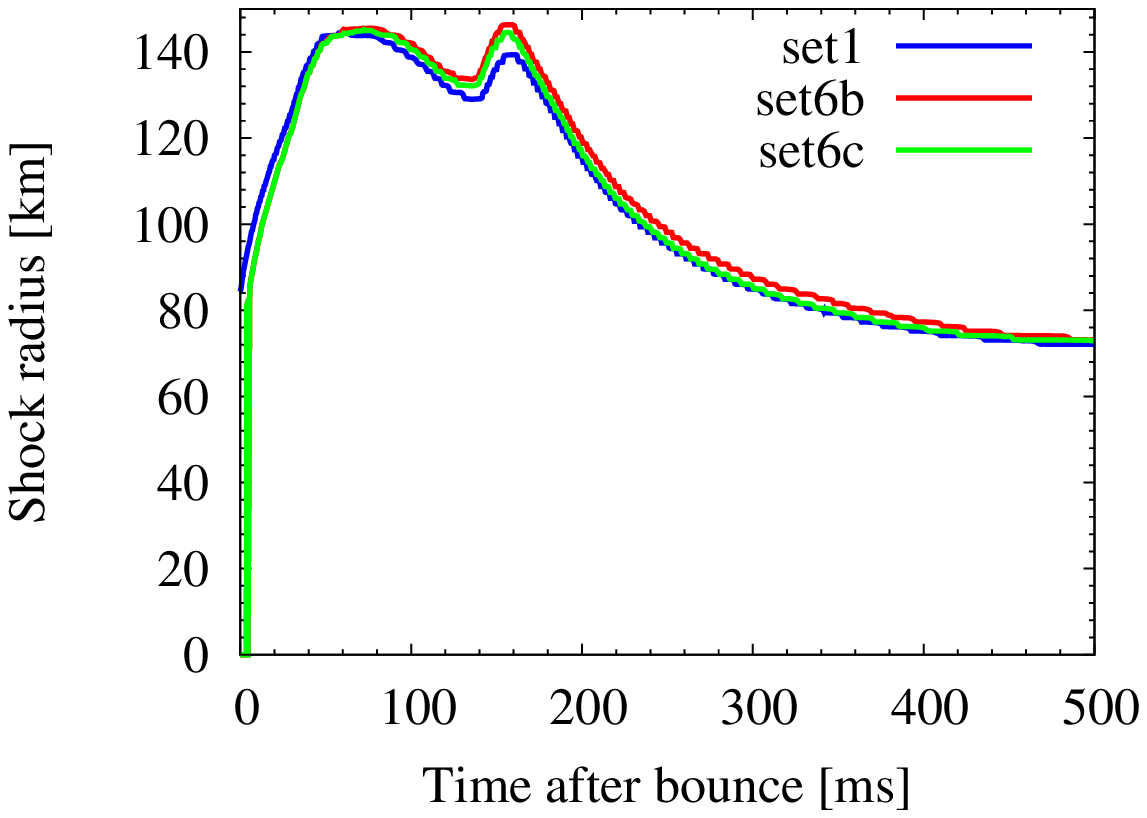}
\includegraphics[width=81mm]{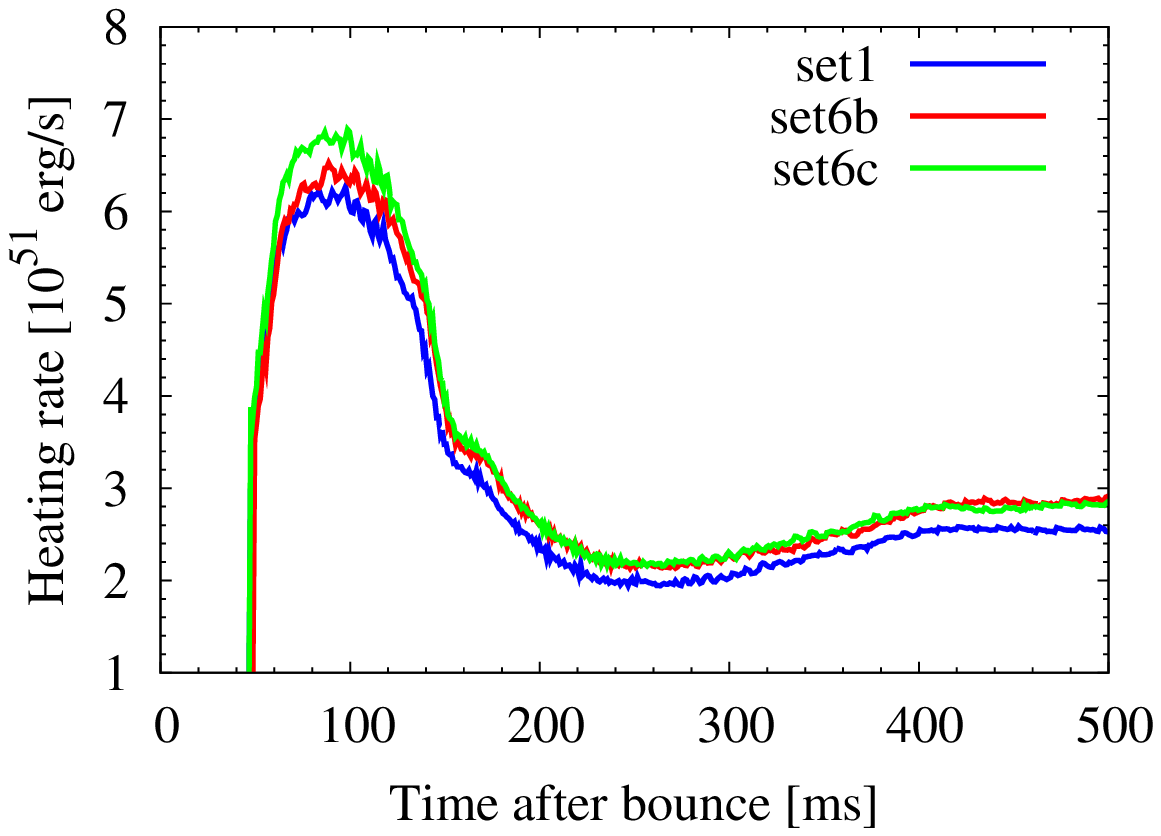}
\end{center}
  \caption{Comparison of the shock radius and the net heating rate in the 
 gain layer between set6b, set6c, and set1, respectively. Note that set6a is not
 shown because of the overlap with set1, which makes the differences in the plots 
difficult to see.}
\label{f14}
\end{figure}

The top left panel of Figure \ref{f14} shows that maximum shock extent becomes 
 by $\sim 5 \%$ bigger for set6b and set6c compared to set1 near the hump region
 ($\sim 160$ ms after bounce). The mentioned 
higher $\nu_e$ and $\bar \nu_e$ luminosities in
 the accretion phase (e.g., Figure \ref{f13}) 
is in line with this feature. In fact, the right panel 
 of Figure \ref{f14} shows that the net heating rate for set6b 
(red line) and set6c (green line) is bigger than that of set1 (blue line). For set6b, 
the increase from set1 (blue line) is about $\sim 6 \%$ 
around 100 ms after bounce and higher at later times. This is in good agreement with 
\citet{horowitz17} (see their Figure 3). Regarding the strange-quark contribution,
 the heating rate (set6c, green line) becomes larger by $\sim 12 \%$ than that of
 set1. This is in accordance with \citet{horowitz17}. Note that significantly 
bigger impact
 ($\sim 20 \%$ increase) was observed in \citet{horowitz17} probably because of 
 the larger value of $g^s_a = - 0.2$ and the use of a more massive 
 20 $M_{\odot}$ progenitor.



\begin{figure}[H]
\begin{center}
\includegraphics[width=81mm]{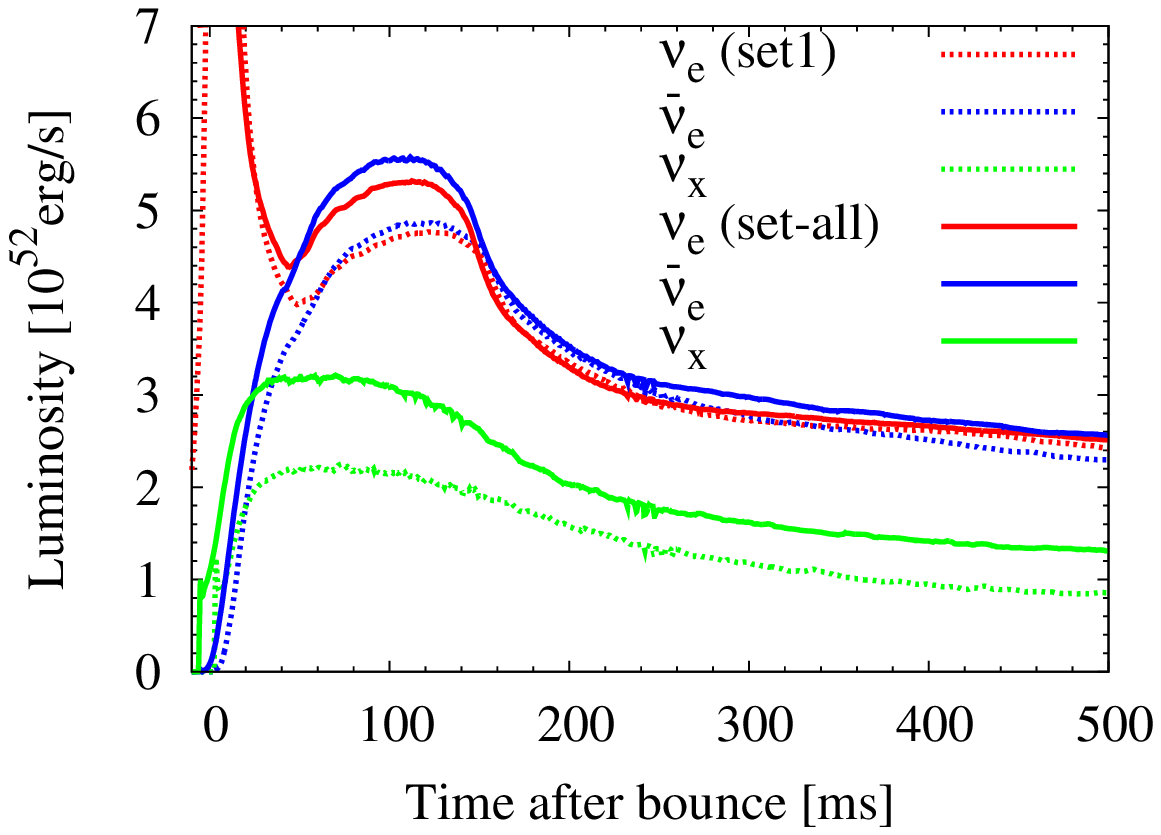}
\includegraphics[width=81mm]{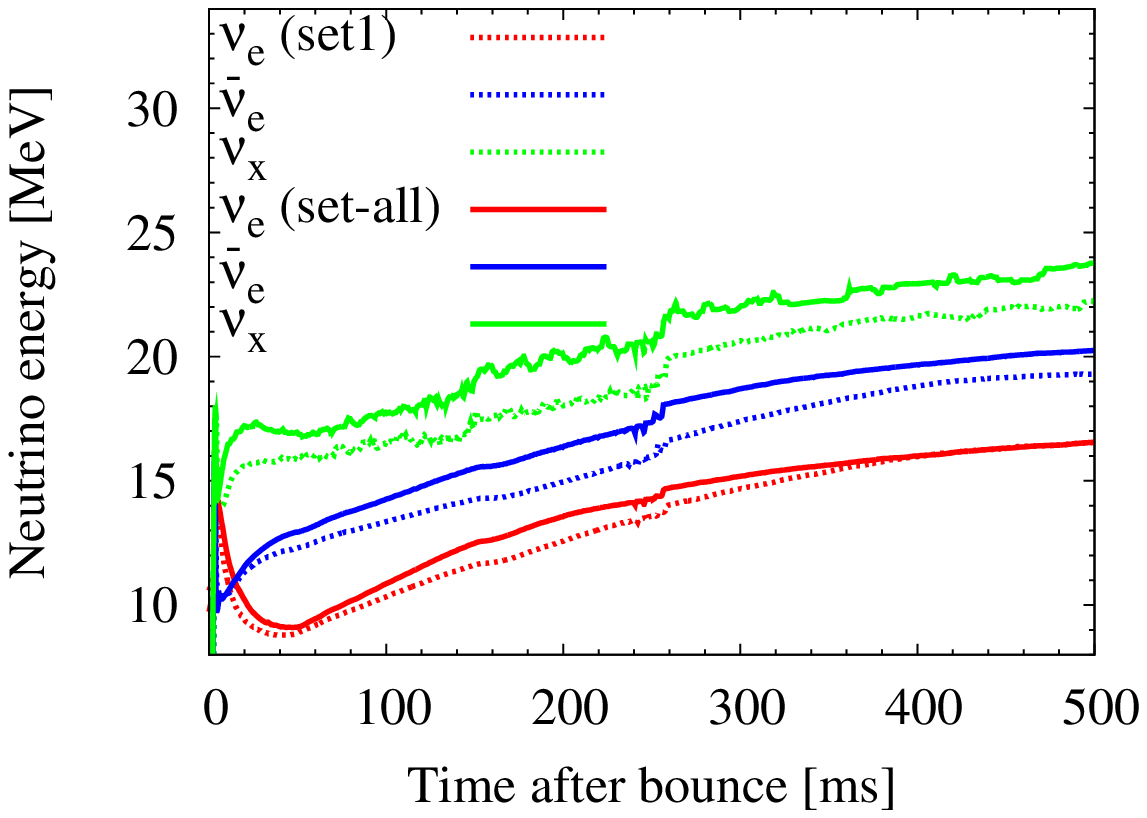}
\includegraphics[width=81mm]{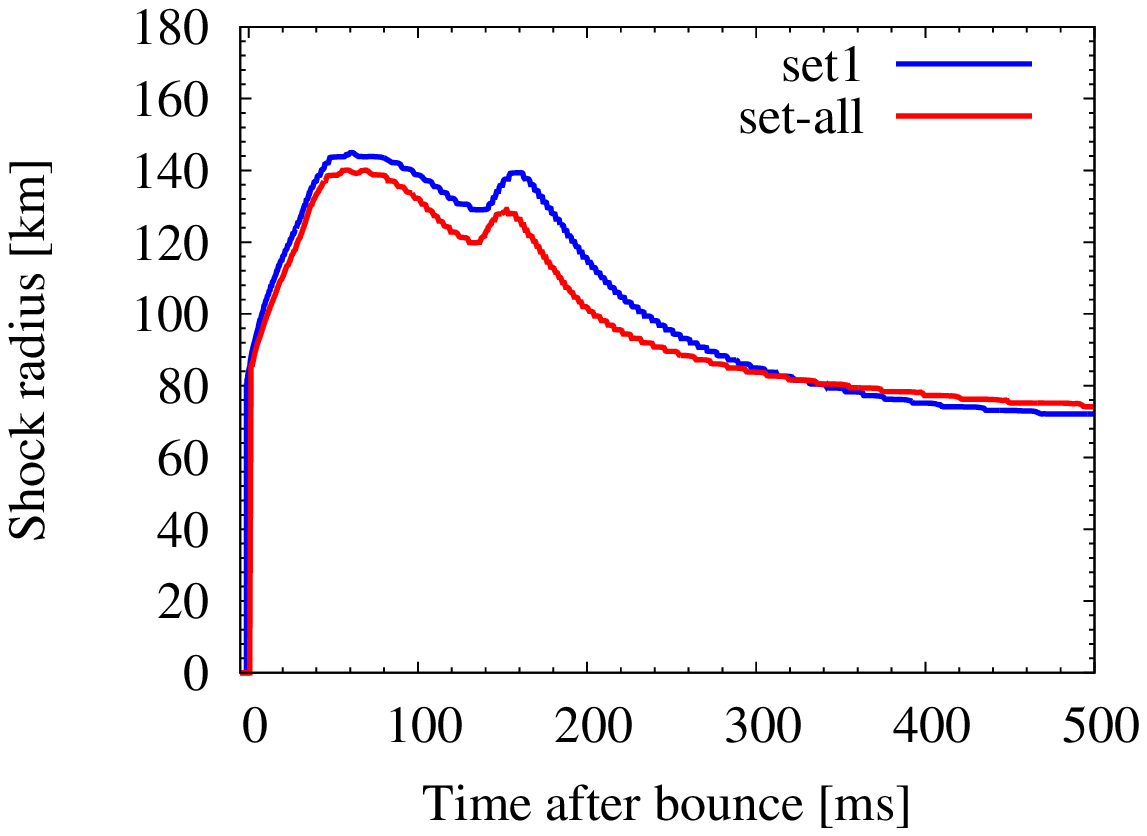}
\includegraphics[width=81mm]{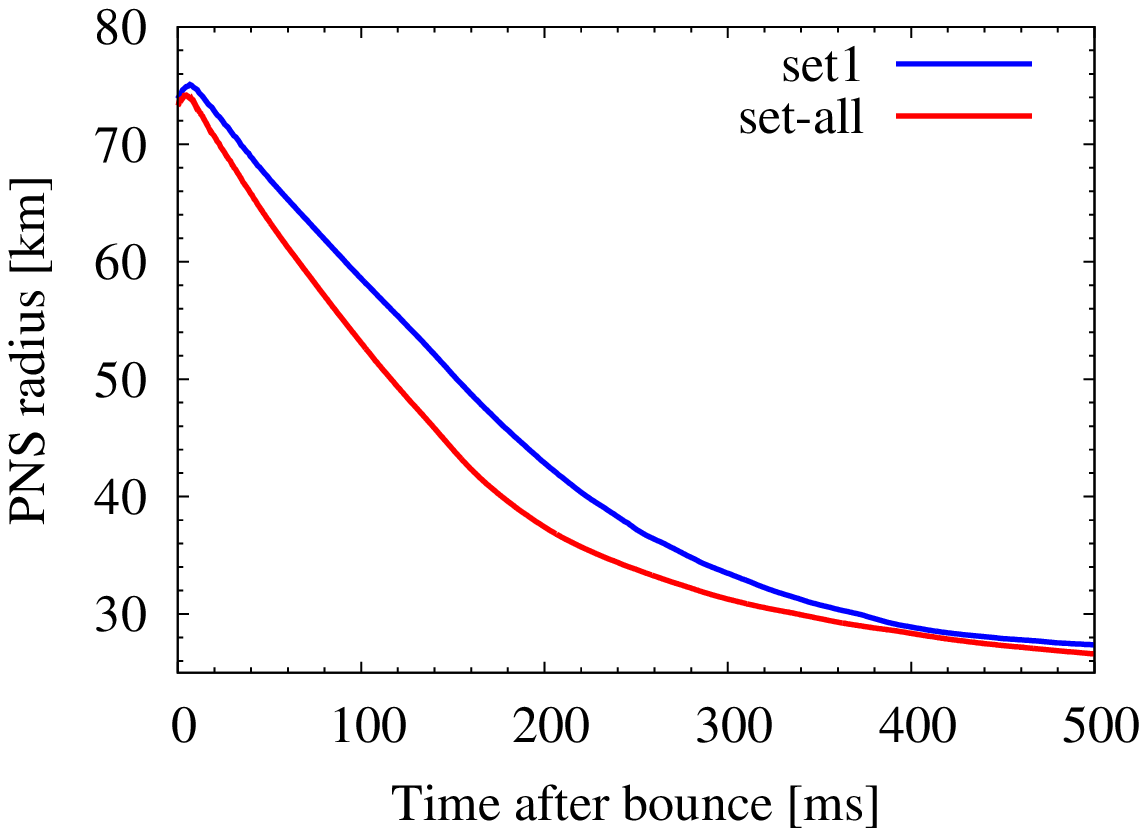}
\includegraphics[width=81mm]{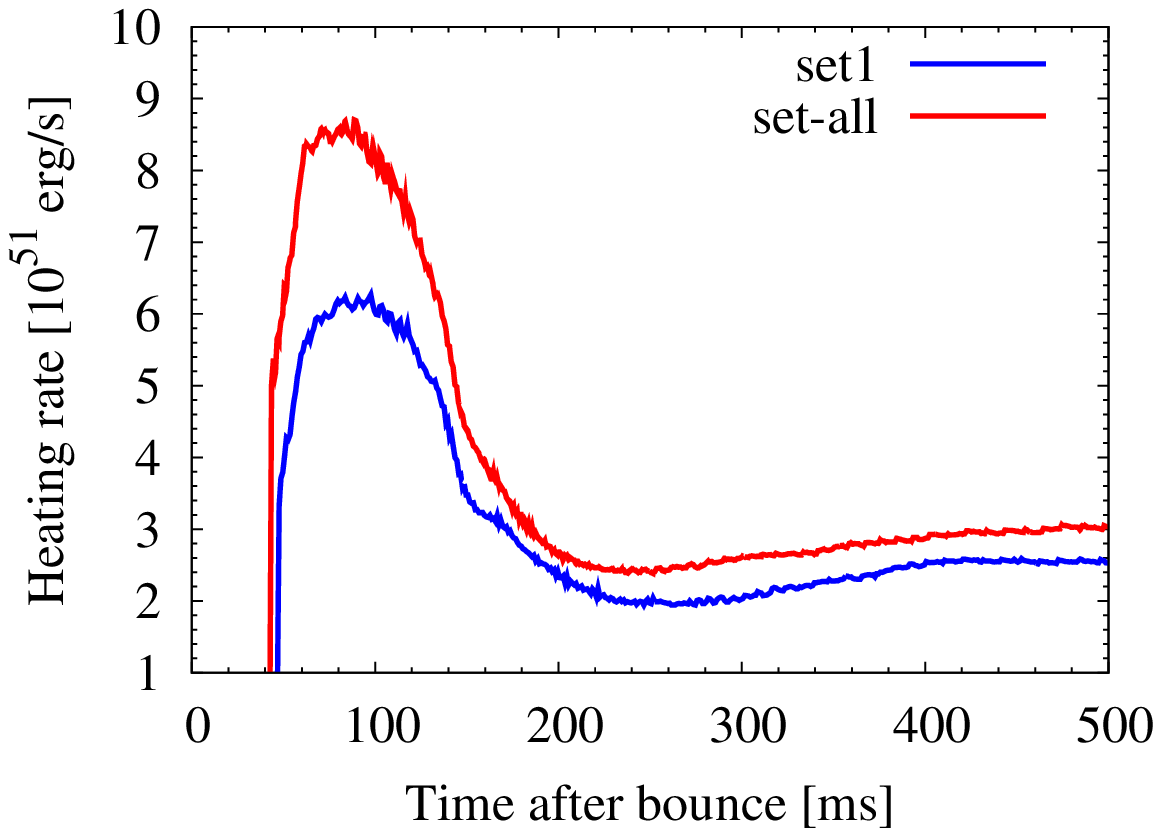}

\caption{Similar to Figure \ref{f5} but for the comparison between set-all and set1.
  Note that set-all includes all of the updates to
 set1 (from set2 to set6c in Table \ref{table1}).}
\label{f15}
\end{center}
\end{figure}

Finally, Figure \ref{f15} compares the model including all of the updates to
 set1 (from set2 to set6c in Table \ref{table1}, labeled as "set-all" in the panels\footnote{Note that the Pauli blocking factor of nucleons have been already included 
in \citet{horowitz17}, that one should not double count.}) 
with set1. Comparing with set1, the top left panel shows that the largest increase 
of the neutrino luminosities is for $\nu_x$ (by $\sim 31 \%$), 
which is followed in order by $\bar \nu_e$ ($\sim 14 \%$) and $\nu_e$ ($\sim 11\%$).
Among the individual updates, the increase of the $\nu_x$ luminosity 
is biggest ($\sim 20\%$) due to the inclusion of electron neutrino pair annhilation
(set3a) as shown 
in the top left panel of Figure \ref{f7}. Note that the increase from the weak magnetism 
 and recoil is $\sim 10\%$ in set5b (top right panel of Figure \ref{f12}),
 from the many-body effects is $\sim 8 \%$ in set6b (middle left panel of Figure \ref{f13}),
 and $\sim 9 \%$
 from the strangeness contribution
 (middle left panel of Figure \ref{f13}).  
 As is expected, by adding
 the contribution from these individual rates (for the $\nu_x$ luminosity)
 does not simply explain the total increase ($\sim 31 \%$). 
 This is not surprising because the individual update affects non-linearly 
 the postbounce evolution that is governed by non-linear neutrino-radiation 
hydrodynamics.

The $\bar\nu_e$ luminosity is bigger ($\sim 5.4 \%$) than the $\nu_e$ luminosity 
at the peak around 100 ms after bounce, 
thereafter the difference becomes smaller towards the final simulation time.
As already seen from Figure \ref{f12} (top left panel), the dominance of
 the $\bar \nu_e$ over the $\nu_e$ luminosity is mainly due to
 the inclusion of the weak magnetism and recoil (set5a). Using the same 15$M_{\odot}$
 progenitor \citep{WW95}, this feature is also
 seen in \citet{BMuller14} (see their Figure 1, the panel labeled with "s15s7b2"), 
where neutrino signals from 2D GR simulations using the {\tt Vertex-CoCoNuT} code
 were investigated. In \citet{BMuller14}, the main difference regarding the 
microphysics inputs from this work is the 
use of LS180 EOS and the inclusion of the non-elastic effects in the 
charged current absorption reactions \citep{BS98,BS99}. The energy-redistribution from
 the latter would make the recoil effect smaller, which would explain the smaller
 difference between the $\bar\nu_e$ and the $\nu_e$ luminosity in \citet{BMuller14}.
Their 2D run of the 15$M_{\odot}$ star (G15) starts to explode at $\sim 570$ ms 
 after bounce and the postbounce dynamics deviates from 1D 
after around 100 ms after bounce \citep{bernhard13}. At the 100 ms after bounce,
 their $\nu_e$ and $\bar\nu_e$ luminosity is $\sim 5 \times 10^{52}$ erg/s, which 
is slightly lower than those in this study $\sim 5.3 - 5.6 \times 10^{52}$ erg/s. 
Note that the neutrino luminosities in \citet{bernhard13} take into account the 
GR effects (their Equation (2) and (3)), which could potentially lead to $\sim 10-20 \%$
 reduction comparing to those without the GR corrections. 
Regarding the $\nu_x$ luminosity, the peak value is $\sim 2.4 \times 10^{52}$ erg/s
 in \citet{bernhard13}, which is ($\sim 25 \%$) lower than that in this work. 
Although we do not have 
 a clear-cut answer, we consider that the reduction of the $\nu_x$ luminosity 
 due to the GR redshift effects could partly explain the discrepancy.
 This is because the neutrinospheric radii of $\nu_x$ are formed deeper inside
 where the GR correction becomes more significant among the other neutrino species.

The top right panel of Figure \ref{f15} shows that all the rms neutrino energies
 become higher for set-all (solid lines) than set1 (dashed lines) 
over the first 500 ms after bounce. The enhancement due to the updated opacity 
 is bigger for $\nu_x$ and $\nu_e$ by $\sim$ 2 MeV compared to $\nu_e$ by $\sim 1$ MeV.
Note in \citet{bernhard13} that not the rms but the mean neutrino energy was plotted.
 At the 100 ms after bounce,
 the mean energy is (probably incidentally) very close,  $\nu_e$, $\bar\nu_e$, and $\nu_x$ is $\sim 9$ MeV, $\sim 13$ MeV,
 and $\sim 15$ MeV in \citet{bernhard13}, which is $\sim 10$ MeV, $\sim 13$ MeV,
 and $\sim 15$ MeV in our work, respectively.
 
The middle left panel of Figure \ref{f15} 
shows that the shock position when the bounce shock stalls 
 (at 100 ms after bounce) is $\sim 3 \%$ smaller for set-all (red line) compared 
 to set1 (blue line). At the hump that marks the passing of the Si-rich layer through 
the shock ($\sim 160$ ms after bounce), the difference of the shock between 
set-all and set1 becomes largest $\sim 9\%$. After $\sim$ 300 ms postbounce, the 
two shock radius approaches very close, but the shock radius of set-all becomes
as big as $\sim 5\%$ compared to set1 toward the final simulation time. The enhanced 
$\nu_e$ and $\bar \nu_e$ luminosities due to the many-body effects (set6b, see the middle
 left panel after $\sim$ 300 ms postbounce) and the extended shock radius 
 (red line in the left panel of Figure \ref{f14}) is reconciled with the above 
features seen in set-all. The middle right panel of Figure \ref{f15} 
shows the more compact PNS radius 
 for set-all (red line) compared to set1 (blue line).
Note that the PNS radius is estimated at a fiducial density 
of $10^{11}\,{\rm g}\,{\rm cm}^{-3}$.
 The difference is biggest at 
 the hump seen in the shock evolution ($\sim 160$ ms after bounce), when the PNS radius is 
 smaller by $\sim 17\%$ for set-all relative to set1.
 Although the difference becomes smaller toward the final simulation time,
  the PNS radius is always smaller for set-all over the entire 500 ms after bounce.
 
The bottom panel of Figure \ref{f15} shows that the maximum enhancement 
 of the net heating rate of set-all (red line) is $\sim 30 \%$ compared to set1
 (blue line) at around 100 ms after bounce. After the 160 ms postbounce,
the net heating rate in the gain region becomes $\sim 10 - 24 \%$ higher 
 for set-all. This is predominantly because of the higher $\bar \nu_e$ and $\nu_x$ 
luminosities and rms energies (top left and right panels) and of the smaller PNS 
radius (middle right panel). As already discussed above, the improved opacities
  add non-linearly and synergetically to increase the net heating 
rate, where each of the individual update amounts to the 
$\lesssim 10 \%$ level (e.g., see Figures \ref{f5} to \ref{f14}).

\section{2D Results}\label{sec4}

\begin{figure}[H]
\begin{center}
\includegraphics[width=170mm]{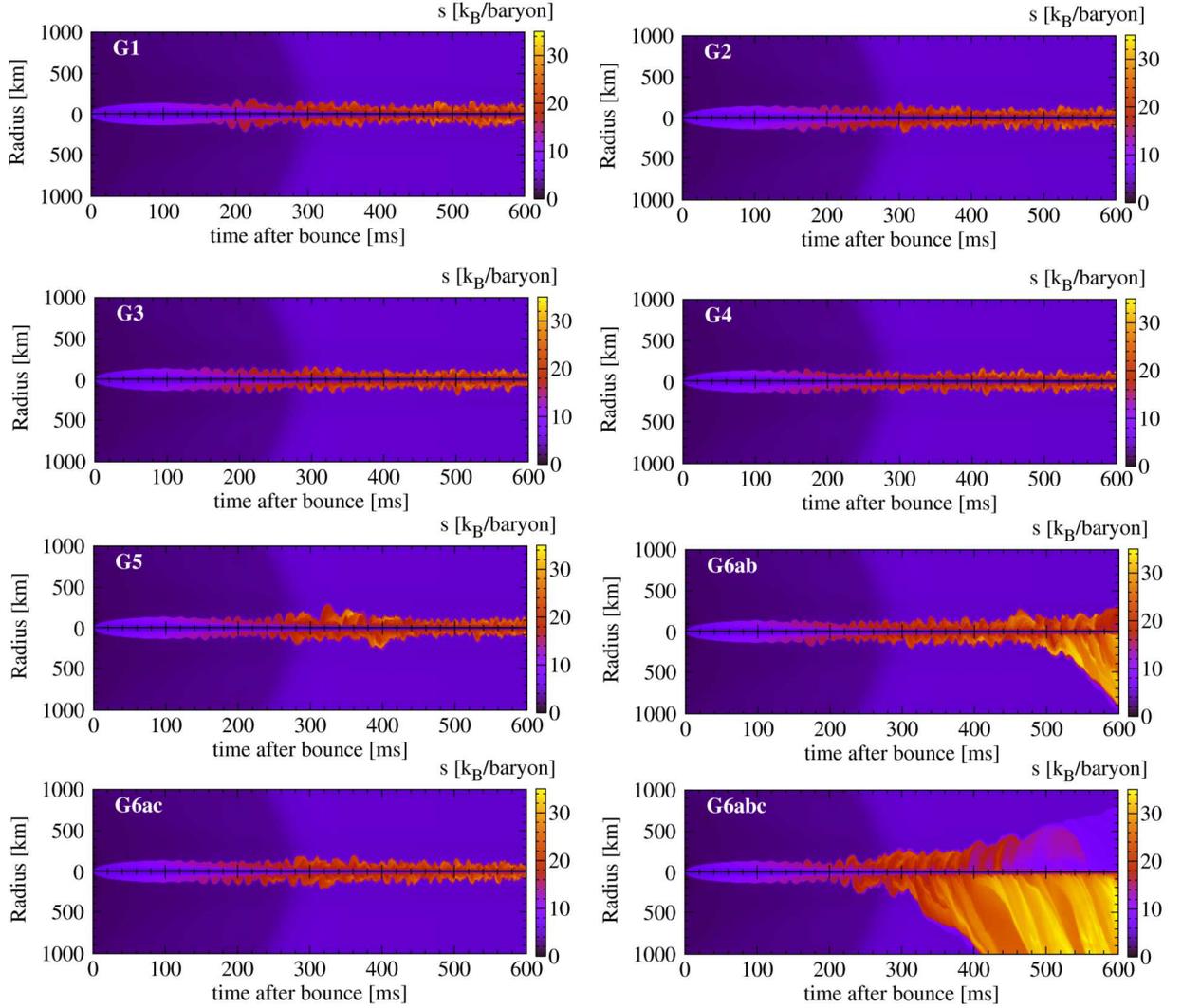}
\caption{Entropy along the north and south polar axis a postbounce time for 
the 2D simulations of G1 to G6abc (from top left to bottom panel). The shock trajectory 
 can be seen as a discontinuity between the dark violet and blue tones in the pre-shock 
region and light colors (red, orange, yellow) in the post-shock region. Here entropy is 
 in the unit of $k_B^{-1}\,{\rm nucleon}^{-1}$ with $k_B$ the Boltzmann constant.
}
\label{f16}
\end{center}
\end{figure}

In this section, we present results of our 2D core-collapse simulations where
 selected sets of the neutrino opacities in Table \ref{table1} are included in set1.
As already mentioned in Section \ref{sec2-1},  we choose 
 a 20$M_{\odot}$ star \citep{Woosley07} in our 2D runs. 

Our 2D run with the Bruenn rate (set1)
 is now called "G1" (meaning group one). "G2" is the model that 
is equivalent to set2 in the previous section. "G3" is the model where set3a and set3b are
 {\it added} to G2. Like this, for "G4", set4a and set4b are added to G3, and for "G5", 
set5a and set5b are added to G4. For "G6ab", set5b, set6a, and set6b are added to G5. 
Note that for G6ab we collectively add the three updates to G5 because set5b and set6a 
have no visible impact in our 1D runs.  The model difference between G6ab and G5 is 
the inclusion of the many-body effects of \citet{horowitz17}. For "G6ac", set5b, set6a, and set6c are added to G5. Finally "G6abc" corresponds to set-all in our 1D models where the strangeness contribution 
is added to G6ab.

Figure \ref{f16} shows a compact overview of all of the 2D runs. Up to the final simulation 
time of $\sim$ 600 ms after bounce, we observe the shock revival only for G6ab and G6abc. This may 
 not be very surprising because in 1D the many-body effects (set6b) and the strangeness
 contribution (set6c) are expected to primarily enhance the explodability (see, e.g.,
 right panel of Figure \ref{f14}). We furthermore explain in detail 
the reason that G6ac that simply includes the strangeness correction to G6a does not 
lead to explosion.

\begin{figure}[H]
\begin{center}
\includegraphics[width=81mm]{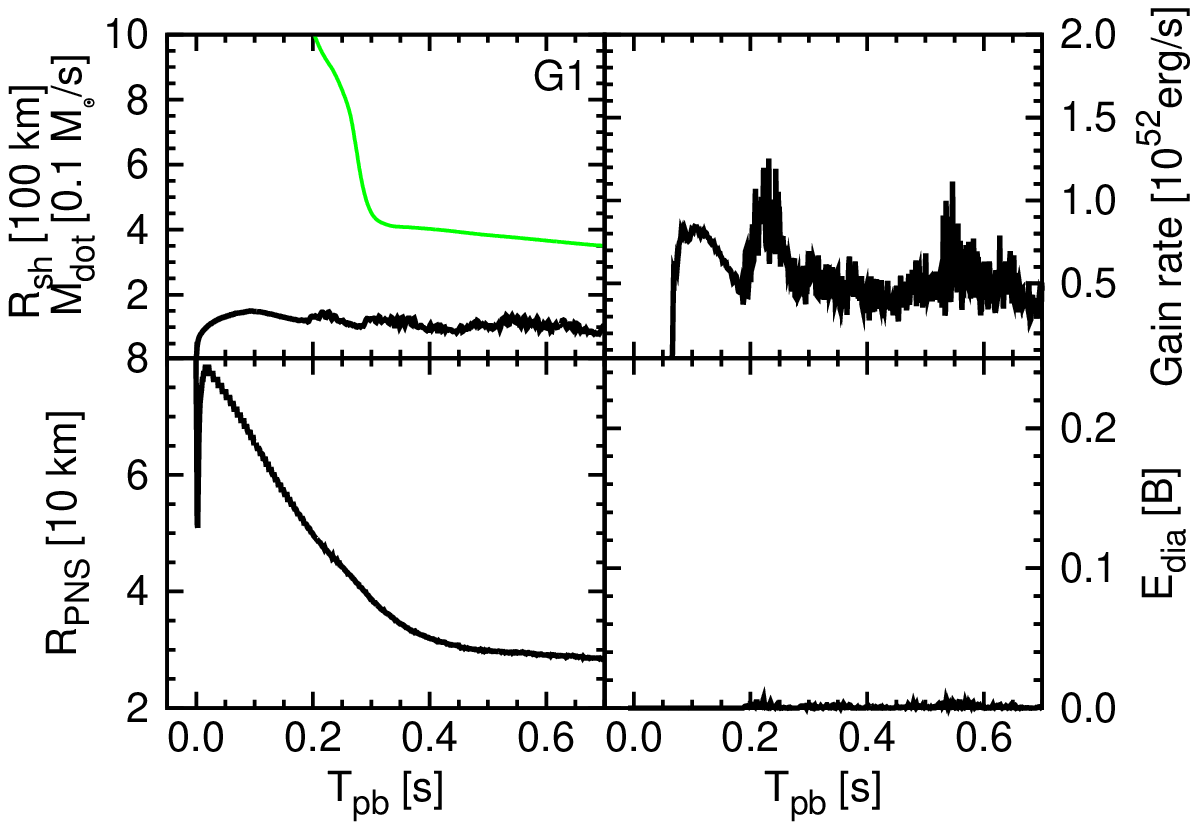}
\includegraphics[width=81mm]{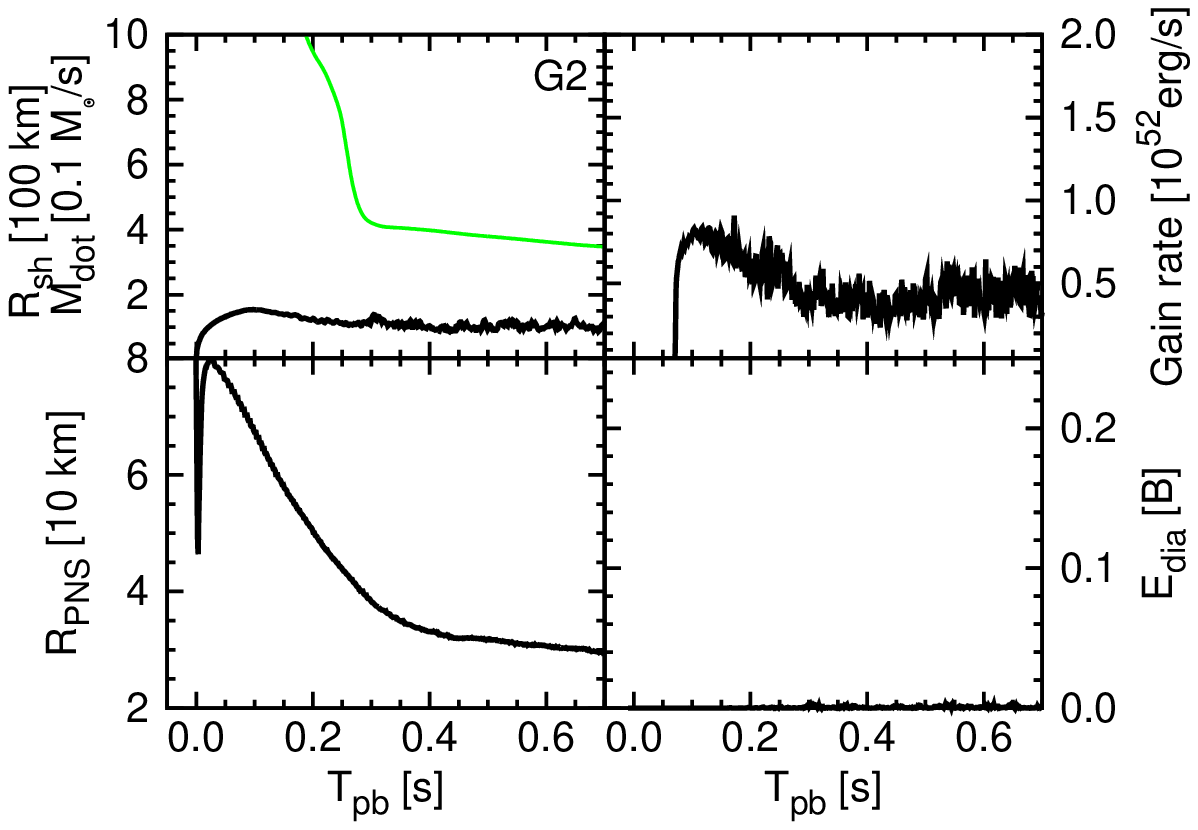}
\includegraphics[width=81mm]{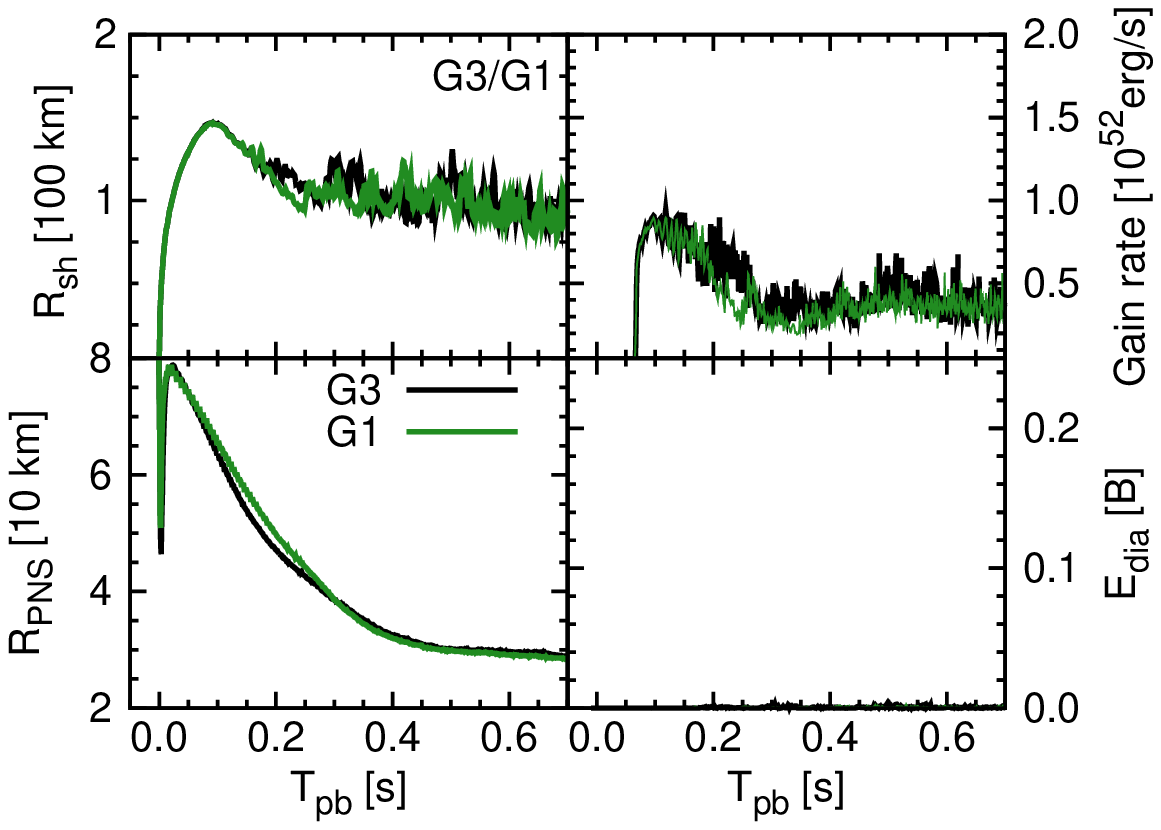}
\includegraphics[width=81mm]{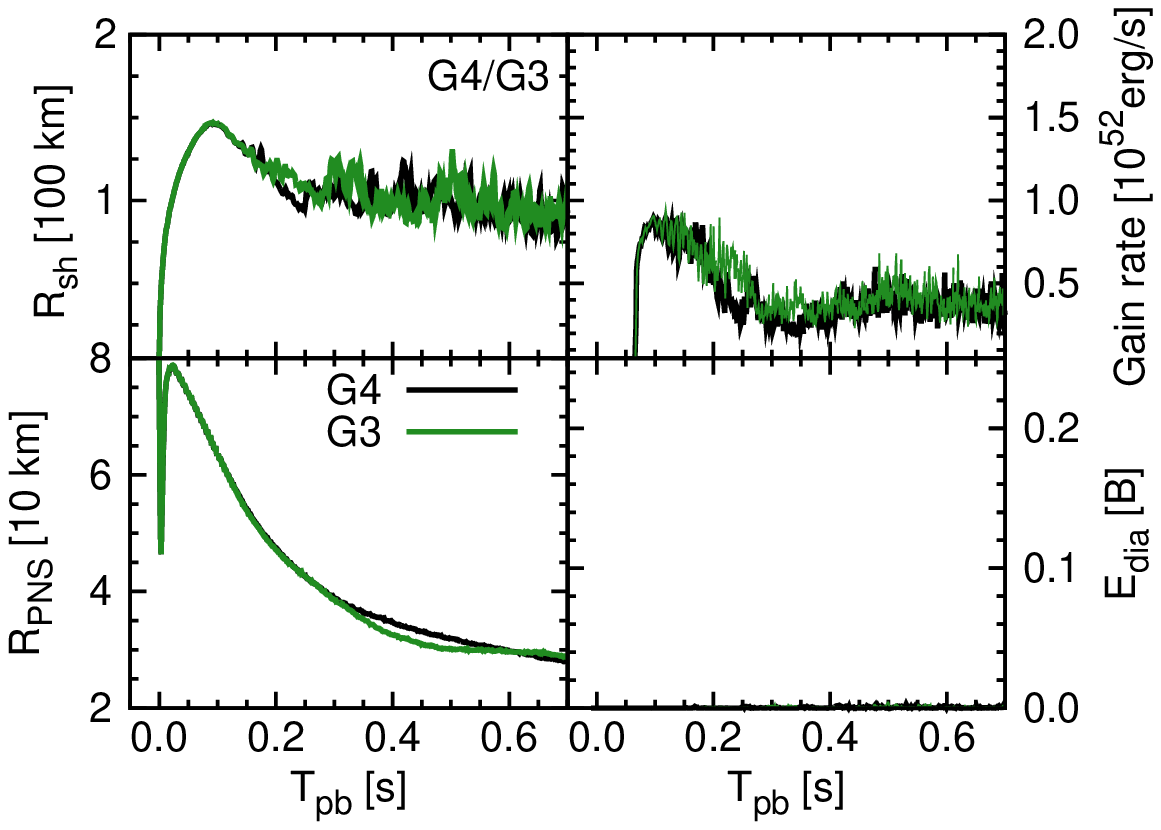}
\includegraphics[width=81mm]{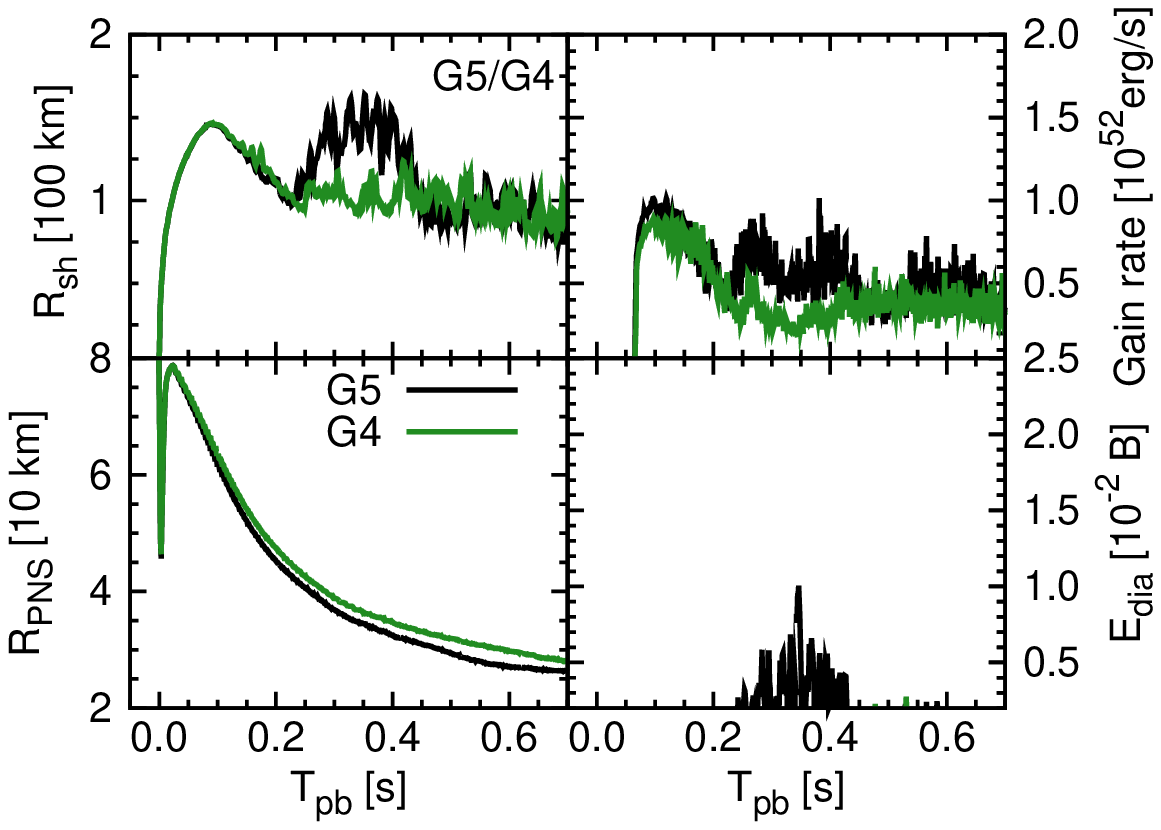}
\includegraphics[width=81mm]{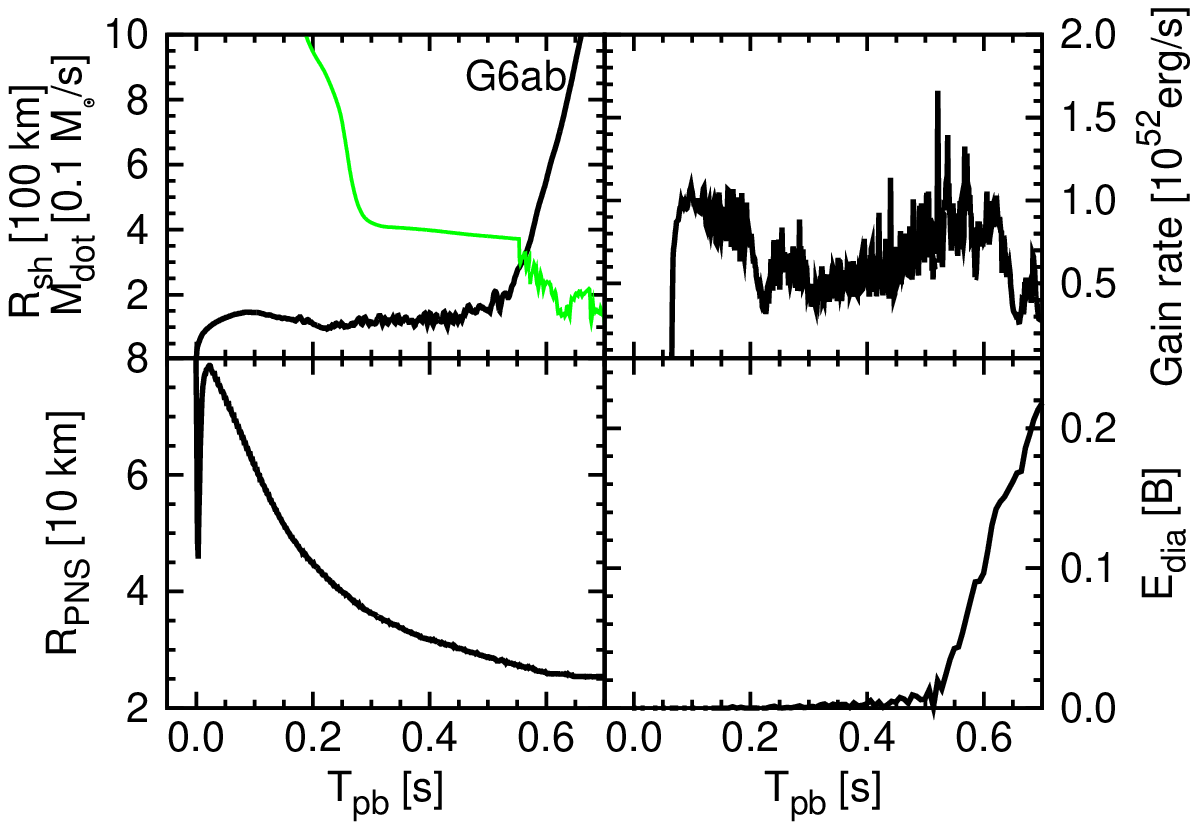}
\includegraphics[width=81mm]{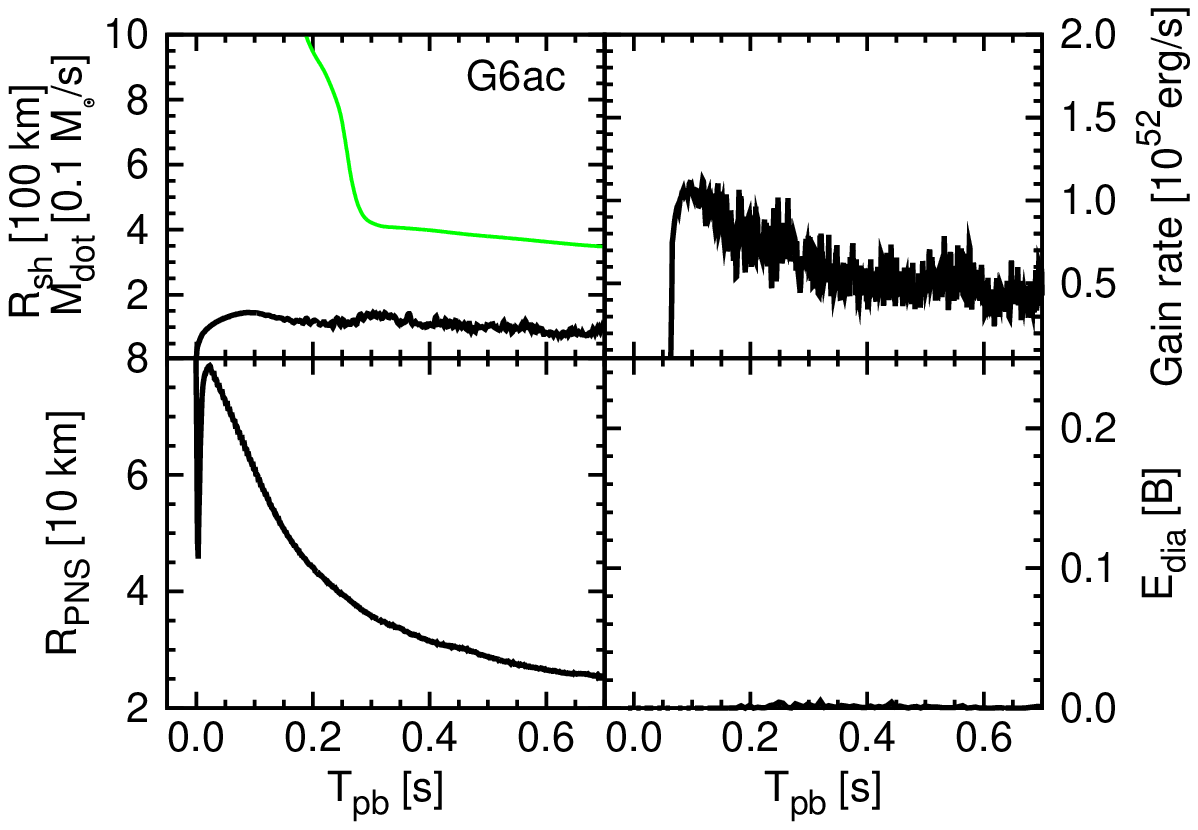}
\includegraphics[width=81mm]{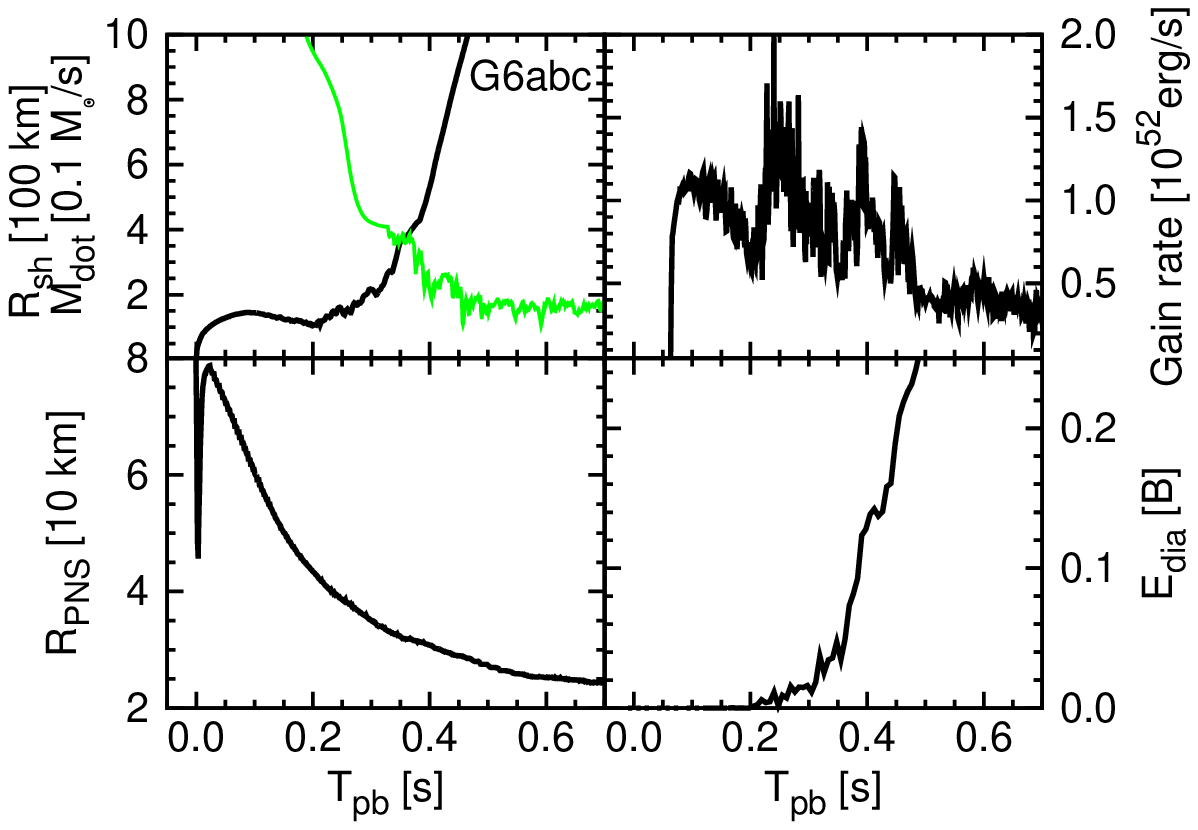}
\caption{Summary of our 2D models (see text).}
\label{f17}
\end{center}
\end{figure}

\begin{figure}[H]
\begin{center}
\includegraphics[width=124mm]{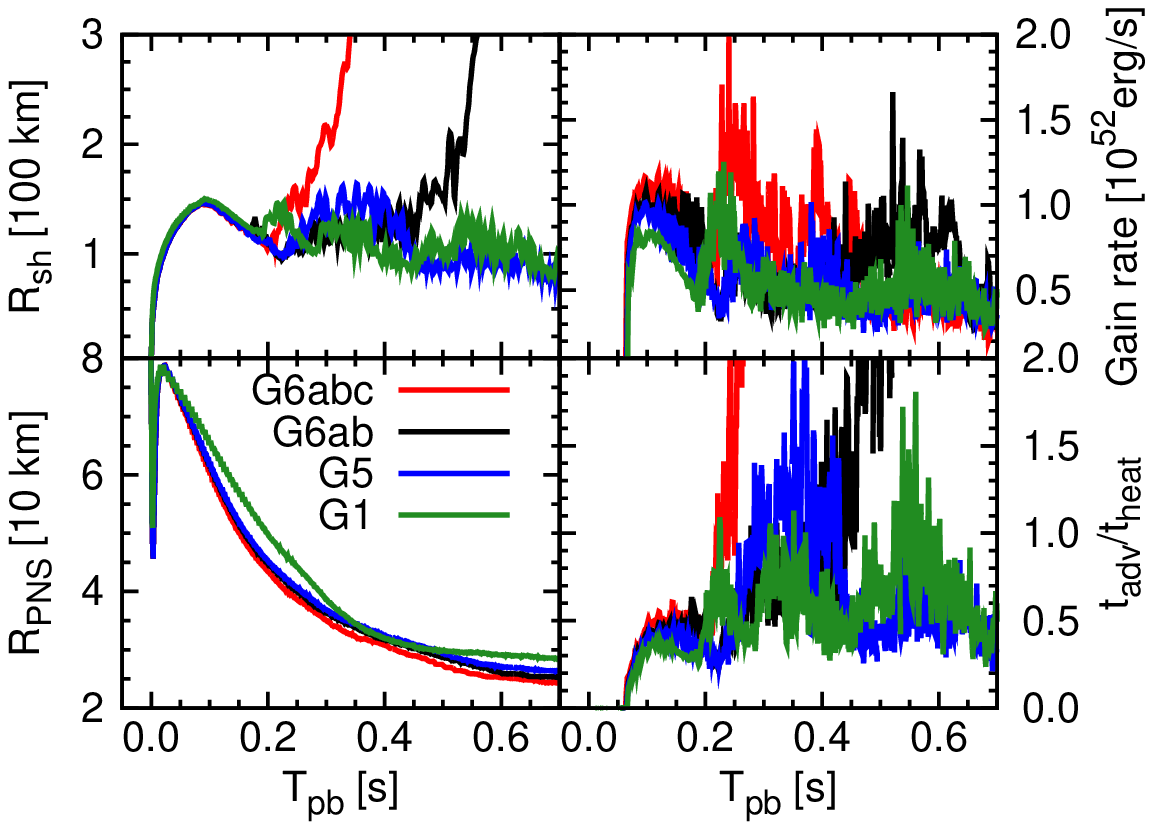}
\includegraphics[width=124mm]{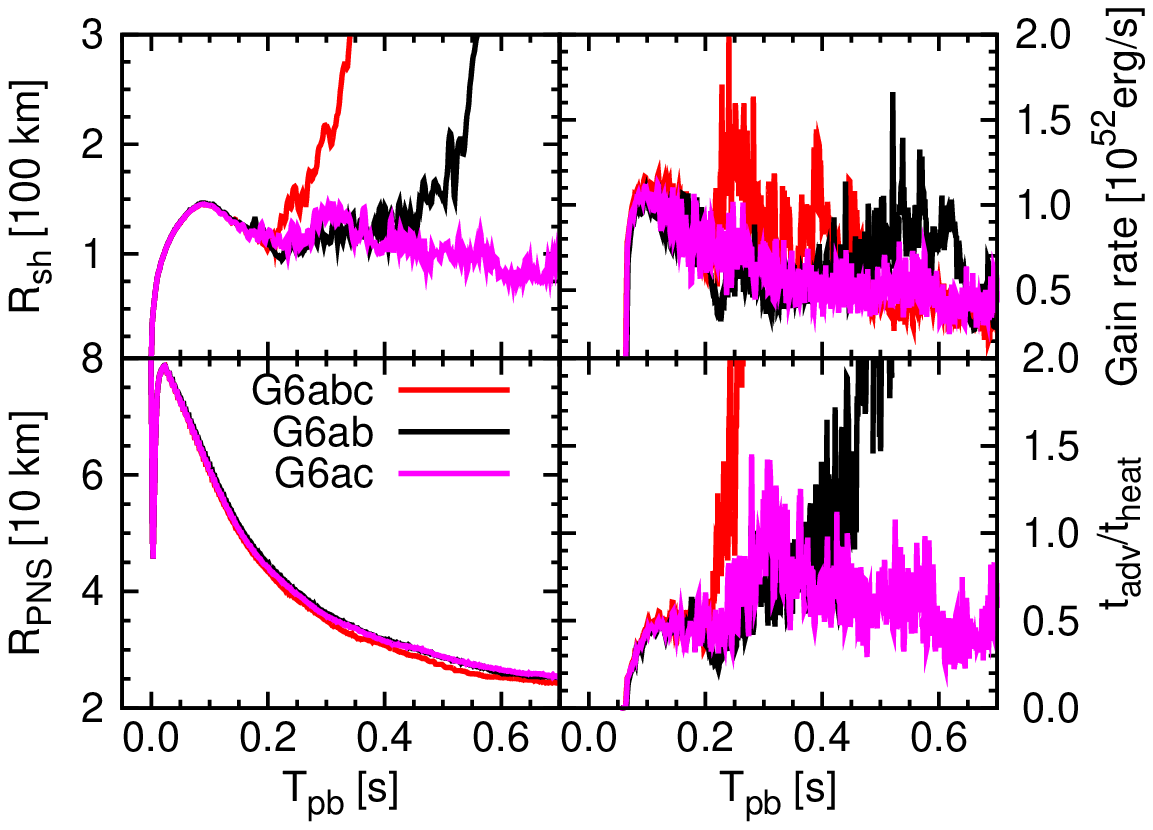}
\caption{Same as Figure \ref{f17} but for the comparison between the 2D exploding and 
 non-exploding models (top panel) and the comparison between models G6ab, G6ac, and G6abc
 (bottom panel). Note that the bottom right panel shows "$t_{\rm adv}/t_{\rm heat}$" which denotes the ratio of advection to heating timescale in the gain region \citep{Buras06a}.}
\label{f18}
\end{center}
\end{figure}

\begin{figure}[H]
\begin{center}
\includegraphics[width=81.5mm]{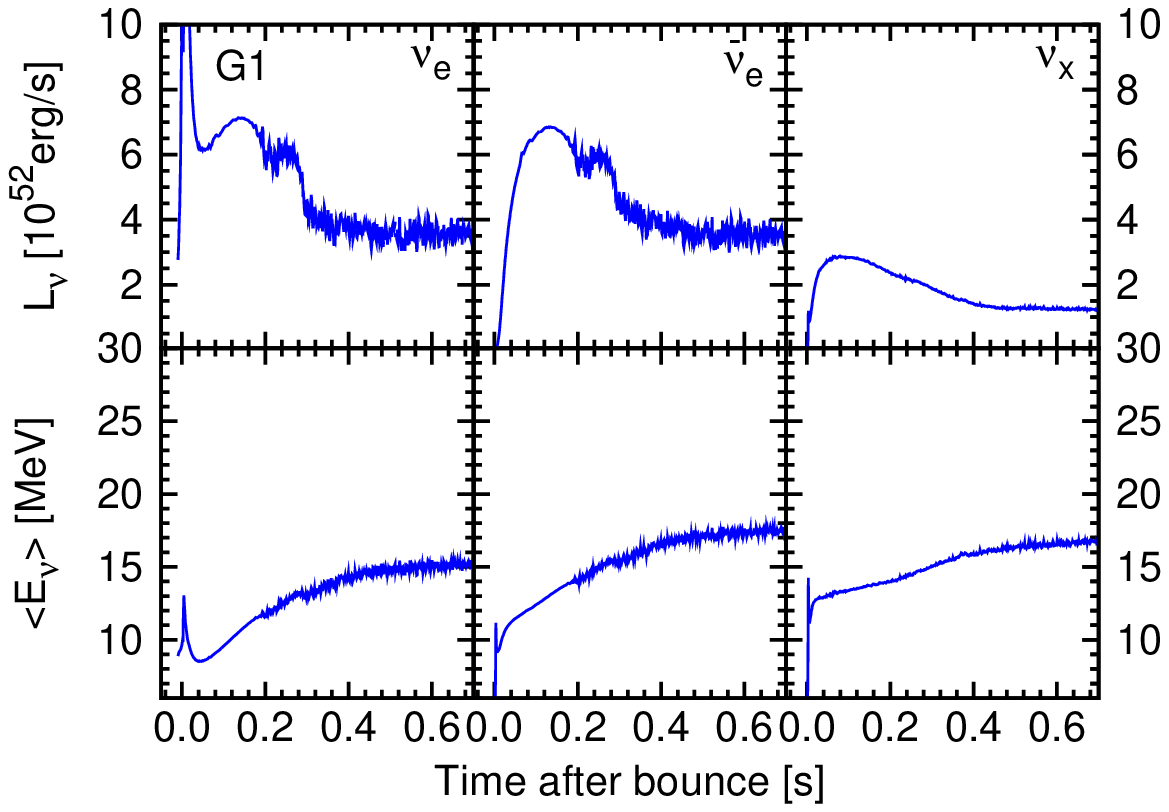}
\includegraphics[width=81.5mm]{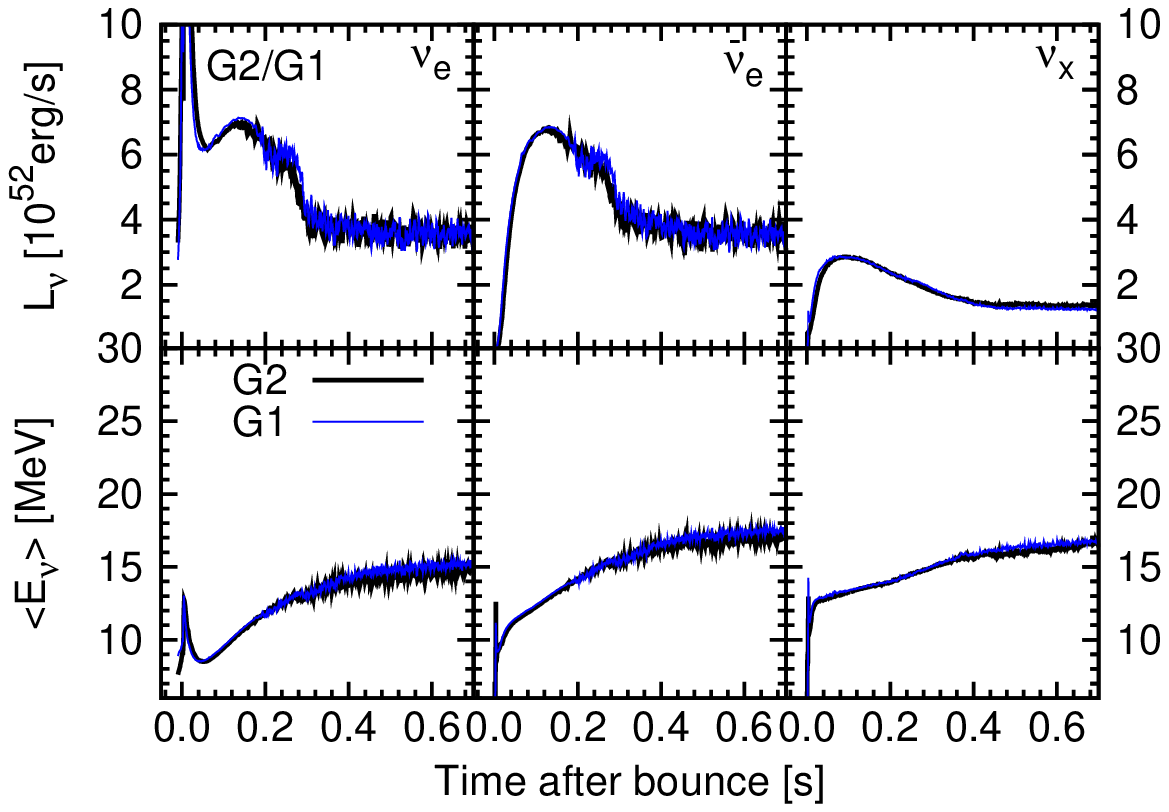}
\includegraphics[width=81.5mm]{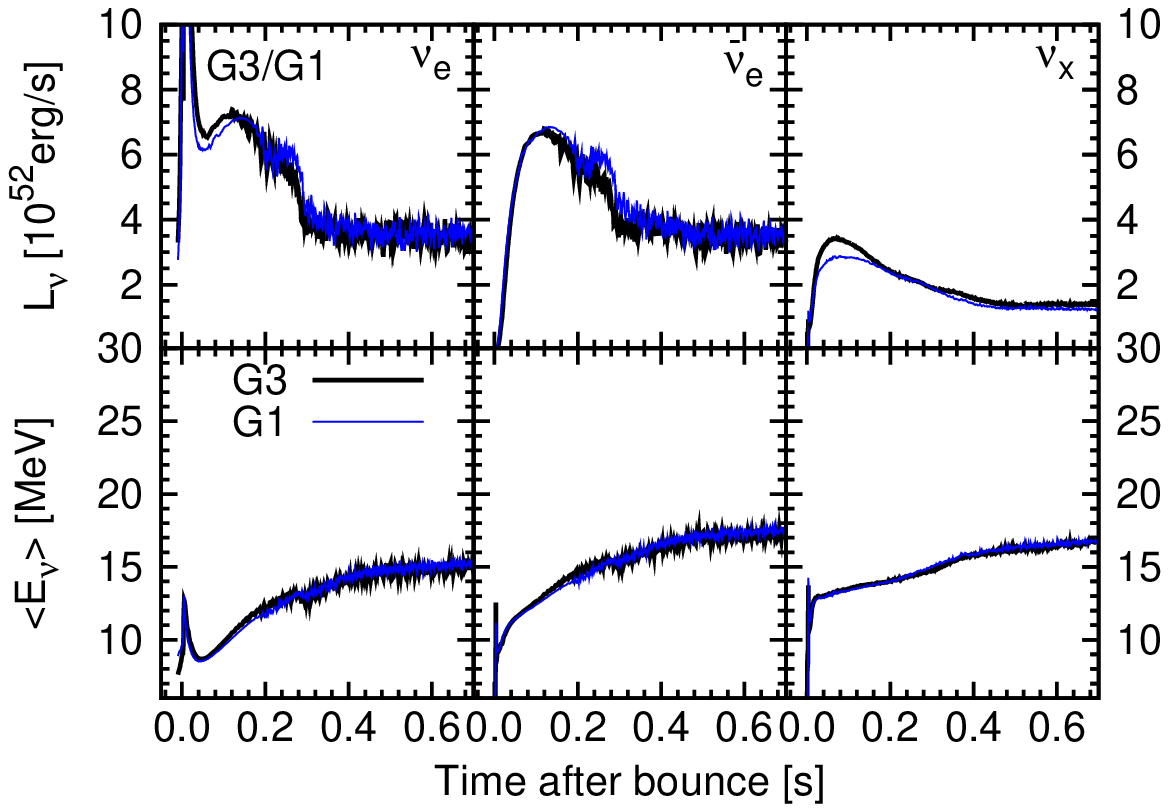}
\includegraphics[width=81.5mm]{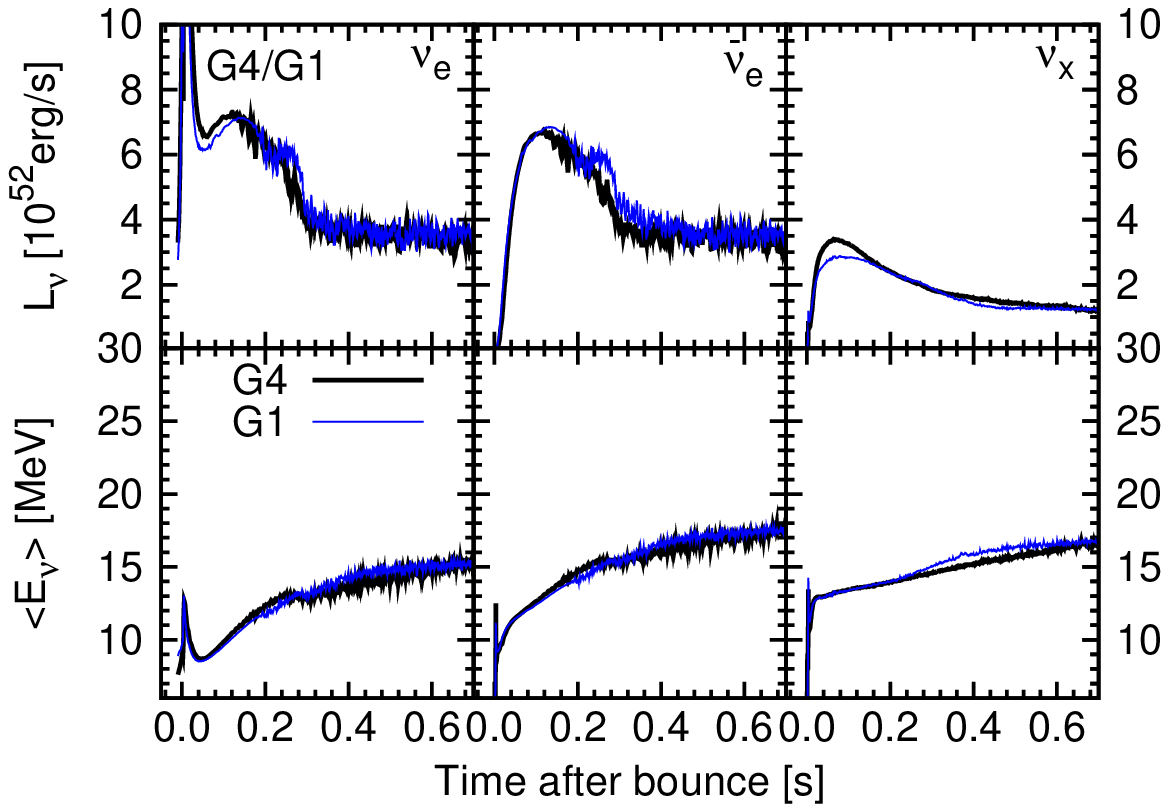}
\includegraphics[width=81.5mm]{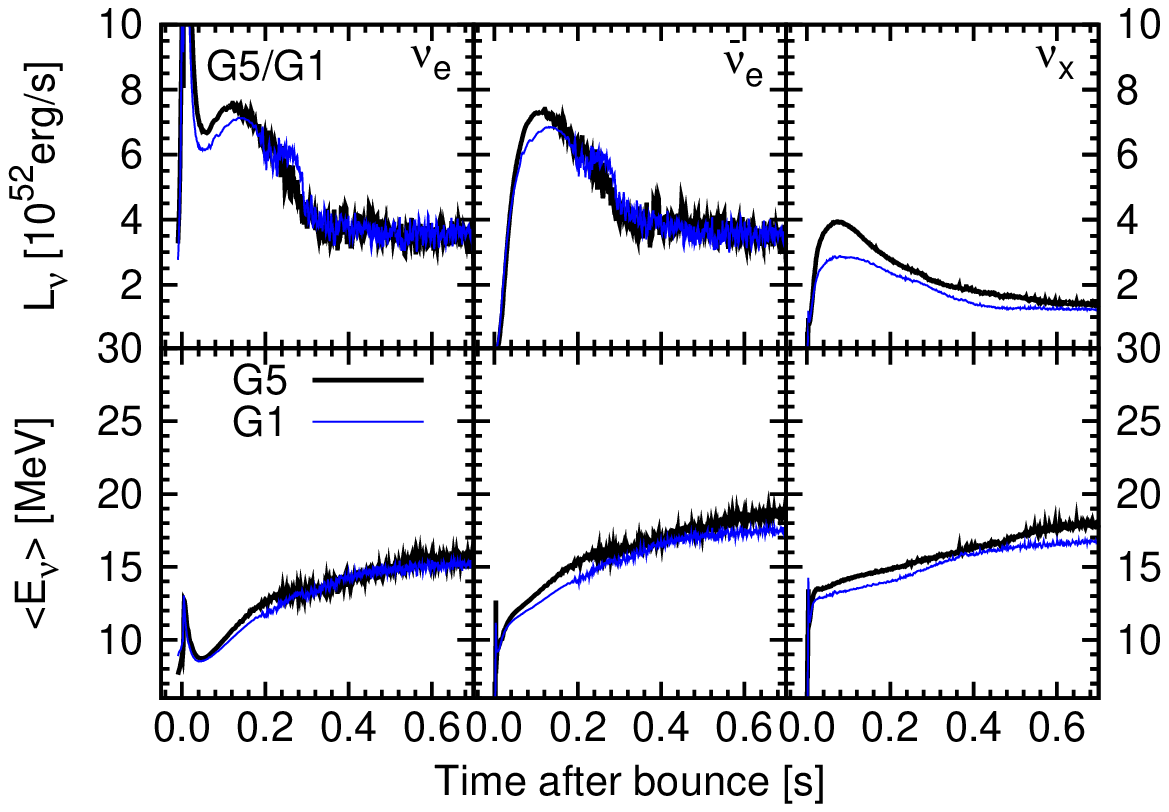}
\includegraphics[width=81.5mm]{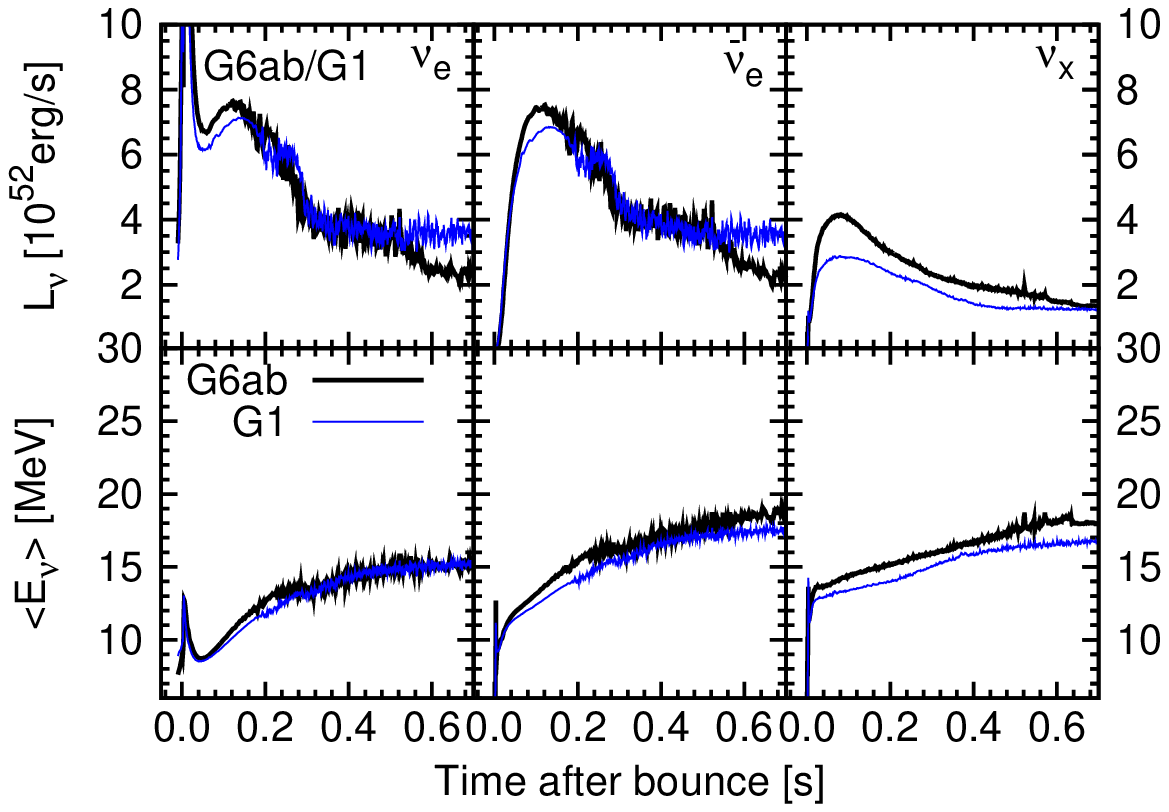}
\includegraphics[width=81.5mm]{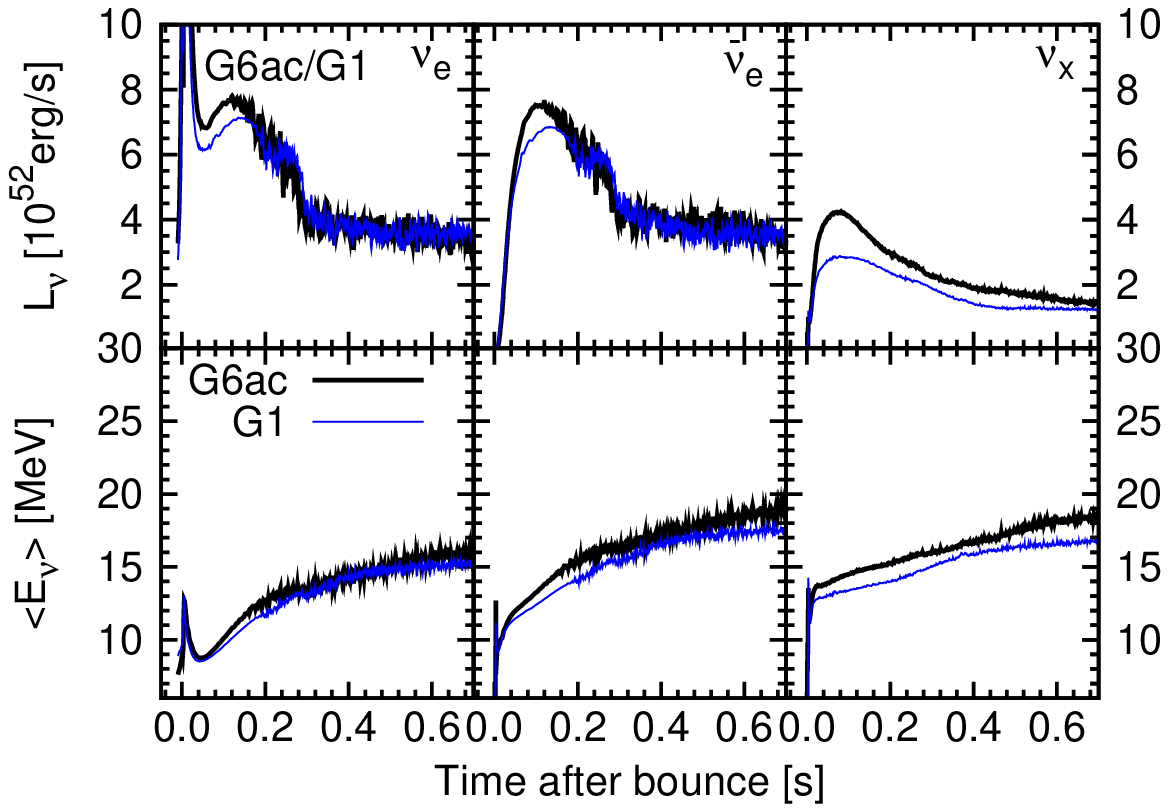}
\includegraphics[width=81.5mm]{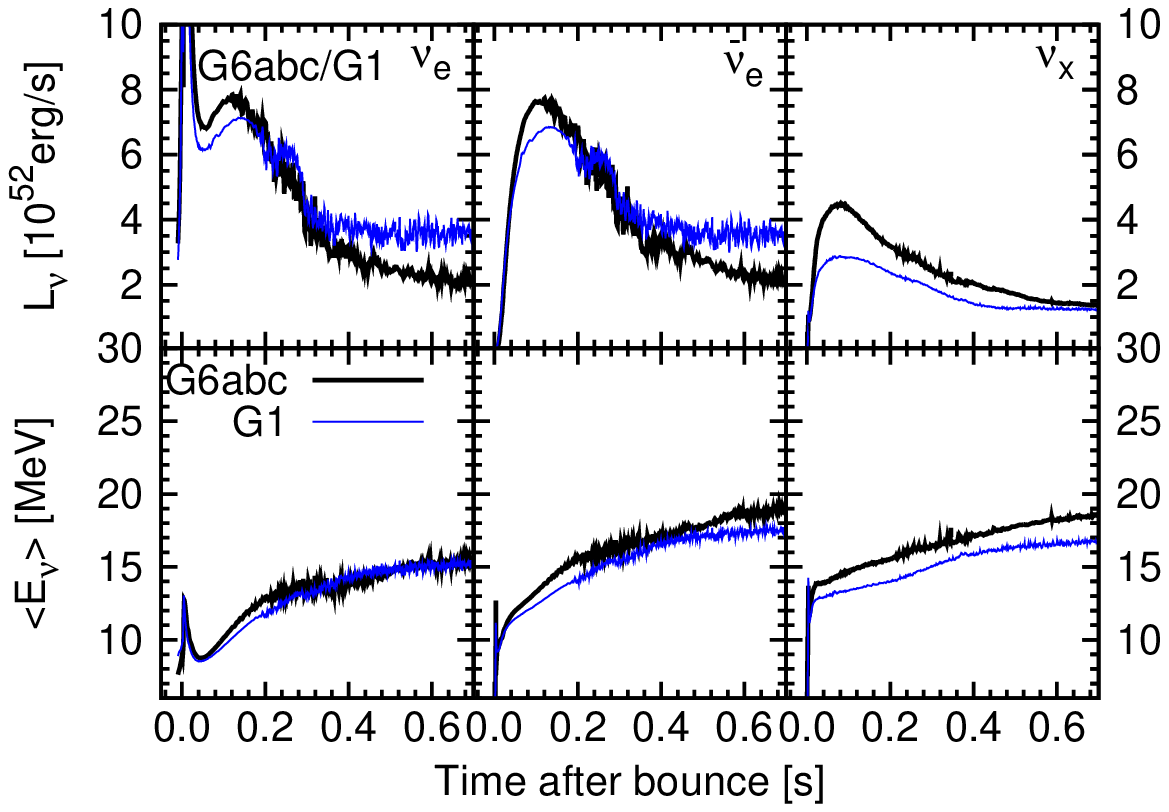}
\vspace{-1cm}
\caption{Comparison of neutrino luminosities (upper panels) and the rms neutrino energies (lower panels) for all the 2D models. }
\label{f19}
\end{center}
\end{figure}

Figures \ref{f17}, \ref{f18}, and \ref{f19} show several key quantities useful 
for our 2D model comparison. In each model in Figure \ref{f17}, 
the top left panel shows time evolution of 
the average shock radius (black line) with the mass accretion rate ($M_{\rm dot}$) 
at 500 km (green line),
 the top right panel shows the net heating rate in the gain region, the bottom
 left panel is the PNS radius, 
 and the bottom right panel is the diagnostic explosion energy. 
The model name is indicated in the upper right part in the top left panel.

Top two panels of Figure \ref{f17} compare G1 (left panel) and G2 (right panel), where the 
 Juodagalvis (electron-capture) rate is implemented in G2 instead of the Bruenn prescription of
 G1. These two panels are almost identical with respect to the shock evolution ({\it top left}), 
PNS contraction ({\it bottom left}), and the non-explodability 
($E_{\rm dia} \sim 0$, {\it bottom
 right}). Here $E_{\rm dia}$ denotes the diagnostic (explosion) energy that 
is calculated following the literature (e.g., \citet{buras06b,Suwa10,Bruenn13}).
Note also from the comparison of the neutrino luminosities 
 and rms energies, any clear differences between G2 (black line) and G1 (blue line) cannot be 
 seen (top panels of Figure \ref{f19}). The only exception is the reduction of the net heating 
rate in the gain region for G2 (Figure \ref{f17}) compared to G1. The reduction of the 
 net heating rate for G2 is more apparent
 in the accretion phase, namely before $\sim 300$ ms postbounce. Note that the timescale
 ($\sim 300$ ms postbounce) coincides with the sudden drop in the mass accretion rate 
(green line).
 Our 1D comparison between set2 and set1 
 (e.g., the bottom panel of Figure \ref{f5}) suggests that the improved electron capture 
 rate on heavy nuclei lowers the explodability. Using the different progenitor model 
(note again the use of $20 M_{\odot}$ in 2D and $15 M_{\odot}$ in 1D), our results show 
 that this feature still remains in 2D.

 Three panels in the second and third raw of Figure \ref{f17} show more 
 detailed comparison between G3 and G1 (labeled with G3/G1), G4 and G3 (with G4/G3), 
G5 and G4 (with G5/G4), respectively. Regarding the shock evolution (second raws), 
the average shock radii of G3 and G4 (black lines) show no remarkable difference 
from G1 (green line). This is in line with the lack of significant change in the net
 heating rate ({\it top right}) relative to G1. As expected in 1D,  the inclusion
 of the nupair reaction (Section \ref{nupair}) makes the PNS radius\footnote{Note that the PNS radius is estimated at a fiducial density 
of $10^{11}\,{\rm g}\,{\rm cm}^{-3}$.} more compact
 compared to G1 (see panel labeled with  G3/G1). The PNS radius of G4 somehow becomes 
slightly bigger between $\sim$ 375 - 600 ms compared to G3, however comes closer to G3
 thereafter. From Figure \ref{f19}, the neutrino luminosities and rms energies show 
 no clear discrepancies between G3 and G4 
compared to those already observed in the corresponding 1D models.
 
More big change can be seen in the shock evolution of G5 (black line, see panel with G5/G4
 in Figure \ref{f17}).
 The shock of G5 (black line) starts to expand at $\sim 214$ ms after bounce
(see the hump in the shock evolution), maximally reaching at $\sim 163$ km, but returns to 
 closely match with the shock trajectory of G4 (green line) after 400 ms
 postbounce (see also Figure \ref{f16}). In fact, the rms energies of $\nu_e$ 
and $\bar\nu_e$ are higher for G5 than G1 (see the panel with G5/G1 in Figure \ref{f19}) and 
the net heating rate (top panel of Figure \ref{f18}) 
becomes clearly larger for G5 (blue line) compared to G1 (green line) during the transient
 shock expansion phase. This is reconciled with the (slightly) higher heating rate 
due to the weak magnetism and recoil effect (as we saw in our 1D model, set5a). 
The higher heating rate is also fingerprinted in the diagnostic explosion energy 
(black line, see the panel with G5/G4 in Figure \ref{f17}).

Using the same progenitor, LS220 EOS, and the 
similar set of neutrino opacities, our G5 run is close to model s20-2007 of \citet{Summa16}.
Their s20-2007 start to explode $\sim 200 - 300$ ms after bounce, whereas our G5 does not.
 For a quantitative discussion, we choose to compare the neutrino luminosities and the rms 
energies at 100 ms after bounce, which closely corresponds to the peak of the luminosities 
(e.g., their Figure 3). For the progenitor, the luminosity of $\nu_e$, $\bar\nu_e$, and 
 $\nu_x$ is $\sim$7.0, 6.5, and 4.2 $\times 10^{52}$erg/s for s20-2007, and 
$7.5$, $7.5$, and $4.0$ times $\times 10^{52}$erg/s for G5, respectively. The rms energy 
  of $\nu_e$, $\bar\nu_e$, and $\nu_x$ is $\sim$12.2, 14.4, and 16.2 MeV for s20-2007, and 
 $11.0$, $13.8$, and $14.9$ $\times 10^{52}$erg/s for G5, respectively. 
 The lower $\nu_e$ and $\bar \nu_e$  rms energies would explain more difficult explosion
  for G5 compared to s20-2007 of \citet{Summa16}.
As already mentioned, 
 our neglect of the non-elastic effects in the charged current reactions, the simplified 
 transport schemes could explain such $\sim 10 \%$ level of discrepancies.
 We cannot unambiguously identify which of the missing sophistication in
 this work could explain the above difference. 
 This apparently manifests that implementation of detailed neutrino opacities and 
accurate treatment of neutrino transport as well as GR are mandatory for quantitative studies of the CCSN mechanism.

 G6ab shows a shock revival at $\sim 500$ ms after bounce (Figure \ref{f16}), 
which we can see also from the clear deviation of the diagnostic energy from zero 
(see the panel labeled by G6ab in Figure \ref{f17}). On the other hand, G6ac is not 
 exploding during the simulation time (Figure \ref{f16}).
 The neutrino luminosities and rms energies show fast time variations in the 
 accretion phase (Figure \ref{f19}), it is not easy to clearly see the increase for G6ab
 relative to G5. 

 In order to understand the above trend, we show
the ratio of advection to heating timescale in the gain region \citep{Buras06a} 
(the bottom panel in Figure \ref{f18} ({\it bottom right})).
 From the panel, one can see that the strangeness contribution 
(magenta line, G6ac) does enhance the chance of explosion at 
around 300 ms postbounce, which can be seen as peaks of the ratio 
("$t_{\rm adv}$/$t_{\rm heat}$")
exceeding unity. In fact, the shock radius (the top left panel) and 
  the net heating rate in the gain region (the top right panel)
 becomes slightly bigger for G6ac (magenta lines) comparing to those of G6ab 
(black lines) at the same time.

 The enhanced chance of explosion in models with the strangeness correction (G6ac and 
G6abc) originates from the reduction of the neutrino opacity, which is most significant 
in the accretion phase. This was already seen in our 1D runs of the $15 M_{\odot}$ star 
(see the green line in the 
right panel of Figure \ref{f14} in the accretion phase). 
Our 2D results show that the inclusion of only the strangeness correction (relative
 to the standard rates) is not sufficient 
to trigger the onset of explosion for the 20 $M_{\odot}$ star. As one can see from the red line 
(G6abc, the bottom panel in Figure 18), our 2D run demonstrates that 
the combination of the strangeness and 
 the many-body correction makes the onset of the explosion easier. 

 As already mentioned in our 1D comparison (Section \ref{set6}), 
 the many-body correction of \citet{horowitz17} is expected to enhance 
the explodability primarily after the accretion phase because the many-body 
corretion reduces the opacity at high densities. Our 2D results are in line with 
the 1D expectation. The net heating rate in the gain region is higher and the PNS radius 
 is slightly smaller for G6ab comparing to those of G5 (and G1) in the post
 accretion phase (e.g., top panel of Figure \ref{f18}). 
These results demonstrate that the many-body correction mainly impacts the 
explodability also in 2D after the accretion phase. 
Using the same EOS (LS220) 
and the same progenitor, model s20.0-LS220 in 
\citet{Bollig17} that includes the many-body correction in addition to
 their standard neutrino opacities leads to explosion after 
 $\sim$ 400 ms postbounce (see, sky-blue line in the top right panel of their Figure 1),
 whereas the corresponding model further including the strangeness correction 
($g^s_a = - 0.1$) leads to more earlier explosion at $\sim$ 200 ms postbounce
(e.g., blue line in their Figure 1). These features are basically consistent with our 
2D runs.

Finally, G6abc (Figure \ref{f16})
 shows the earliest runaway shock expansion
(starting at $\sim 200$ ms after bounce) among our 2D models. As already seen in 
 set6c (Section \ref{set6}), the strangeness effects contribute to enhance the 
$\nu_e$ and $\bar \nu_e$ luminosity before the accretion phase ends 
($\sim 300$ ms after bounce). Quantitatively, the peak of the accretion luminosity of 
 $\nu_e$ and $\bar \nu_e$ is enhanced by $\sim 3.8 \%$ and $\sim 1.3 \%$ for G6abc compared
 to G6ab (e.g., Figure \ref{f19}).
 The slightly enhanced heating rate in the accretion phase due to the 
strangeness correction works synergetically with the many-body correction
 to revive the stalled shock into explosion for the 20 $M_{\odot}$ star 
(e.g., Figure \ref{f18}). The diagnostic 
energy when the shock reaches at 1000 km is 0.24 B and 0.2 B with B representing 
"Bethe" = $10^{51}$ erg.
 To get the saturated value, long-term simulations are needed, which is beyond 
the scope of this work. Using the same progenitor and the LS220 EOS, the onset time of 
 an explosion ($\sim 200$ ms after bounce) is close to that seen in the 2D model of 
 \citet{Bollig17} with the strangeness contribution ($g^s_a = - 0.1$). However, 
 the match may be simply incidental because the contributions from muons are not yet
 included in this work.



\section{Conclusions and Discussion} \label{sec6}
In this study, we have explored impact of updated
 neutrino opacities in CCSN simulations where spectral neutrino transport is 
solved by the three-flavor IDSA scheme. To verify our code,
 we first presented 1D results following core-collapse, bounce, and 
up to $\sim 250 $ ms postbounce of a $15 M_{\odot}$ star using the standard
set of neutrino opacities by \citet{Bruenn85} and made a comparison with the 
seminal work by \citet{Liebendorfer05}. A good agreement of the code comparison 
supports the reliability of our three-flavor IDSA scheme with 
the standard opacity set. Then we investigated in 1D runs
 how the individual updated rate
 could lead to the difference from the base-line run with the 
 standard opacity set. By making a detailed comparison with 
 previous literature, we 
have checked the validity of our each implementation in a step-by-step manner.
As previously identified, we have confirmed that adding up the individual rates impacts 
non-linearly the 
 neutrino luminosities and energies.
 In our 2D runs,  we implemented selected sets of the neutrino opacities because 
a full investigation of 
the individual rates is currently too computationally expensive to do
 even in 2D with our improved IDSA scheme.
 Regarding the explodability, our results showed that several expectations 
from the individual update in 1D are indeed correct in 2D. 
 Among the updates considered in this work, 
 the inclusion of both the strangeness-dependent contribution and the many-body correction
 to the neutrino-nucleon scattering has the largest impact in enhancing the explodability 
in our 2D models.

Using the same progenitor, the same EOS, and the similar set of 
 the neutrino opacities, there are $\sim 10 \%$ levels of discrepancies in the 
 neutrino luminosities and the rms energies between our results and the results
 from the codes with more accurate neutrino transport schemes 
 (e.g., {\tt Agile-BOLTZTRAN} and {\tt Vertex}).
 Our neglect of energy-bin/flavor coupling in the transport equation, 
 non-isoenergetic effects in the charged current reactions, and 
the partial implementation of the Doppler-shift terms could solve the 
 mismatch. Moreover, our approximate GR treatment as well as in the neutrino 
 transport should be also improved. In this respect, the microphysical
 update we have done in this work is nothing but among the first steps
 toward more sophisticated CCSN modeling.

\acknowledgements
We are thankful to M. Hempel for providing a table for calculating 
the nucleon potential difference ($\Delta U$) for LS220 EOS.
KK is full-heartedly thankful to H.T. Janka, E, M\"uller, R. Bollig, A. Lohs,
 T. Foglizzo, T. Kuroda, E. Abdikamalov, and R. Kazeroni for stimulating discussions
during his six-month stay in Max Planck Institute for Astrophysics in 2017
 that was supported by JSPS KAKENHI Grant Number JP15KK0173.
TF acknowledges support from the Polish National Science Center (NCN) 
under grant number UMO-2016/23/B/ST2/00720.
GMP acknowledges partial
support by the Deutsche Forschungsgemeinschaft through grant
SFB~1245 ("Nuclei: From Fundamental Interactions to Structure 
and Stars").
Numerical computations were carried out in part on 
XC30 and general common use computer system at the center for 
Computational Astrophysics, CfCA, 
the National Astronomical Observatory of Japan, and also on XC40 at
 YITP at Kyoto University.
This study was also supported by JSPS KAKENHI Grant Number
 (JP15H00789, JP15H01039, JP17H01130, JP17H06364), and by the Central Research Institute of 
Fukuoka University (Nos.171042, 177103), and JICFuS as a priority issue to be tackled by using
 the Post `K' Computer.

\appendix
\label{app_nupair}
\section{Implementing electron neutrino pair annihilation} 
\label{appA}

\subsection{$\nu_e + \bar\nu_e \rightarrow \nu_x +\bar\nu_x$: evolution equation of $\nu_x$}
\label{appA1}
Following \citet{Buras03}, the scattering kernels of the electron-neutrino pair-annihilation can be calculated essentially in the same way as electron-positron annihilation, $e^- + e^+ \rightarrow \nu_x +\bar\nu_x$. It is convenient to define the $\nu_x$ pair-production kernels labelled (p),
\begin{equation}
 \mathcal{R}_{\nu_x\bar\nu_x}^{\rm p}(\cos\theta_{\nu_x\bar\nu_x},E_{\nu_x}+E_{\bar\nu_x})
=
\int\frac{d^3p_{\nu_e}}{(2\pi\hbar)^3}\frac{d^3p_{\bar\nu_e}}{(2\pi\hbar)^3}
2 f_{\nu_e}(p_{\nu_e}) 2 f_{\bar\nu_e}(p_{\bar\nu_e}) 
\left\vert \mathcal{M} \right\vert^2
\delta^4(p_{\nu_e}+p_{\bar\nu_e}-p_{\nu_x}-p_{\bar\nu_x})~,
\label{eq:pair-kernel}
\end{equation}
with the initial-particle's distribution functions $f_{\nu_e}$ and $f_{\bar\nu_e}$, for which we assume local thermodynamic equilibrium, i.e. $\mu_{\nu_e}=\mu_e-(\mu_n-\mu_p)$. The pair-production kernel~\eqref{eq:pair-kernel} depends on the incident scattering angle $\theta_{\nu_x\bar\nu_x}$ between $\nu_x$ and $\bar\nu_x$ \citep[for the definition, cf.,][]{tony93a} as well as on the sum of $\nu_x$ and $\bar\nu_x$ energies, $E_{\nu_x}$ and $E_{\bar\nu_x}$ respectively. Moreover, the spin-averaged and squared matrix element, $\left\vert \mathcal{M} \right\vert^2$, is obtained from $e^--e^+$-annihilation \citep[cf.][]{Bruenn85} with the following replacements for the weak coupling constants, $C_V=C_A=+1/2$. Since within the IDSA no explicit angle-dependence of weak processes is employed, we perform a Legendre expansion of the pair-production kernel~\eqref{eq:pair-kernel} in terms of $\cos\theta$,
\begin{equation}
\mathcal{R}_{\nu\bar\nu}^{\rm p}(\cos\theta_{\nu\bar\nu},E_{\nu}+E_{\bar\nu})
\longrightarrow
\frac{1}{2} \Phi_{0,\nu\bar\nu}^{\rm p}(E_{\nu}+E_{\bar\nu}) + \mathcal{O}(\cos\theta)~,
\end{equation}
such that the corresponding collision term for the the zeroth component of the distribution function, $f_{\nu_x}^{(0)}$, reads as follows,
\begin{eqnarray}
\left.\frac{\partial f_{\nu_x}^{(0)}(E_{\nu_x})}{c\partial t}\right\vert_{\rm coll}
&=&
\frac{2\pi}{c(2 \pi\hbar c)^3}
\left (1 - f_{\nu_x}^{(0)}(E_{\nu_x}) \right )
\int E_{\bar\nu_x}^2 d E_{\bar\nu_x} 
\left (1 - f_{\bar\nu_x}^{(0)}(E_{\bar\nu_x}) \right )
\Phi_{0,\nu_x\bar\nu_x}^{\rm p}(E_{\nu_x}+E_{\bar\nu_x}) \nonumber
\\
&-&
\frac{2\pi}{c(2 \pi\hbar c)^3}
f_{\nu_x}^{(0)}(E_{\nu_x})
\int E_{\bar\nu_x}^2 d E_{\bar\nu_x} 
 f_{\bar\nu_x}^{(0)}(E_{\bar\nu_x})
\Phi_{0,\nu_x\bar\nu_x}^{\rm a}(E_{\nu_x}+E_{\bar\nu_x})~,
\label{eq:nupair1}
\end{eqnarray} 
where $\Phi_{0,\nu_x\bar\nu_x}^{\rm a}$ denotes the zeroth-order Legendre coefficient of the $\nu_x$-pair absorption kernel. It is related to $\Phi_{0,\nu_x\bar\nu_x}^{\rm p}$ via the relation of detailed balance,
\begin{equation}
\Phi_{0,\nu_x\bar\nu_x}^{\rm a}(E_{\nu_x}+E_{\bar\nu_x}) = \exp\left\{\frac{E_{\nu_x}+E_{\bar\nu_x}}{T}\right\} \Phi_{0,\nu_x\bar\nu_x}^{\rm p}(E_{\nu_x}+E_{\bar\nu_x})~,
\label{eq:db}
\end{equation}
which is realized straight forward in a similar fashion as expression~\eqref{eq:pair-kernel}. Now, Eq.~\eqref{eq:nupair1} can be rewritten as follows,
\begin{equation}
\left.\frac{\partial f_{\nu_x}^{(0)}(E_{\nu_x})}{c\partial t}\right\vert_{\rm coll}
=
C^0_{\rm nupair}(E_{\nu_x}) +  A^0_{\rm nupair}(E_{\nu_x}) \,f_{\nu_x}^{(0)}(E_{\nu_x})~,
\label{eq:nupair2}
\end{equation} 
with
\begin{eqnarray}
C^0_{\rm nupair}(E_{\nu_x})
&=&
\frac{2\pi}{c(2 \pi\hbar c)^3}
\int E_{\bar\nu_x}^2 dE_{\bar\nu_x} \left (1 - f_{\bar\nu_x}^{(0)}(E_{\bar\nu_x})\right ) \Phi_{0,\nu_x\bar\nu_x}^{\rm p}(E_{\nu_x}+E_{\bar\nu_x})~,
\label{eq:nupair3}
\\
A^0_{\rm nupair}(E_{\nu_x})
&=&
-\frac{2\pi}{c(2 \pi\hbar c)^3}
\int E_{\bar\nu_x}^2 d E_{\bar\nu_x} 
\left[
\left (1 - f_{\bar\nu_x}^{(0)}(E_{\bar\nu_x})\right )
\Phi_{0,\nu_x\bar\nu_x}^{\rm p}(E_{\nu_x}+E_{\bar\nu_x})
\right.
\nonumber
\\
&& \;\,+
\left.
f_{\bar\nu_x}^{(0)}(E_{\bar\nu_x}) \Phi_{0,\nu_x\bar\nu_x}^{\rm a}(E_{\nu_x}+E_{\bar\nu_x})
\right]~,
\label{eq:nupair4}
\end{eqnarray} 
which denote the production Eq.~\eqref{eq:nupair3} and annihilation rates Eq.~\eqref{eq:nupair4} of this process, respectively. In the IDSA, the terms $C^0_{\rm nupair}$ and $A^0_{\rm nupair}$ are added to the Eqs.~\eqref{eq1} and \eqref{eq2} for the streaming neutrinos and to Eq.~(15) of \citet{idsa} for the trapped neutrinos.

\subsection{ $\nu_e + \bar\nu_e \rightarrow \nu_x +\bar\nu_x$: evolution equations of $\nu_e$ and $\bar{\nu}_e$}
\label{appA2}
The collision term for $f_{\nu_e}^0$ associated with this process is obtained in a similar fashion as for the production of $\nu_x$-pairs \ref{appA1},
\begin{eqnarray}
\left.\frac{\partial f_{\nu_e}^{(0)}(E_{\nu_e})}{c\partial t}\right\vert_{\rm coll}
&=&
\frac{2\pi}{c(2 \pi\hbar c)^3}
\left (1 - f_{\nu_e}^{(0)}(E_{\nu_e}) \right )
\int E_{\bar\nu_e}^2 d E_{\bar\nu_e} 
\left (1 - f_{\bar\nu_e}^{(0)}(E_{\bar\nu_e}) \right )
\Phi_{0,\nu_e\bar\nu_e}^{\rm p}(E_{\nu_e}+E_{\bar\nu_e}) \nonumber
\\
&-&
\frac{2\pi}{c(2 \pi\hbar c)^3}
f_{\nu_e}^{(0)}(E_{\nu_e})
\int E_{\bar\nu_e}^2 d E_{\bar\nu_e} 
 f_{\bar\nu_e}^{(0)}(E_{\bar\nu_e})
\Phi_{0,\nu_e\bar\nu_e}^{\rm a}(E_{\nu_e}+E_{\bar\nu_e})~.
\label{eq:nupair5}
\end{eqnarray} 
Note that the expression for the collision integral for $\bar\nu_e$ is obtained equivalently by replacing the labels $\nu_e\leftrightarrow\bar\nu_e$ in above expression. Moreover, due to the following symmetry considerations,
\begin{equation}
\Phi_{0,\nu_e\bar\nu_e}^{\rm p}=\Phi_{0,\nu_x\bar\nu_x}^{\rm a}~,
\;\;\;\;\;
\Phi_{0,\nu_e\bar\nu_e}^{\rm a}=\Phi_{0,\nu_x\bar\nu_x}^{\rm p}~,
\end{equation}
expression~\eqref{eq:nupair5} can be reduced to a similarly simple form as expression~\eqref{eq:nupair2}, with equivalent definitions for $C^0$ and $A^0$. Due to the relation of detailed balance~\eqref{eq:db}, which holds here as well, it becomes clear that it is necessary to obtain only one pair-reaction kernel, e.g., $\phi_{0,\nu_x\bar\nu_x}^{\rm p}$. Therefore, we follow Eq.~(C62)--(C74) in \citet{Bruenn85} for the computation presented in this work. Note that also here we assume that $\nu_x$ obey local thermodynamic equilibrium, i.e. $\mu_{\nu_x} = 0$ \citep[this was also assumed in][]{Buras03,tobias09}. For practical reasons we monitor the $\nu_e$-distribution function which must not differ by more than 10\% from the corresponding Fermi-Dirac distribution. This treatment was tested to work well in {\tt Agile-BOLTZTRAN} simulations by \citet{tobias09}.

\subsection{Integrated pair production rates}
\label{appA3}
Here we provide definitions of quantities shown in the main part of the present paper. Therefore, the $(\nu_x,\bar\nu_x)$ pair production rate is defined as follows,
\begin{eqnarray}
Q_{\nu_x\bar\nu_x}(E_{\nu_x})
&=&
\frac{1}{(2 \pi\hbar c)^3} \int E_{\bar\nu_x}^2 d E_{\bar\nu_x} \left(1 - f_{\bar\nu_x}^{(0)}(E_{\bar\nu_x})\right) 2\pi \Phi_{0,\nu_x\bar\nu_x}^{\rm p}(E_{\nu_x}+E_{\bar\nu_x})
\;\;\;\;\;\;
[{\rm s}^{-1}]~,
\end{eqnarray}
such that the total number production rate of $\nu_x$ is given as follows,
\begin{eqnarray}
\frac{\partial n_{\nu_x}}{\partial t}
&=&
\frac{1}{(2 \pi\hbar c)^3} \frac{1}{n_{\nu_x}} \int E_{\nu_x}^2 d E_{\nu_x} \left(1 - f_{\nu_x}^{(0)}(E_{\nu_x})\right) 
{4\pi}Q_{\nu_x\bar\nu_x}(E_{\nu_x})~,
\end{eqnarray}
normalized to the number density of neutrinos $n_{\nu_x}$. Then, we obtain the number production spectra, 
\begin{eqnarray}
\frac{\partial^2 n_{\nu_x}}{\partial E_{\nu_x}\partial t}
&=&
\frac{1}{(2 \pi\hbar c)^3} E_{\nu_x}^2 \left(1 - f_{\nu_x}^{(0)}(E_{\nu_x})\right) 4\pi Q_{\nu_x\bar\nu_x}(E_{\nu_x})
\;\;\;\;\;\;
[{\rm s}^{-1}~{\rm MeV}^{-1}~{\rm cm}^{-3}]~,
\label{number_nupair}
\end{eqnarray}
which is shown in the left panel of Fig.~\ref{f6} for some selected conditions, and the energy production spectra,
\begin{eqnarray}
\frac{\partial^2 \varepsilon_{\nu_x}}{\partial E_{\nu_x}\partial t}
&=&
\frac{1}{(2 \pi\hbar c)^3} E_{\nu_x}^3 \left(1 - f_{\nu_x}^{(0)}(E_{\nu_x})\right) 4\pi Q_{\nu_x\bar\nu_x}(E_{\nu_x})
\;\,\;\;\;\;
[{\rm s}^{-1}~{\rm cm}^{-3}]~,
\label{energy_nupair}
\end{eqnarray}
which is shown in the right panel of Fig.~\ref{f6}.

\section{$\nu_x + \nu_e (\bar\nu_e) \leftrightarrows \nu'_e (\bar\nu'_e) +\nu'_x$}
\label{appB}
For the calculation of the scattering kernel we follow the same procedure as for neutrino-electron(positron) scattering \citep{Buras03},
\begin{equation}
 \mathcal{R}_{\nu_x}^{\rm in}(\cos\theta_{\nu_x\nu'_x},E_{\nu_x}-E_{\nu'_x})
=
\int\frac{d^3p_{\nu_e}}{(2\pi\hbar)^3}\frac{d^3p_{\nu'_e}}{(2\pi\hbar)^3}
2 f_{\nu'_e}(p_{\nu'_e}) \left(1- f_{\nu_e}(p_{\nu_e}) \right)
\left\vert \mathcal{M} \right\vert^2
\delta^4(p_{\nu_x}+p_{\nu_e}-p_{\nu'_x}-p_{\nu'_x})~,
\label{eq:pair-kernel}
\end{equation}
which is evaluated under the assumption of local thermodynamic equilibrium, exactly as for the neutrino-pair processes in appendix~\ref{appA}. Here the spin-averaged and squared matrix element, $\left\vert \mathcal{M} \right\vert^2$, is obtained from neutrino-electron scattering \citep[cf.][]{Bruenn85} with the following replacements for the weak coupling constants, $C_V=C_A=+1/2$ for $\nu_x$-scattering on $\nu_e$ and $C_V - 1$ and $C_A - 1$ for $\nu_x$-scattering on $\bar\nu_e$. With the Legendre expansion of the scattering kernel in terms of $\cos\theta_{\nu\nu'}$, we obtain the zeroth-order term of the collision integral for $f_{\nu_x}^0$ as follows, 
\begin{eqnarray}
\frac{\partial f_{\nu_x}^{(0)}(E_{\nu_x})}{\partial t}\Big|_{\rm coll}
&=&
\frac{2 \pi}{c(2 \pi\hbar c)^3}
\left (1 - f_{\nu_x}^{(0)}(E_{\nu_x}) \right )
\int E_{\nu'_x}^2 d E_{\nu'_x} 
 f_{\nu'_x}^{(0)}(E_{\nu'_x}) \Phi_{0,\nu_x}^{\rm in}(E_{\nu_x}-E_{\nu'_x})
\nonumber
\\
&-&
\frac{2 \pi}{c(2 \pi\hbar c)^3}
f_{\nu_x}^{(0)}(E_{\nu_x})
\int E_{\nu'_x}^2 d E_{\nu'_x} 
 (1 - f_{\nu'_x}^{(0)})(E_{\nu'_x})
\Phi_{0,\nu_x}^{\rm out}(E_{\nu_x}-E_{\nu'_x})~,
\label{nupair6}
\end{eqnarray}
where the out-scattering kernel, $\Phi_{0,\nu_x}^{\rm out}$, is obtained via the relation of detailed balance,
\begin{equation}
\Phi_{0,\nu_x}^{\rm out}(E_{\nu_x}-E_{\nu'_x}) = \exp\left\{\frac{E_{\nu_x}-E_{\nu'_x}}{T}\right\} \Phi_{0,\nu_x}^{\rm in}(E_{\nu_x}-E_{\nu'_x})~.
\label{eq:db}
\end{equation}
Similarly as Equations \eqref{eq:nupair3} and \eqref{eq:nupair4}, one can obtain $C^0_{\rm NNS}$ and $A^0_{\rm NNS}$ for the neutrino-neutrino scattering (NNS) processes here. These are added to the evolution equation of the streaming and trapped neutrinos, accordingly. Note that the expression for $\nu_x$-scattering on $\bar\nu_e$ is obtained equivalently by replacing the labels $\nu_e\leftrightarrow\bar\nu_e$ in the above expressions. For the calculations of the scattering kernels, we follow Eq.~(C50) in \citet{Bruenn85} for the computation presented in this work. Then, the inverse mean-free path $1/\lambda_{\rm mfp}$ shown in Fig.~\ref{f8} is obtained as follows,
\begin{eqnarray}
A_{\rm NNS, \nu_x}^0(E_{\nu_x}) 
&=& 
-\frac{2\pi}{c(2\pi\hbar c)^3}
\int E_{\nu_x'}^2 d E_{\nu_x'} 
\left(
f_{\nu_x'}^{(0)}(E_{\nu_x'})
\Phi_{0,\nu_x}^{\rm in}(E_{\nu_x}-E_{\nu_x'})
\right.
\nonumber
\\
&&
\left.+
\left( 1- f_{\nu_x'}^{(0)}(E_{\nu_x'}) \right )
\Phi_{0,\nu_x}^{\rm out}(E_{\nu_x}-E_{\nu_x'})
\right)
\label{IMFP}
\end{eqnarray}
where $1/\lambda_{\rm mfp}=A_{\rm NNS, \nu_x}^0$ in units of [s$^{-1}$], which holds also for the other processes \citep[cf., Equation~(139) of][]{KurodaT16}.


\bibliographystyle{apj}
\bibliography{mybib}
\end{document}